\documentclass[sigconf, nonacm]{acmart}

\newcommand\vldbdoi{10.14778/3611479.3611492}
\newcommand\vldbpages{2845 - 2857}
\newcommand\vldbvolume{16}
\newcommand\vldbissue{11}
\newcommand\vldbyear{2023}
\newcommand\vldbauthors{\authors}
\newcommand\vldbtitle{\shorttitle} 
\newcommand\vldbavailabilityurl{https://github.com/decisionbranches/decisionbranches}
\newcommand\vldbpagestyle{empty} 

\usepackage{import,transparent}
\usepackage{nicefrac}
\usepackage{makecell}
\usepackage{algorithm}
\usepackage{algorithmic}
\usepackage{subfigure}
\usepackage{xspace}
\usepackage{pgfplots}
\usepackage{pgfplotstable}
\usepackage{booktabs}
\usepackage{multirow}
\usepackage{wrapfig}
\usepackage{paralist}
\usepackage{balance}
\usepackage[aboveskip=1ex, belowskip=-1ex]{caption}
\usetikzlibrary{patterns}
\tikzset{every mark/.append style={mark size=1pt}}
\usetikzlibrary{pgfplots.colorbrewer}
\usepgfplotslibrary{groupplots}

\pgfplotsset{height=3.5cm, 
    label style={font=\footnotesize},
    tick label style={font=\footnotesize}, 
    title style={font=\footnotesize, yshift=-1.2ex},
    ylabel shift=-1ex,
    xlabel shift=-0.8ex,
    axis lines*=left,
    grid=major,
    grid style={line width=.1pt, draw=gray!20},
    legend style={
        font = \scriptsize,
        fill = white!90, 
        draw opacity = 1,
        text opacity = 1,
        very thin,
        legend columns = 3,
        draw=none, % remove legend border
    },
    compat=newest}

\pgfplotsset{
    barPlotStyle/.style={
        enlarge y limits = 0,
        enlarge x limits =.1,
        x tick label style={rotate=45, anchor=east},
	    cycle list/.style={
            colormap/Set1,
            cycle list/Set1,
        },
        every axis plot/.append style={fill},
        label style={font=\footnotesize},
    }
}

\pgfplotsset{
    linePlotStyle/.style={
    every axis plot/.append style={line width=1pt},
    legend style = {legend columns = -1},
    colormap/Paired,
    cycle list/Paired,
    cycle multiindex* list={
        mark list*\nextlist
        Paired\nextlist
        linestyles\nextlist
    },
    label style={font=\footnotesize},
}}

\pgfplotsset{
    linePlotStyleSet1/.style={
    every axis plot/.append style={line width=1pt},
    legend style = {legend columns = -1},
    colormap/Set1,
    cycle list/Set1,
    cycle multiindex* list={
        mark list*\nextlist
        Set1\nextlist
        linestyles\nextlist
    },
    label style={font=\footnotesize},
}}
\usepackage{fontawesome5}
\usepackage{tikz}
\usetikzlibrary{shapes, shapes.geometric, shapes.arrows, arrows, patterns, positioning, calc, arrows.meta, bending}
\usetikzlibrary{mindmap,trees}
\definecolor{mygreen}{rgb}{0.553,0.682,0.063}
\definecolor{myblue}{rgb}{0.0,0.208,0.376}
\definecolor{mygray}{rgb}{0.906,0.906,0.906}

\definecolor{darkblue}{rgb}{0.0,0.0,0.8}
\definecolor{mydarkgreen}{rgb}{0.0,0.5,0.0}
\definecolor{tablehighlight}{rgb}{0.0,0.5,0.0}

\newcommand*{\mycolorbox}[1]{%
\tikzstyle{mybox} = [draw=black, rectangle, inner sep=1pt, inner ysep=2pt, inner xsep=2pt, fill=white]
\tikzstyle{fancytitle} = [fill=white, text=black]
\begin{tikzpicture}
\node [mybox] (box){%
     #1
};
\end{tikzpicture}%
}

\newcommand{\ie}{i.\nolinebreak[4]\hspace{0.01em}\nolinebreak[4]e.\@\xspace}
\newcommand{\eg}{e.\nolinebreak[4]\hspace{0.01em}\nolinebreak[4]g.\@\xspace}

\newcommand{\Reals}{{\mathbb R}}
\newcommand{\RealsInfinity}{{\overline{\mathbb R}}}

\newcommand{\tdim}{d}
\newcommand{\tsize}{n}
\newcommand{\testsize}{N}
\newcommand{\indexStructuresQueryPoints}{q}
\newcommand{\numberfeaturetested}{\mu}

\newcommand{\tdimsubsets}{D}
\newcommand{\dtbox}{B}
\newcommand{\dtnumberbounded}{n_b}
\newcommand{\dtnumberunbounded}{n_u}

\newcommand{\decisionbranchbox}{B}
\newcommand{\decisionbranchmodel}{\mathcal{B}}
\newcommand{\decisionbranchset}{\mathcal{T}}
\newcommand{\featuresubset}{F}
\newcommand{\featuresubsetsequence}{F_s}
\newcommand{\featuresubsets}{\mathcal{F}}
\newcommand{\tsizesubsets}{k}
\newcommand{\numberfeaturesubsetstest}{p}
\newcommand{\trainset}{T}
\newcommand{\trainsetnonrare}{T_0}
\newcommand{\trainsetrare}{T_1}
\newcommand{\trainsetnonrareremoved}{R_0}
\newcommand{\trainsetrareremoved}{R_1}
\newcommand{\trainsetsubset}{S}
\newcommand{\greedygain}{g}
\newcommand{\greedytrainsubset}{S}
\newcommand{\greedytrainsubsetsize}{m}
\newcommand{\greedymaxpoints}{p_m}
\newcommand{\ndbensemble}{M}
\newcommand{\predfunc}{h}

\newcommand{\sizedecisionbranchset}{l}

\newcommand{\forestNTrees}{M}

\renewcommand{\vec}[1]{\mathbf{#1}}
\def\O{{\mathcal{O}}}

\newcommand{\Yspace}{\mathcal{Y}}

\renewcommand{\vec}[1]{\mathbf{#1}}
\newcommand{\nclasses}{K}

\newcommand{\kdtree}{\emph{k}-d tree\@\xspace}
\newcommand{\kdtrees}{\emph{k}-d trees\@\xspace}

\newcommand{\decisionbranch}{DBranch\@\xspace}
\newcommand{\dbensemble}{DBEns\@\xspace}
\newcommand{\decisiontree}{DTree\@\xspace}
\newcommand{\randomforest}{RForest\@\xspace}
\newcommand{\extratrees}{ExTrees\@\xspace}
\newcommand{\nnb}{NNB\@\xspace}

\newcommand{\tdecisionbranch}{decision branch\@\xspace}
\newcommand{\tdecisionbranches}{decision branches\@\xspace}

\newcommand{\multiplicationfactor}{\tau}

\newcommand{\fscore}{F_1}
\newcommand{\trainingtime}{$T_{train}$\@\xspace}
\newcommand{\querytime}{$T_{query}$\@\xspace}
\newcommand{\totaltime}{$T_{total}$\@\xspace}

\newcommand{\dbb}{\decisionbranch}
\newcommand{\dbbts}{\decisionbranch\textsubscript{[\textit{T\textsubscript{s}}]}}
\newcommand{\dbbta}{\decisionbranch\textsubscript{[\textit{T\textsubscript{a}}]}}
\newcommand{\esb}{\dbensemble}
\newcommand{\esbts}{\dbensemble\textsubscript{[T\textsubscript{s}]}}
\newcommand{\esbta}{\dbensemble\textsubscript{[T\textsubscript{a}]}}

\newcommand{\precision}{{\mathcal{P}}}
\newcommand{\recall}{{\mathcal{R}}}

\newcommand{\knearestneighbors}{\mathcal{K}}
\begin{document}
%\title{Fast Search-By-Classification for Large-Scale Databases}
\title{Fast Search-By-Classification for Large-Scale Databases Using Index-Aware Decision Trees and Random Forests}

%%
%% The "author" command and its associated commands are used to define the authors and their affiliations.
\author{Christian Lülf}
\affiliation{%
  \institution{University of Münster}
  %\streetaddress{P.O. Box 1212}
  \city{Münster}
  %\state{Ohio}
  \country{Germany}
  \postcode{48153}
}
\email{christian.luelf@uni-muenster.de}

\author{Denis Mayr Lima Martins}
\affiliation{%
  \institution{University of Münster}
  %\streetaddress{P.O. Box 1212}
  \city{Münster}
  %\state{Ohio}
  \country{Germany}
  \postcode{48153}}
\email{denis.martins@uni-muenster.de}

\author{Marcos Antonio Vaz Salles}
\affiliation{%
  \institution{Independent Researcher}
  %\streetaddress{P.O. Box 1212}
  %\city{Copenhagen}
  %\state{Ohio}
  \country{Portugal}
}
\email{msalles@acm.org}
\authornote{Work was primarily performed while the author was at the University of Copenhagen.}

\author{Yongluan Zhou}
\affiliation{%
  \institution{University of Copenhagen}
  %\streetaddress{P.O. Box 1212}
  \city{Copenhagen}
  %\state{Ohio}
  \country{Denmark}
}
\email{zhou@di.ku.dk}

\author{Fabian Gieseke}
%  \affiliation{%
%   \institution{University of Copenhagen}
%   \city{Copenhagen}
%   \country{Denmark}
% }
\affiliation{%
  \institution{University of Münster}
  \city{Münster}
  \country{Germany}
  \postcode{48153}}
\email{fabian.gieseke@uni-muenster.de}

%%
%% The abstract is a short summary of the work to be presented in the
%% article.
\begin{abstract}
%With the increasing volume of data collected in many fields such as remote sensing or astronomy, a common scenario in modern data analysis is that users aim at finding special target objects in such massive data collections.
The vast amounts of data collected in various domains 
%volume of data collected in many fields such as remote sensing or astronomy, where the data gathered by satellites and telescopes amount to several petabytes, 
pose great challenges to modern data exploration and analysis. To find ``interesting'' objects in large databases, 
users typically define a query using positive and negative example objects and train a classification model to identify the objects of interest in the entire data catalog. However, this approach requires a scan of all the data to apply the classification model to each instance in the data catalog, making this method prohibitively expensive to be employed in large-scale databases serving many users and queries interactively. %, even given powerful computers.
%Instead, such search queries are usually addressed by means of nearest neighbor search or machine learning models. The former approach allows to quickly return a set of similar items, but might yield undesired results. The latter approach, in turn, allows to accurately specify the desired rare objects via a corresponding classification data set, but requires scanning all the data, which can easily take hours even given powerful computers.
%In this work, we combine the benefits of both search strategies.
%by providing a novel construction scheme that outputs ``index-aware'' classification models that allow to transfer the application phase of the model to a set of database operations that can be efficiently implemented.
%and to apply an appropriately trained machine learning model by scanning all the data.
%So to identify the desired objects, this search strategy requires scanning the entire data set.
%
In this work, we propose a novel framework for such search-by-classification scenarios that allows users to interactively search for target objects 
%in fields such as remote sensing or astronomy 
by specifying queries through a small set of positive and negative examples. Unlike previous approaches, our framework can rapidly answer such queries at low cost without scanning the entire database. 
Our framework is based on an index-aware construction scheme for decision trees and random forests that transforms the inference phase of these classification models into a set of range queries, which in turn can be efficiently executed by leveraging multidimensional indexing structures.
%in low-dimensional feature subspaces. The range queries, in turn, are efficiently be supported by multidimensional indexing structures such as \kdtrees. 
%Given a large collection of data and indexing structures pre-built for a set of feature subspaces, 
%The resulting search-by-classification framework can be used to rapidly answer recurrent user search queries for target objects, efficiently and at low cost. % by avoiding scans of all the data.
%that are independent of the specific search query, the resulting framework can be used to retrieve rare objects in massive collections of data efficiently and at low cost without scanning the whole data.
%Instead of applying machine learning models -- in particular decision trees and random forests -- trained based on these examples to all instances in the database, we transform these examples automatically to a set of efficient database operations that approximate the predictions of such models without scanning the data. To this end, we develop a novel multidimensional indexing and search strategy, called decision branches, that can be efficiently applied to any set of input examples. %The main advantage of our technique is that answers equivalent to a time-consuming application of these models will be obtained without actually scanning all the data.
%This allows to answer corresponding queries in seconds instead of minutes or even hours. 
%We assess the performance of our implementation of decision branches on several real-world data sets. 
Our experiments show that queries over large data catalogs with hundreds of millions of objects can be processed 
in a few seconds using a single server, compared to hours needed by classical scanning-based approaches.
%on similar hardware.
%), while still yielding as accurate results.
%, amounting to a search speed-up of over two orders of magnitude compared with the typical classification-based strategy based on scans, while yielding as accurate results. %which takes several minutes or even hours given powerful compute devices.
%

%we propose an efficient implementation for a particular query-by-examples approach, namely the application of a decision tree or an ensemble of decision trees to classify objects into ``rare'' or ``not rare''. In particular, we show how the application of such models can be transformed to standard SQL queries in low-dimensional search spaces, which can be efficiently supported appropriate spatial index structures.
\end{abstract}

\maketitle

%%% do not modify the following VLDB block %%
%%% VLDB block start %%%
\pagestyle{\vldbpagestyle}
\begingroup\small\noindent\raggedright\textbf{PVLDB Reference Format:}\\
\vldbauthors. \vldbtitle. PVLDB, \vldbvolume(\vldbissue): \vldbpages, \vldbyear.\\
\href{https://doi.org/\vldbdoi}{doi:\vldbdoi}
\endgroup
\begingroup
\renewcommand\thefootnote{}\footnote{\noindent
This work is licensed under the Creative Commons BY-NC-ND 4.0 International License. Visit \url{https://creativecommons.org/licenses/by-nc-nd/4.0/} to view a copy of this license. For any use beyond those covered by this license, obtain permission by emailing \href{mailto:info@vldb.org}{info@vldb.org}. Copyright is held by the owner/author(s). Publication rights licensed to the VLDB Endowment. \\
\raggedright Proceedings of the VLDB Endowment, Vol. \vldbvolume, No. \vldbissue\ %
ISSN 2150-8097. \\
\href{https://doi.org/\vldbdoi}{doi:\vldbdoi} \\
}\addtocounter{footnote}{-1}\endgroup
%%% VLDB block end %%%

%%% do not modify the following VLDB block %%
%%% VLDB block start %%%
\ifdefempty{\vldbavailabilityurl}{}{
\vspace{.3cm}
\begingroup\small\noindent\raggedright\textbf{PVLDB Artifact Availability:}\\
The source code, data, and/or other artifacts have been made available at \url{\vldbavailabilityurl}.
\endgroup
}
%%% VLDB block end %%%

\section{Introduction}
\label{sec:introduction}
The data volumes produced and processed in many domains are massive. In remote sensing and astronomy, for instance, the amount of data is currently rising to a new level due to technical advancements in satellites and telescopes~\cite{cheng2020remotesensing,songnian2016biggeodata,zhang2015astronomy}. In particular, the unprecedented pace of increase in spatial and temporal resolutions is leading to the assembly of large-scale surveys consisting of exabytes of valuable data, as it is the case for the data catalogs associated with the Sentinel missions operated by the European Space Agency~(ESA)~\cite{DRUSCH201225} or the Large Synoptic Survey Telescope~\cite{ivezic2019lsst}. %~\cite{kremer2017biguniverse,ivezic2019lsst}. 
A common task in these and other domains is the search for ``interesting'' objects~\cite{weiss2004miningrare,bailerjones2008rare,metcalf2019strong} in such massive databases, \ie, objects of particular value for a specific application. 
For example, an astronomer might be interested in a special type of galaxy, while a data analyst in the energy sector might be interested in wind turbines visible in satellite imagery.
%In the domains of databases and information retrieval, such problems are often formulated as {query-by-example}~\cite{fariha2019squid,mottin2019examples}
%or {search-by-multiple-examples} tasks~\cite{zhu2014sbme}. %,lissandrini2018example

\begin{figure*}
    \centering
    % \includegraphics[width=1.0\columnwidth]{figures/intro_figure/Figure1_V12.pdf}
    %\vspace{-0.8cm}
    \def\svgwidth{2.0\columnwidth}
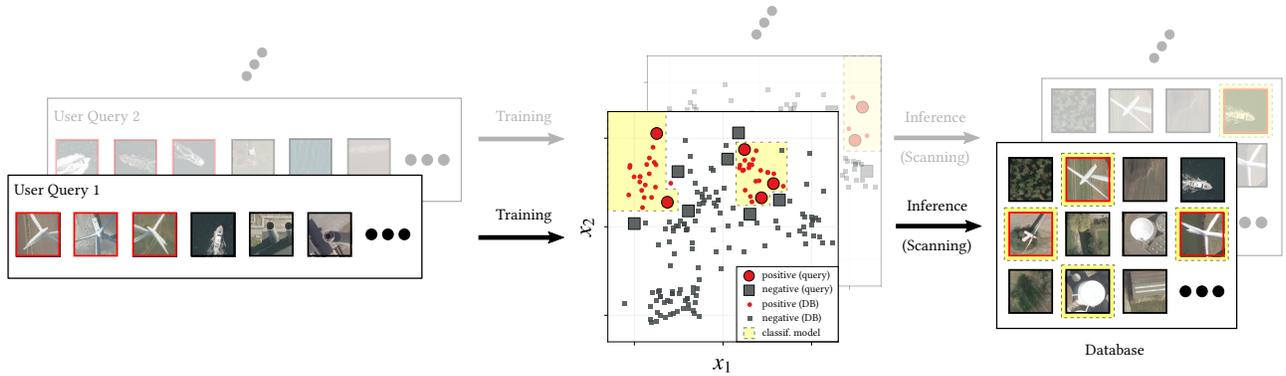
    %\vspace{-0.4cm}
    \caption{Traditional search-by-classification trains models for each query set, causing latency or high costs when scanning the entire database. Our framework leverages a co-design of indexes and models for efficient retrieval of positive instances, significantly reducing search time.}
    %\caption{Traditional search-by-classification approaches train a classification model for each given query set of positive and negative examples (\ie, per user query). However, identifying the corresponding positive objects in a large data catalog requires scanning the complete database and applying the model to each object, which either induces a high latency or high costs (due to parallel scans). Our proposed framework solves this problem by leveraging a co-design of multidimensional indexes and classification models, and returns the desired (as positive classified) instances in a fraction of the time.}
    %\caption{Traditional nearest neighbor search (left) does not allow users to fully specify their intent, but leverages mechanisms for fast data retrieval (e.g., via spatial indexes). In contrast, search-by-classification (right) improves retrieval accuracy via additional user input, but generally requires scanning the entire data catalog. Our approach combines the benefits of both approaches, i.e., our novel search framework yields as accurate results as search-by-classification schemes that scan the entire data catalog while providing the results as fast as traditional nearest neighbor based approaches.
    %\vspace{-0.1cm}
    \label{fig:motivation}
   % \vspace{-0.2cm}
\end{figure*}

There are two prominent methodologies to address such search tasks from a technical perspective. The first one is based on nearest neighbor search, where a user query is specified via a query object, and objects similar to the query object are returned as query result~\cite{KEISLER2019visualsearch}.
These query scenarios are known under the guise of content-based (image) retrieval~\cite{wan2014cbir} and have led to popular search engines such as Google's reverse image search or geospatial search engines~\cite{KEISLER2019visualsearch}.
In general,
%queries are often defined via exemplary objects and the search for similar rare objects is conducted by means of efficient (approximate) nearest neighbor search methods. 
%With this method, 
the corresponding user queries can be answered efficiently using index structures or approximation techniques~\cite{Bentley1975,pmlr-v80-keivani18a}.
However, the restriction of nearest neighbor queries to a single search item per query may lead to incomplete and inaccurate query results, as it is challenging to model the user's intent based on a single query instance. This restriction is particularly problematic when it is necessary to identify \emph{all} instances of an interesting object in a given database, such as finding all wind turbines in satellite imagery to estimate their quantity. In such scenarios, the use of nearest neighbor search methods may prove insufficient.
%Therefore, online search engines, such as Google reverse-image search, adopt this approach. 
%, the results are often not in line with the actual user intent. % (\eg, many false positives).

The second methodology is based on the application of machine learning models, in particular classification models such as decision trees.
%or neural networks. 
Such a search-by-classification approach involves using a (small) labeled data set containing both positive and negative instances. Based on this data set---which can be seen as a ``user query''---a classification model is trained. During the inference phase, the entire database is scanned, and the trained model is applied to classify each object.
% Such a search-by-classification approach resorts to a labeled data set (in the following named as user query) provided by the user, consisting of both the positive and negative instances, over which a model is trained. Then in the inference phase, the database will be scanned, and the trained model will be applied to classify each object. Objects of the target class will then be returned.
Only the objects classified as belonging to the target class (\eg, class ``special galaxy'') will be returned as the search result. As users can specify both types of objects---the ones the search \emph{should} return and those it \emph{should not} return---search-by-classification approaches can capture the user's query intent more precisely and, hence, generally produce search results with higher quality compared to single-object queries~\cite{branco2016surveyimbalanced,cheng2020remotesensing,wang2019metarare}. 
However, despite its superiority,
%of search-by-classification, 
%online 
search engines serving many users cannot employ this approach since scanning the entire database per user query is too time-consuming and induces a high latency.

Figure~\ref{fig:motivation} illustrates the concept of search-by-classification on a database of satellite imagery
%such as the ones of the Large Synoptic Survey Telescope~\cite{ivezic2019lsst}, 
over which many user queries have to be executed. Here, a data analyst can specify both ``interesting'' objects, such as different wind turbines, as well as ``non-interesting'' examples, such as ships or chimneys, via a binary classification data set (left: red and black borders indicate positive and negative examples, respectively). A classification model (middle: yellow regions) is then trained using suitable features extracted from the data (for the sake of illustration, a two-dimensional feature space is shown; typically, more features are considered). 
The classification model is then applied to the entire database and the objects classified as ``interesting'' are returned to the user (right: yellow dashed border; false positives/negatives are possible). 
%\todot{We are looking for a small portion of the data (e.g., rare objects).}
Note that \emph{each} user query requires a new training and a new scan of the entire database (\eg, ``User Query 2'' induces a new model and different results). 
Although the training phase might not take much time for small query data sets (\eg, training a decision tree might take a tenth of a second or less), the inference phase generally requires scanning \emph{all} the data for \emph{each} incoming user query, which can become very time-consuming for large data catalogs, with each query taking hours or even more to complete. 
Hence, the search-by-\-classifi\-cation strategy either implies high latency---\ie, a long time necessary to scan the entire database---or high costs---\ie, the costs caused by the parallelization of scans over a massively parallel infrastructure to reduce the latency. 
Note that a typical user query aims at finding a relatively small subset of ``interesting'' objects in the database, \ie, the answers sets are often very small compared with the totality of the objects (\eg, in a database of satellite imagery, only one out of one million image patches might show a wind turbine). Thus, we would ideally wish to devise a method with per-query costs that do not grow in proportion to the input database size.

%In contrast, traditional search queries, especially in the case of low-dimensional range queries, can typically be efficiently executed via multidimensional indexes~\cite{Bentley1975,pmlr-v80-keivani18a} or describe or approximation schemes (\eg, locality-sensitive hashing). However, as illustrated above, the answers to such queries are typically less well-defined.

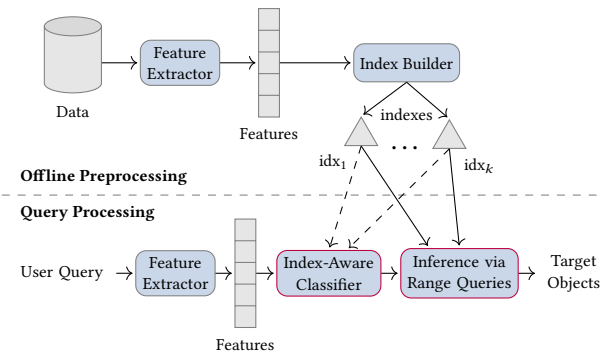
\begin{figure}
	\centering
 %\vspace{-0.1cm}
	% Requires \usetikzlibrary{shapes, shapes.geometric, shapes.arrows, arrows, patterns, positioning, calc}
\definecolor{modblue}{RGB}{166,189,219}
\tikzstyle{module}=[
    rectangle, 
    rounded corners,
    draw = gray, 
	fill = modblue!60,
	minimum height = 2em,
	align=center,
]

\tikzstyle{asset}=[
    rectangle,
    draw = gray!20, 
	fill = gray!20,
	minimum height = 2em,
	minimum width = 2em,
	text width=3em, 
	align=center,
]

\tikzstyle{database}=[
    cylinder,
    cylinder uses custom fill, 
    cylinder body fill = gray!20, 
    cylinder end fill = gray!30,
    aspect = 1.5,
    shape border rotate = 90,
    minimum height = 4.0em,
    minimum width  = 3.0em,
    draw = gray,
    anchor=west,
]

\tikzstyle{index}=[
    isosceles triangle,
    isosceles triangle apex angle = 60,
    rotate = 90,
    draw = gray, 
	fill = gray!20,
    minimum size = 1.5em,
    anchor = apex,
]

\tikzstyle{featvector}=[
    rectangle split, 
    rectangle split parts=5,
    draw = gray, 
	fill = gray!20,
    draw,
]

\tikzstyle{arrow} = [->,-To, shorten >=1pt]
\tikzstyle{customlabel} = [color=black]

\begin{tikzpicture}[scale=0.8, transform shape, font=\small]
    \node[database] (db) at (-0.3, 0) {};
    \node[module, text width=3.5em, align=center, right=2em of db] (featurizer) {Feature Extractor};
    \node[featvector, right=2em of featurizer] (fvectors1) {};
    \node[module, right=7em of featurizer] (builder) {Index Builder};
    \node[index, below left=2.5em and 0.0em of builder] (idx1) {};
    \node[index, below right=3.5em and -1.0em of builder] (idx2) {};
    
    \draw[gray, dashed] (-1.0, -2.2) -- (9.0, -2.2);    
    
    \node[text width=5em, align=center] (query) at (0, -3.5)
    {User Query};
    \node[module, text width=3.5em, align=center, right=1em of query] (fextractor2){Feature Extractor};
    \node[featvector, right=1em of fextractor2] (fvectors2) {};
    \node[module, draw=purple, text width=4.8em, align=center, right=3.2em of fextractor2] (cls) {Index-Aware Classifier};
    \node[module, draw=purple, text width=5.5em, align=center, right=1em of cls] (rq) {Inference via\\Range Queries};
    \node[text width=3.1em, align=center, right=1em of rq] (result) {Target Objects};
    
    \draw[arrow] (db.east)--(featurizer.west);
    \draw[arrow] (featurizer.east)--(fvectors1.west);
    \draw[arrow] (fvectors1.east)--(builder.west);
    \draw[arrow] (builder.south)--(idx1.apex);
    \draw[arrow] (builder.south)--(idx2.apex);
    \draw[arrow, dashed] (idx1.west)--(cls.north);
    \draw[arrow, dashed] (idx2.west)--($(cls.north) + (0.3, 0.0)$);
    \draw[arrow](query.east)--(fextractor2.west);
    \draw[arrow](fextractor2.east)--(fvectors2.west);
    \draw[arrow](fvectors2.east)--(cls.west);
    \draw[arrow](cls.east)--(rq.west);
    \draw[arrow](rq.east)--(result.west);
    \draw[arrow](idx1.west)--($(rq.north) + (-0.5, 0.0)$);
    \draw[arrow](idx2.west)--(rq.north);

    \node[customlabel, anchor=west] at (-0.8, -1.9) {\textbf{Offline Preprocessing}};
    \node[customlabel, anchor=west] at (-0.8, -2.5) {\textbf{Query Processing}};
    \node[customlabel, below=3em of builder] {\huge \ldots};
    \node[customlabel, below=1em of builder] {indexes};    
    \node[customlabel, below left=0.1em and 0.8em of idx1] {idx$_1$};
    \node[customlabel, below right=1.0em and 0em of idx2] {idx$_k$};
    \node[customlabel, below=0.2em of fvectors1] {Features};
    \node[customlabel, below=0.2em of fvectors2] {Features};
    \node[customlabel, below=0.3em of db] {Data};
\end{tikzpicture}
    %\vspace{-0.2em}
   % \vspace{-0.16cm}
	\caption{Fast search-by-classification with an index-aware classifier and pre-built indexes. The classifier uses range queries on indexes for efficient inference and obtains target objects quickly via spatial lookups, bypassing data scans. 
 %via an index-aware classifier and indexes pre-built for the data. The index-aware classifier allows implementing its inference phase via a series of range queries, which are efficiently supported by the indexes. The target objects can then be obtained via fast spatial lookups instead of a time-consuming scan of all the data.
	%via collaborative design of machine learning models and multi-dimensional index structures. of ML models and indexes: indexes are built to enable accurate search-by-classification, while ML models are designed to efficiently leverage indexes during inference for fast result retrieval.
	}
	\label{fig:framework}
 %\vspace{-0.2cm}
\end{figure}
%\subsection{Contributions}
\textbf{Contributions:} In this work, we introduce a novel search-by-classification framework that allows us to return an answer set 
%\emph{without} a full scanning of the data catalog for each new user query 
efficiently with low latency and at low cost by exploiting pre-built index structures, see Figure~\ref{fig:framework}. %in the database, 
More precisely, we propose a co-design of machine learning models and indexes for fast search-by-classification. As detailed in Section~\ref{sec:approach}, a set of multidimensional index structures is built in an offline phase, where each index corresponds to a random subset of features (which are extracted from the data). These multidimensional indexes are constructed \emph{only once} for the entire database and are \emph{independent} of a particular user query. Afterwards, in the query processing phase, a new classification model is trained for each new user query, where the construction of this model is ``index-aware'', \ie, during the construction, information about the indexes built in the offline phase is taken into account. Instead of scanning the entire data catalog, the instances being classified as positive  (``Target Objects'') can then be efficiently retrieved via range queries, \ie, the inference phase is supported by means of fast lookups over the pre-built index structures. 
Hence, for a typical user query with a small answer set, one can quickly return \emph{all} the desired instances with minimal computing resources, which is essential for interactive search engines. Of note, this paper primarily focuses on databases with infrequent changes. In summary, 
our main contributions are as follows: 
%Accommodating changing data requires an additional step of updating the pre-built indexes (as we discuss in Section~\ref{sec:space_consumption}).}
%\footnote{It is worth stressing that our framework can also efficiently handle queries with larger answer sets. However, the runtime benefits become less and less prominent since returning the instances takes time at least proportional to the size of the answer set.}
%Hence, different user queries can be answered with both a short response time and with little computing resources, which is crucial for interactive search engines.\footnote{\red{One important ingredient of our approach is the fact that the answer sets are often relatively small compared with the totality of the objects.}}
%With this framework, online search engines can adopt advanced classification models rather than simple nearest-neighbor models to provide high-quality search services. 
%In this work, we combine the benefits of both strategies by providing a novel framework for fast search-by-classification. 
%. We propose a co-design of ML models and indexes, where indexes are chosen to enable accurate model-based classification, while ML models are designed to efficiently utilize indexes for fast result retrieval.
%In essence, we provide the basis for conceptually novel search-by-classification engines that allow users to quickly search for interesting objects over massive data catalogs. 
%In this work, the following contributions are made:

\begin{enumerate}[(i)]
    \item We propose a novel search-by-classification framework, which resorts to a co-design of index structures and machine learning models. Users can formulate queries with positive and negative examples and the retrieval of target objects in large databases can be done in a few seconds instead of the hours that scanning-based approaches would typically need. 
    \item %\todot{stress quality of random forests; link to experiments same as random forests performance} 
    %\todot{construction of tree fragments, novel concept, index-aware, decision branches, which can be seen as fragments of decision trees} 
    To realize this framework, we propose an ``index-aware'' construction scheme for decision trees and random forests.\footnote{
 Leveraging multidimensional indexes in the framework requires classifiers that are efficient to construct and can be represented as range predicates. These requirements are naturally fulfilled by decision trees and random forests, which belong to the most powerful classification models in machine learning~\cite{delgado:14}. Incorporating other models, such as deep neural networks or boosted trees, is subject of future research. % and is out of this paper's scope.     
    } %, which belong to the most popular and powerful classification models in machine learning~\cite{delgado:14}. 
    Our central idea involves the construction of decision-tree-like models based on low-dimensional feature subsets that match multidimensional index structures. More precisely, we build fragments of the trees
    %which we call decision branches, 
    following a novel bottom-up construction instead of the well-known and commonly applied top-down construction scheme. The resulting models exhibit two important properties:~(a)~the classification quality of the models created bottom-up is close to that of their original top-down counterparts, and (b)~the set of database instances assigned to the positive class in the inference phase can be efficiently retrieved via range queries. % in low-dimensional spaces.
    %be efficiently obtained with the support of indexing structures pre-built for the data at hand.
    %In particular, our approach only yields lower parts of the trees, which we call \emph{\tdecisionbranches}
%The construction of the \tdecisionbranches is restricted in such a way that the inference phase is efficiently supported by indexed accesses in low-dimensional feature subspaces of the data to be classified.
    %More precisely, during the construction phase, we take information about multi-dimensional indexes into account, which are built for the available data catalog in a preprocessing phase. 
    %The resulting classification models exhibit a similar performance in terms of F1 scores, 
    %We introduce a novel index-aware bottom-up construction algorithm for tree-based classification model construction that can natively use and benefit from multidimensional index structures during inference.
    \item We provide a prototype of our search-by-classification framework,\footnote{\url{https://github.com/decisionbranches/decisionbranches}} implemented in Python with Cython optimizations. We also implemented a graphical search engine on top of the prototype (accessible at \url{https://web.rapid.earth}).
    \item  We conduct an extensive experimental evaluation of our prototype, which includes a case study in the geospatial domain with more than a billion image patches, a detailed analysis of the involved parameters, and an extensive comparison with competing search strategies.
    The results show that our search-by-classification framework yields accuracy comparable with traditional search-by-classification schemes that repeatedly scan the entire database while returning the results in a fraction of the time needed by those methods.
    %as fast as traditional nearest neighbor search. % (\ie, fast response). 
    %More precisely, the answer set for each query can be retrieved via range queries in low-dimensional space, which can be efficiently implemented via indexing structures. 
    %In practice, the results can be computed in seconds instead of hours that are needed by full scans of the data.
    %\todot{Denis/Christian: sub-linear test time see sec 2.2 $\O(\tdim \testsize^{1-\nicefrac{1}{\tdim}} + \indexStructuresQueryPoints)$} %\todot{remove? ->} All our source code will also be made publicly available as open-source software upon acceptance of this manuscript.
    %Our approach yields search results up to $715\times$ faster than decision trees and up to $195\times$ faster than random forests, without compromising result quality in terms of $\fscore$-score.
\end{enumerate}

\section{Background}
\label{sec:background}  
We consider search tasks with user queries consisting of positive and negative examples, see again ``User Query 1'' and ``User Query~2'' in Figure~\ref{fig:motivation}. 
That is, each user query gives rise to a binary classification task and a user query is provided in the form of a data set $\trainset=\{(\vec{x}_1,y_1), \ldots, (\vec{x}_\tsize, y_\tsize)\} \subset \Reals^\tdim \times \Yspace$ with $\Yspace=\{0,1\}$, where a $(\vec{x}_i,y_i)$ with $y_i=1$ corresponds to a positive example (target class) and $y_i=0$ to a negative example. Here, $\tdim$~corresponds to the number of features that are either given or that are extracted per instance. For instance, in the case of image data, a feature extractor can be applied to extract meaningful features, see again Figure~\ref{fig:framework}. 
\subsection{Decision Trees and Tree Ensembles}
%The searches for rare objects described above can be formulated as highly imbalanced, binary classification tasks with training sets of the form $\trainset=\{(\vec{x}_1,y_1), \ldots, (\vec{x}_\tsize, y_\tsize)\} \subset \Reals^\tdim \times \Yspace$ with $\Yspace=\{0,1\}$ and where $(\vec{x}_i,y_i)$ with $y_i=1$ correspond to the rare objects and $\tdim$ to the dimensionality of the feature space.

% \begin{figure}
% 	\includegraphics[width=0.55\columnwidth]{figures/Splitting_2D.png}
% 	\hfill 
% 	\includegraphics[width=0.35\columnwidth]{figures/Decision_Tree.png}
% 	\caption{A simple decision tree algorithm with the induced decision surface. All points that are contained in one of the two red rectangles would be classified as ``rare'' during the inference phase. Hence, given such a low-dimensional feature space, all rare objects in a large catalog could be obtained via two simple range queries. For low-dimensional feature spaces, such queries can be efficiently supported via spatial index structures.}
% 	\label{fig:sketch_approach}
% \end{figure}

%\subsubsection{Decision Trees}
Decision trees are simple yet powerful models and remain among the most popular methods in machine learning~\cite{HastieTF2009,Breiman2001}. % Murphy2012,
Typically, decision trees are built recursively in a top-down manner, see Algorithm~\ref{alg:topdown}. The root of the tree to be built corresponds to a subset $\trainsetsubset \subseteq \trainset$ of the available data instances (\eg, a subset of the data sampled uniformly at random with replacement).
%The subset $\trainset'$ can contain the same instances as $\trainset$ or can be a so-called \emph{bootstrap sample}, which is, for instance, obtained by drawing $\tsize$ samples with replacement. 
During the recursive construction, the data are typically split into two subsets at each node, which form the basis for the recursive construction of two subtrees that are the children of that node. The recursion stops as soon as an associated subset is {pure}, which means that all the instances belong to the same class, or as soon as some other stopping criterion is fulfilled. %The corresponding nodes at this point form the leaves of the tree. %, such a node becomes a leaf. 
\begin{algorithm}[t]
\caption{\textsc{TopDownConstruct}($S$, $\numberfeaturetested$)}
\label{alg:topdown}
\small
\begin{algorithmic}[1]
\REQUIRE Set $S \subset \Reals^\tdim \times \Yspace$ and $\numberfeaturetested\in\{1,\ldots,\tdim\}$.
\ENSURE Decision tree $\mathcal{T}$ built for $S$
\IF{$S$ is pure (or some other criterion fulfilled)}
\STATE \textbf{return} leaf node
\ENDIF
\STATE $(i^*,\theta^*) = argmax_{i\in \{i_1, \ldots, i_\numberfeaturetested\} \subseteq \{1,\ldots,d\}, \theta} G_{i,\theta}(S)$
\STATE $\mathcal{T}_l$ = \textsc{TopDownConstruct}($L_{i,\theta}$)
\STATE $\mathcal{T}_r$ = \textsc{TopDownConstruct}($R_{i,\theta}$)
\STATE Generate node storing the pair $(i^*,\theta^*)$ and pointers to its subtrees $\mathcal{T}_l$ and $\mathcal{T}_r$. Let $\mathcal{T}$ denote the resulting tree.
\STATE \textbf{return} $\mathcal{T}$
\end{algorithmic}%
\end{algorithm}%
Each internal node corresponds to a subset $S \subseteq \trainset$
of instances, which is split via a splitting dimension~$i \in \{i_1,\ldots, i_\numberfeaturetested \}$ and
an appropriately chosen threshold~$\theta \in \Reals$, where $\numberfeaturetested \leq \tdim$ is a user-defined parameter that determines the number of random features that are considered per split. Typically, one aims at maximizing the so-called {information gain}, defined as:
\begin{equation}
 G_{i,\theta}(S) = Q(S) - \frac{|L_{i,\theta}|}{|S|} Q(L_{i,\theta}) -\frac{|R_{i,\theta}|}{|S|} Q(R_{i,\theta})
\end{equation}
Here, $L_{i,\theta}=\{ (\vec x, y)\in\ S \,|\,   x_i \leq \theta \}$ is the subset of $S$ containing the elements whose $i$-th feature is less or equal to the threshold~$\theta$ and $R_{i,\theta}=\{ (\vec x, y)\in\ S \,|\,   x_i > \theta \}$ the subset of $S$ containing the remaining instances. Maximizing the information gain corresponds to minimizing the (weighted) ``impurities'' of the two subsets above, which is typically quantified by an {impurity measure}~$Q$~\cite{HastieTF2009,Breiman2001}. % Murphy2012,
For binary classification scenarios---on which we focus in this work---one typically resorts to the %so-called
{Gini index}, which is defined as
\begin{equation}
\label{eq:gini}
 Q(S)=\sum_{c=0}^1 p_{S}^c (1-  p_{S}^c),
\end{equation}
where $p_{S}^c$ corresponds to the fraction of points in~$S$ belonging to class $c \in \{0,1\}$. Given a decision tree, the predicted class for a new data point $\vec{x} \in \Reals^\tdim$ is obtained by traversing the tree from top to bottom and by resorting to the information stored in that particular leaf (\eg, dominant class of the training instances assigned to that leaf).
To avoid overfitting, fully-grown decision trees are often pruned in a post-processing phase.
%to reduce their complexity.% (this is usually not done for tree ensembles though).
%, this step is usually omitted though, as detailed next.

% \begin{algorithm}[tb]
%   \caption{\textsc{TopDownConstruct($\trainset$)}}
%   \label{alg:topdownconstruct}
% \begin{algorithmic}
% \REQUIRE Training data $\trainset=\{(\vec{x}_1,y_1), \ldots, (\vec{x}_\tsize, y_\tsize))\} \subset \Reals^\tdim \times \{0,1\}$
% \ENSURE A decision tree ...
%   \STATE $\trainset^0 \gets \{(\vec{x},y) \in \trainset | y=0\}$
% \end{algorithmic}
% \end{algorithm}

%\subsubsection{Tree Ensembles}
While decision trees often exhibit a good performance in practice, their quality can be typically improved by considering ensembles. Random forests are a prominent example of this class of techniques~\cite{Breiman2001}. A random forest consists of a user-defined number~$\forestNTrees$ of decision trees $\mathcal{T}_1,\ldots,\mathcal{T}_\forestNTrees$, which are built independently from each other. % (see Algorithm~\ref{alg:forest} for the case of binary classification scenarios). 
To obtain different individual models, one can consider so-called bootstrap samples that are drawn uniformly at random (with replacement) from the data $\trainset$, different subsets $\{i_1, \ldots, i_\numberfeaturetested\}$ of features for each node split, and different (random) splitting thresholds. A prediction $\predfunc(\vec{x})$ for a new point $\vec{x} \in \Reals^\tdim$ is then obtained by combining the individual predictions, that is
\begin{equation}
 \predfunc(\vec{x}) = \mathcal{C} \left( \predfunc_1(\vec{x}), \ldots, \predfunc_\forestNTrees(\vec{x}) \right),
\end{equation}
where $\mathcal{C}:\Reals^\forestNTrees \rightarrow \Reals$ depends on the
learning scenario. For classification scenarios, a common choice is the majority vote~\cite{HastieTF2009,Breiman2001}, %Murphy2012,
\ie, 
$\mathcal{C} \left( \predfunc_1(\vec{x}),
  \ldots, \predfunc_\forestNTrees(\vec{x}) \right)
=\operatorname{argmax}_{c\in \Yspace}|\{i\,|\,\predfunc_i(\vec{x})=c\}|$.
%\begin{algorithm}[t]
%\caption{\textsc{RandomForest($T$, $\forestNTrees$, $\numberfeaturetested$)}}
%\label{alg:forest}
%\small
%\begin{algorithmic}[1]
%\REQUIRE $T=\{(\vec{x}_{1},y_{1}), \ldots, (\vec{x}_{\tsize}, y_{\tsize})\} \subset \Reals^\tdim %\times \Yspace$, $\forestNTrees \in \Nats$, and $\numberfeaturetested\in\{1,\ldots,\tdim\}$.
%\ENSURE Trees $\mathcal{T}_1,\ldots,\mathcal{T}_\forestNTrees$ for $T$.
%\FOR{$b=1,\ldots,\forestNTrees$}
%\STATE Draw bootstrap sample $T'$ from $T$
%\STATE $\mathcal{T}_b$ = \textsc{TopDownConstruct}($T'$, \numberfeaturetested)
%\ENDFOR
%\STATE \textbf{return} $\mathcal{T}_1, \ldots, \mathcal{T}_\forestNTrees$
%\end{algorithmic}%
%\end{algorithm}%
For standard random forests~\cite{Breiman2001}, the optimal splitting threshold is computed as in Line 4 of Algorithm~\ref{alg:topdown}. A well-known alternative choice is to instead select a random threshold between the minimum and the maximum feature value, which yields so-called {extremely randomized trees}~\cite{GeurtsEW2006}.
It is worth stressing that such potentially ``suboptimal'' thresholds often yield competitive if not superior tree ensembles.

\subsection{Multidimensional Indexes}
\label{sec:multidim_indexes}
% \begin{itemize}
%     \item \todot{Denis: write short overview over spatial search structures }
%     \item spatial search structures, k-d trees, B-trees, ...
%     \item range queries ...
%     \item asypmtotic runtimes; given an index, the search is very fast in low-dimensional search spaces
%     \item \todot{writing text, literature search, runtime guarantees as mentioned above}
% \end{itemize}

%Both Random Forest and Decision Tree models involve scanning and filtering data based on the splitting decisions learned during training. Despite the rich research corpus on making these models more efficient and faster, applying such models on extremely large datasets comprising billions of data objects is still time-consuming. 
% \begin{itemize}
%     \item \todot{Orthogonal range searching: \url{https://en.wikipedia.org/wiki/Range_searching}; our bounds might also be infinity per dimension. Do the runtime bounds still hold?}
% \end{itemize}
Our novel index-aware classifier transforms the inference phase of decision trees and random forests to a set of range queries in low-dimensional spaces. Such queries are efficiently supported by spatial index structures, such as %\Btrees~\cite{Bayer1972btrees} or
\kdtrees~\cite{Bentley1975}, which are commonly used in the context of, \eg, database systems or nearest neighbor search~\cite{GaedeG98}. %For the sake of completeness, 
We therefore briefly summarize a few important concepts behind these index structures that are necessary for this work. % and report corresponding runtime bounds.

For multidimensional data, a \kdtree is a popular choice to speed up range queries and nearest neighbor search~\cite{Bentley1975}. A \kdtree is a balanced binary search tree that recursively partitions the
%a $k$-dimensional 
search space into two half-spaces at each node. Given $\testsize$ points in the $\tdim$-dimensional Euclidean space, such a tree can be built in $\O(\testsize \log \testsize)$ time using linear-time median finding and occupies $\O(\testsize)$ additional space~\cite{Bentley1975}.
%For relatively low-dimensional spaces (\eg, $\tdim < 30$), \kdtrees can be used to find nearest neighbors in logarithmic time in practice (linear time is needed in the worst case). 
%Search operations over a \kdTree are performed in $O(\log n)$ time for low-dimensional spaces~\cite{Bentley1975}. 
%Similarly, 
%Orthogonal range queries with multi-dimensional axis-parallel rectangles defining the regions of interest can be accelerated via \kdtrees.
% search includes two factors: (1) traversing a subtree to report data stored in its leaf nodes; and, (2) checking nodes intersecting the query region but not fully contained in it. 
%Factor (1) is done in linear time $O(q)$, where $q$ is the number of reported data points (i.e., the query result size), whereas (2) is bounded by $O(n^{1-\nicefrac{1}{k}})$, where $k$ is the number of dimensions or attributes of the data. 
%Here,
The time complexity of %for orthogonal range queries
orthogonal range queries with multidimensional axis-parallel boxes defining the regions of interest over a \kdtree can be upper-bounded, in the worst case, by $\O(\tdim \testsize^{1-\nicefrac{1}{\tdim}} + \indexStructuresQueryPoints)$, where $\indexStructuresQueryPoints$ is the number of points returned~\cite{Lee77worstcase}. Hence, sublinear time is needed to answer such queries. In practice, a \kdtree exhibits good performance for small~$\tdim$, \eg, $\tdim=3$, and can be constructed as to store data objects in external memory while the tree is kept in main memory~\cite{Bentley79}. %\footnote{We adopt this scheme to efficiently answer range queries.}
%, and defer a comparison with other multidimensional index structures 
%for large datasets %, such~as the R-Tree~\cite{QiTCZ20}, 
%to future work.} % omitted Bentley1979datastructures from complexity result, kept only Lee and Wong.
%The creation of \kdtrees takes in $\O(\kdtreesNumberPoints \log \kdtreesNumberPoints)$ time~\cite{Bentley1979datastructures} and requires additional storage space, usually in the order of $\O(\kdtreesNumberPoints)$. 
%\todot{Paper with bounds for range tree missing} 
Another popular index structure for multidimensional data is the so-called range tree, which, along with fractional cascading, enables queries to be answered in $\O(\log^{\tdim -1} \testsize + \indexStructuresQueryPoints )$ time at the cost of an increased construction time and space consumption of $\O(\testsize \log^{\tdim -1} \testsize)$~\cite{BergCKO08}. 

Care must be exercised when employing index structures over high-dimensional data. In particular, the difference in distances of a point to its nearest and farthest points drops dramatically as $\tdim$ is increased, making indexing less and less effective~\cite{BeyerGRS99}. At dimensionality as low as $\tdim=5$, this effect is already highly pronounced, while indexes are expected to be outperformed by a full scan of the data at~$\tdim$ values as low as 10~\cite{BeyerGRS99}.

\subsection{Decision Trees and Range Queries}
\label{sec:rare_leaves_range_queries}

%Decision trees are typically built in a top-down manner.
During the inference phase of a standard decision tree built for a classification data set, a new instance is classified as positive in case the tree traversal ends up in a leaf containing only or mostly positive instances (\eg, if the decision is based on majority vote).
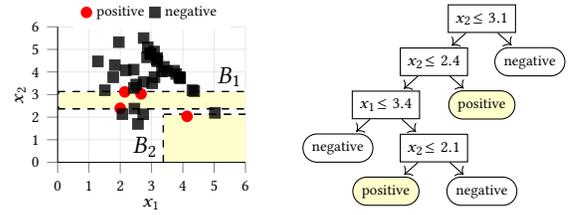
\begin{figure}[t]
%\vspace{-0.1cm}
    \newcommand{\xfeature}{$x_1$}
\newcommand{\yfeature}{$x_2$}
\newcommand{\nodewidth}{6ex}
\newcommand{\nodeheight}{2ex}
\newcommand{\nodedistance}{4.5ex}
\newcommand{\leveldistance}{10ex}

\definecolor{yellow2}{HTML}{ffff00}

\tikzstyle{treearrow} = [->,-To, shorten >=1pt]

\tikzstyle{treenode} = [
  draw,     
  shape=rectangle,
  color=black,
  solid, 
  minimum width=\nodewidth,
  minimum height=\nodeheight,
]

\tikzstyle{rareclass} = [
  draw,
  shape=rounded rectangle,
  color=black,
  fill=yellow2!20,
  solid, 
  minimum width=\nodewidth,
  minimum height=\nodeheight,
]

\tikzstyle{nonrareclass} = [
  draw,     
  shape=rounded rectangle,
  color=black,
  solid, 
  minimum width=\nodewidth,
  minimum height=\nodeheight,
]
\begin{tikzpicture}[
    every mark/.append style={mark size=0.55ex},
    level distance=\nodedistance,
	level 1/.style={sibling distance=\leveldistance},
	level 2/.style={sibling distance=\leveldistance},
    node distance=2.2cm and 4cm,
    font=\scriptsize,
    scale=0.94,
    every node/.style={transform shape},
    ]
    \begin{axis}[
        at = {(0\linewidth,0)},
        width=0.5\linewidth,
        xmin=0, xmax=6,
        ymin=0, ymax=6,
        tick align=outside,
        xtick distance=1,
        ytick distance=1,
        tick label style={font=\scriptsize}, 
        xlabel={\xfeature},
        ylabel={\yfeature},
        label style={font=\footnotesize},
        legend style={
          at={(0.5,1.23)},
          anchor=north,
          legend columns=-1,
        },
    ]
     \addplot [
        scatter, 
        only marks,
        scatter src=explicit,
        scatter/classes={
            1={mark=*, fill=red, draw=red},%
            0={mark=square*, fill=black, draw=black, opacity=0.8}
        },
    ]
    table [
        x=x,
        y=y,
        meta=label,
    ] {%
    x  y label
    4.13353701507816 2.03743899035789 1
    2.66531709639346 3.02973229670563 1
    2.14397063539591 3.11875609246825 1
    1.99769850031271 2.39407401567857 1
    3.0945314721689 3.7702152126602 0
    3.39279501670053 4.2302340125135 0
    3.72092961493132 4.05642015776681 0
    2.95344837344581 4.83288178496143 0
    3.16344277501011 4.53800314164617 0
    4.36380975976508 3.15155061608832 0
    2.87857879055556 5.09160537837457 0
    1.7756870183853 3.76615461505307 0
    3.78486492736914 3.91420023248269 0
    1.8254799539577 4.30216675997238 0
    2.50918318446944 3.44569774324521 0
    1.28571823212024 4.46885064330025 0
    3.33161630291041 4.02787705347493 0
    2.85656885534383 4.65104803878442 0
    5.02464880501632 2.19006930756427 0
    2.42809461325994 4.10491315918182 0
    2.74721600669124 5.5032583962494 0
    2.73482726932059 2.12779667928195 0
    3.18598464597997 4.47619871388183 0
    3.48145256596113 4.04260164580873 0
    2.45211608392477 2.38148184276365 0
    2.17772221093714 4.07568830609915 0
    2.43815735523012 3.29095073208471 0
    3.01831874718029 3.50184210780366 0
    2.74598842862603 3.56095613306554 0
    2.97430966925448 4.78629981921542 0
    1.95475546031246 5.32104272294127 0
    3.18093448053639 4.62236345692378 0
    4.30535651558841 3.19410511986788 0
    3.91556974420768 3.72005587139902 0
    3.90483015224456 3.7645002618933 0
    2.06807899097868 2.13496399821754 0
    3.07915150009021 4.90510686299592 0
    3.88294775568109 3.77391846594387 0
    1.50763318228028 3.19213378633646 0
    2.56607920266634 1.70158151437481 0
    };
    \legend {positive, negative};
    \draw[dashed, very thick] (axis cs:0,2.4) -- (axis cs:6,2.4);
    \draw[dashed, very thick] (axis cs:0,3.1) -- (axis cs:6,3.1);
    \draw [draw=none, fill=yellow2!20] (0.05,2.4) rectangle (6.15,3.1);
    \draw[dashed, very thick] (axis cs:3.4,0.0) -- (axis cs:3.4,2.1);
    \draw[dashed, very thick] (axis cs:3.4,2.1) -- (axis cs:6,2.1);
    \draw [draw=none, fill=yellow2!20] (3.4,0.05) rectangle (6.5,2.1);
    \node[draw=none] at (5.5, 3.8) {\large $B_1$};
    \node[draw=none] at (2.8, 0.7) {\large $B_2$};
    \end{axis}
    \node[treenode] (dtree) at (6, 0.24\linewidth) {\yfeature $\le 3.1$}
    child { node[treenode] {\yfeature $\le 2.4$} edge from parent [treearrow]
      child { node[treenode] {\xfeature $\le 3.4$}
      child { node[nonrareclass] {negative} }
      child { node[treenode] {\yfeature $\le 2.1$}
          child { node[rareclass] {positive} }
          child { node[nonrareclass] {negative} } 
        }
      }
      child { node[rareclass] {positive} }
    }
    child { node[nonrareclass] {negative} edge from parent [treearrow] };
\end{tikzpicture}
%    \vspace{-0.1cm}
	\caption{A decision tree partitions instances for two features. Inference phase classifies all instances in the yellow boxes as "positive." Retrieving positive instances from a large database is achieved through two orthogonal range queries.
 %A decision tree along with the induced partitioning into boxes for two features. All instances contained in one of the two yellow boxes would be classified as ``positive'' during the inference phase. Hence, all positive instances in a large database could be obtained via two orthogonal range queries. %Black dashed lines indicate if the boxes are bounded, half-bounded, or unbounded w.r.t. the corresponding dimension. 
	%For low-dimensional feature spaces, such queries can be efficiently supported via spatial index structures.
	}
	\label{fig:range_queries}
% \vspace{-0.1cm}
\end{figure}
%\subsubsection{Fast Inference via Range Queries for Low-Dimensional Data}
A simple yet crucial observation is that orthogonal range queries can be employed---namely, one range query per leaf of the tree that corresponds to the positive class---to obtain all instances in the database that would be classified as positive. % via a full scan of the data. 
Figure~\ref{fig:range_queries} illustrates this observation for $\tdim=2$ features. %Here, all instances being contained in the two yellow boxes will be classified as ``positive'' in the inference phase. %The two leaves corresponding to the positive class are associated with the boxes $\dtbox_1$ and $\dtbox_2$. %in a two-dimensional feature space. 
Without any optimization, these range queries would necessitate a full scan of the database to determine if instances fall within the bounds of a positive leaf. 
%As previously discussed in Section ~\ref{sec:framework_background}, range queries can be optimized in low-dimensional feature spaces (\eg, up to $\tdim=4$) through the use of spatial index structures. 
Instead, given a corresponding index structure pre-built for the database, this would allow to rapidly return the desired instances per user query \emph{without} the need of a full scan in case one is given a low-dimensional feature space (\eg, $\tdim \leq 4$). %\footnote{For instance, all the positive catalog instances (wind turbines) for user query~1 in Figure~\ref{fig:motivation} could be efficiently obtained via four orthogonal range queries, executed with the help of a multidimensional index built beforehand for the two features $x_1$ and $x_2$.} 
%In particular, when utilizing a k-d tree as the underlying indexing structure, the user query can be fulfilled in a sub-linear time of $\O(\tdim \testsize^{1-\nicefrac{1}{\tdim}} + \indexStructuresQueryPoints)$ as opposed to linear time $\O(\tdim\testsize)$ required for a full scan of the data. 
%Such an approach works efficiently in case few features are given (\eg, $\tdim \leq 4$).

However, in practice, more than two features are typically needed in order to achieve a satisfactory classification performance, \ie, decision trees are usually based on splits in many dimensions (\eg, $\tdim=100$). There are two simple but suboptimal ways to conduct the range queries for such high-dimensional feature spaces:
\begin{enumerate}[(i)]
\item \emph{Complete feature space:} A naive approach would be to build a single $\tdim$-dimensional index and to conduct a range query for each positive leaf. However, as discussed in Section~\ref{sec:multidim_indexes}, conducting such range queries would generally not be efficient, \ie, the increase in dimensionality, in general, negates the benefits of index structures. %, since the latter could even end up being outperformed by a full scan of the data.
\item \emph{Restricted feature space:} An alternative approach 
%to avoid this increase in latency 
would be to consider only a small subset of the features (\eg, only 4 features) when constructing a decision tree, so that the underlying feature space can efficiently be indexed. However, as shown in our experiments, this variant leads to models exhibiting a worse classification quality, especially given classification tasks that require many features to be taken into account. Also, considering multiple such low-dimensional feature spaces and picking the one that leads to the best classification performance generally leads to significantly worse results compared to a construction based on all $\tdim$ features.
\end{enumerate}
Next, we show how to efficiently implement the search for target objects 
%given high-dimensional feature spaces
without sacrificing classification quality.
\section{Efficient Search-by-Classification}
%\section{Index-Aware Classification}
\label{sec:approach}
%\todot{Check this paragraph to make a link to the previous section.}
%Our search-by-classification framework builds upon the idea of the previously introduced \emph{restricted feature space} by training on a large number of randomly sampled small feature subsets, rather than a single subset. This way, we can achieve a broad coverage of all features while still having low-dimensional subsets that are efficient to index. The combination of multiple models from different feature subsets allows us to cover all positive training instances and achieves classification accuracy comparable to that of decision trees, as demonstrated in our experiments. 
%Prior to describing the details of how we construct our classification model with respect to the indexes, we present the overall architecture of our search-by-classification framework. 
%\subsubsection{Offline Preprocessing}
%\begin{enumerate}
    %\item 
    %\emph{Offline Preprocessing.} 
We start by outlining the overall search-by-classification framework, prior to providing the details of the algorithmic building blocks.

\subsection{Overall Framework}
The framework is based on two phases, (1) an offline preprocessing phase and (2) a query processing phase, see again Figure~\ref{fig:framework}.

\begin{enumerate}[(i)]
    \item \emph{Offline preprocessing:}
    The offline preprocessing phase is only conducted \emph{once} for the entire database to pre-build a set of index structures. First, the data are processed via a feature extractor that outputs a $\tdim$-dimensional feature vector per instance. 
The features should capture general characteristics of the data and are not tailored to any specific query.\footnote{For images, these features could be based on information on color, shape, edges, and orientation. Such features can be extracted in an automatic manner via, \eg, pre-trained or unsupervised deep neural networks, see Section~\ref{sec:experiments}. In case features are already available, no feature extractor has to be applied.}
%In an offline phase, we transform raw data in the original database into features that help the downstream classifier to accurately distinguish data objects. 
Afterwards, $\tsizesubsets$ index structures $\text{idx}_1, \ldots, \text{idx}_\tsizesubsets$ are built via an index builder. 
More precisely, $\tsizesubsets$ random feature subsets $\featuresubset_1, \ldots, \featuresubset_\tsizesubsets \subset \{1,\ldots,\tdim\}$ with $|\featuresubset_i| = \tdimsubsets$ for $i=1,\ldots,\tsizesubsets$ are considered, where $1 \leq \tdimsubsets \leq \tdim$ is some small constant (\eg, $\tdimsubsets=4$). For each of these subsets $\featuresubset_i$, a corresponding multidimensional index $\text{idx}_i$ is built. 
For instance, given $\tdim=100$ features and $\tdimsubsets=4$, $\text{idx}_1$ could be an index for the feature subset $F_1=\{17,51,60,80\}$, $\text{idx}_2$ an index for $F_2=\{11,45,49,99\}$, and so on.
\item \emph{Query processing:}
In the query processing phase, each user query is formulated via a binary classification data set.
%each user query---defined via a small set of positive and negative training examples (\eg, image patches with wind turbines and other objects, respectively). 
First, the feature extractor used in the offline preprocessing phase is also used to extract a feature vector for each instance of the data set.
%The main task our framework has to perform is to learn a classifier that properly distinguishes rare from non-rare training examples included in user queries. 
%\textbf{Index-Aware Classifier}. 
Next, an adapted version of a decision tree/random forest is built for the given data set. As detailed below, information about the indexes built in the offline preprocessing phase are considered during the construction of these classifiers (dashed lines in Figure~\ref{fig:framework}) such that the database instances being classified as ``positive'' can be efficiently retrieved via orthogonal range queries, supported by a corresponding index structure (solid lines).
\end{enumerate}
Note that, since $\tdimsubsets$ is small, the index structures built in the offline preprocessing phase will support efficient range queries in the spaces defined by the feature subsets. Typically, quite many index structures have to be built for our approach (\eg, $\tsizesubsets=100$ or more).\footnote{
While sufficiently many feature subsets need to be available, large values for $\tsizesubsets$ also lead to an increased space consumption. This trade-off will be discussed in Section~\ref{sec:experiments}.}
Next, in Section~\ref{sec:bottom_up_constrution}, we provide the details of
our novel construction scheme for decision trees and random forests used in the query processing phase.
%that makes use of information about the multidimensional indexes.
%, each involving only a moderate subset of features. 
In a nutshell, we build fragments of the decision trees following a new bottom-up approach instead of the well-known and commonly applied top-down construction scheme. 
%These fragments, , correspond to the lower parts of the trees.
These decision branches allow for efficient retrieval of the desired instances from the database, as we will show in Section~\ref{sec:fast_rare_object_query_processing}.

%so that the inference phase can efficiently be implemented via the pre-built indexes while achieving high prediction quality. 

%Before providing these details, we briefly illustrate the connection between (normal) decision trees and range queries.

\subsection{Decision Branches}
\label{sec:bottom_up_constrution}

%One way to obtain such boxes would be to restrict the number of features that can be used by the construction algorithm for a decision tree. That is, one could keep track of the number of features already used from the root to the given node to be split, and in case this number exceeds a pre-defined threshold, no additional features may be used, \ie, the remaining splits would have to be done via the features already used. Such an approach, however, might yield suboptimal results. 
Our approach is based on an alternative construction scheme for decision trees, which essentially (only) yields the bottom parts of decision trees, which we call \emph{decision branches}. Each of the decision branches is associated with a $\tdim$-dimensional box $\dtbox$, similar to a leaf of a decision tree. We define $\dtbox$ as
$
 \dtbox = (l_1, r_1] \times \cdots \times (l_\tdim, {r}_\tdim] \subset \RealsInfinity^\tdim   
$
with $\RealsInfinity = \Reals \cup \{-\infty, +\infty\}$. If the dimension~$i$ of a box~$\dtbox$ is left- and right-bounded (\ie, $l_i > -\infty$ and $r_i < +\infty$), we call it bounded w.r.t. dimension~$i$, whereas if it is left- or right-bounded (\ie, $l_i > -\infty$ or $r_i < +\infty$), we call it half-bounded w.r.t. dimension~$i$; otherwise, we call it unbounded w.r.t. dimension~$i$.\footnote{
Note that, if a feature $i \in \{1,\ldots,\tdim\}$ is not used in any of the nodes from the root to a leaf of a classical decision tree, then the dimension~$i$ of the box associated with that leaf is unbounded, \ie, $l_i=-\infty$ and $r_i=+\infty$.}
Furthermore, we denote the number of half-bounded and bounded dimensions of a box~$\dtbox$ by~$\dtnumberbounded(\dtbox)$ and the number of unbounded dimensions by $\dtnumberunbounded(\dtbox)$, respectively. For instance, in Figure~\ref{fig:range_queries}, the first box $\dtbox_1$ is bounded w.r.t. the second dimension and unbounded w.r.t. the first dimension (\ie, $\dtnumberbounded(\dtbox_1)=1$ and $\dtnumberunbounded(\dtbox_1)=1$),
whereas the second box~$\dtbox_2$ is half-bounded w.r.t. both dimensions (\ie, $\dtnumberbounded(\dtbox_2)=2$ and $\dtnumberunbounded(\dtbox_2)=0$).

A crucial aspect of our approach is that each box $\dtbox$ associated with a decision branch will be bounded in a few dimensions only (\eg, $\dtnumberbounded(\dtbox) \leq 4$) and that it will contain ``mostly'' positive instances. Also, the union of all these boxes will cover all the positive instances of a given query data set. Finally, the construction of the branches and the boxes will be conducted in such a way that, for each box~$\dtbox$, the bounded dimensions will correspond to one of the index structures that were built for the entire database in the offline preprocessing phase. This will allow to efficiently retrieve all the database instances belonging to the boxes. 
The construction scheme detailed next is, hence, ``index-aware'' in the sense that the generation of the decision branches and associated boxes takes the index structures into account that were built in the offline preprocessing phase, see again Figure~\ref{fig:framework}.

\subsubsection{Constructing Decision Branches}

\begin{algorithm}[t]
   \caption{\textsc{DecisionBranches($\trainset$, $\featuresubset_1, \ldots, \featuresubset_\tsizesubsets$, $\numberfeaturesubsetstest$, $\numberfeaturetested$)}}
   \label{alg:bottom_up_construct}
\begin{algorithmic}[1]
\REQUIRE Data $\trainset=\{(\vec{x}_1,y_1), \ldots, (\vec{x}_\tsize, y_\tsize)\} \subset \Reals^\tdim \times \{0,1\}$, $\featuresubset_1, \ldots, \featuresubset_\tsizesubsets \subset \{1,\ldots,\tdim\}$ with $|\featuresubset_i| = \tdimsubsets$, $i=1,\ldots,\tsizesubsets$, for some $1 \leq \tdimsubsets \leq \tdim$,  $1 \leq \numberfeaturesubsetstest \leq \tsizesubsets$, and $\numberfeaturetested\in\{1,\ldots,\tdim\}$.
%, and $\numberfeaturetested \in \{1,\ldots,\tdim\}$
\ENSURE Set $\decisionbranchset = \{\left(\decisionbranchbox_1, \decisionbranchmodel_1\right), \ldots, \left(\decisionbranchbox_\sizedecisionbranchset,\decisionbranchmodel_\sizedecisionbranchset\right)\}$ of $\tdim$-dimensional boxes $\decisionbranchbox_1,\ldots,\decisionbranchbox_\sizedecisionbranchset$ with $\dtnumberbounded(\decisionbranchbox_i) \leq \tdimsubsets$ for $i=1,\ldots,\sizedecisionbranchset$ along with associated decision branches $\decisionbranchmodel_1,\ldots,\decisionbranchmodel_\sizedecisionbranchset$.
  \STATE $\trainsetnonrare \gets \{(\vec{x},y) \in \trainset | y=0\}$; $\trainsetrare \gets \{(\vec{x},y) \in \trainset | y=1\}$
%   \STATE $\trainsetrare \gets \{(\vec{x},y) \in \trainset | y=1\}$
   \STATE $\decisionbranchset \gets \{\}$
   \REPEAT \label{alg:line:start_loop}
   \STATE Let $(\vec{x'},y')$ by any positive instance in $\trainsetrare$ 
   \STATE $\greedygain_{opt} \gets0$; $\decisionbranchbox_{opt} \gets None$   
%   \STATE $\decisionbranchbox_{opt} \gets None$   
   % Do not test all k feature subsets
   \STATE %Pick random subset
   $\featuresubsets=\{\featuresubset_{i_1},\ldots,\featuresubset_{i_\numberfeaturesubsetstest}\} \subseteq \{\featuresubset_1, \ldots, \featuresubset_\tsizesubsets\}$
   \FOR{each $\featuresubset \in \featuresubsets$}
   \STATE $\decisionbranchbox, \greedygain \gets \textsc{GreedyMaxGainBox}(\trainsetnonrare \cup \trainsetrare, \vec{x'}, \featuresubset)$
   \IF{$\greedygain > \greedygain_{opt}$}
   \STATE $\greedygain_{opt} \gets \greedygain$; $\decisionbranchbox_{opt} \gets \decisionbranchbox$
%   \STATE 
   \ENDIF
   \ENDFOR
   \STATE $\trainsetrare, \trainsetrareremoved \gets \textsc{RemoveInstances}(\trainsetrare, \decisionbranchbox_{opt})$
     \STATE $\trainsetnonrare, \trainsetnonrareremoved \gets \textsc{RemoveInstances}(\trainsetnonrare, \decisionbranchbox_{opt})$
   %\STATE $\trainset^0, R^0 \gets \textsc{RemovePoints}(\trainset^0, \decisionbranchbox_{opt}, 0)$ // \todot{commment not done, Fabian: Why not again? To "block" this area?}
   \STATE $\decisionbranchmodel$ $\gets$ \textsc{TopDownConstruct($\trainsetnonrareremoved \cup \trainsetrareremoved$, $\numberfeaturetested$)}
   % I would suggest to include this here; otherwise, we only have boxes, this way, we get branches. The details can be described later ...
   %\todot{topdownconstruct is described at a later stage! still include and describe?} \todot{Is it still a box or a model? depending on the topdown algorithm -> full/+TD}
   \STATE $\decisionbranchset \gets \decisionbranchset \cup \left(\decisionbranchbox_{opt},\decisionbranchmodel\right)$
   \UNTIL{$\trainsetrare$ is empty} 
   \STATE {\bfseries return} $\decisionbranchset$
\end{algorithmic}
\end{algorithm}
Our construction scheme for \tdecisionbranches is implemented via the function \textsc{DecisionBranches} shown in Algorithm \ref{alg:bottom_up_construct}. 
Its input is composed of a training set~$\trainset$ (corresponding to a single user query), the $\tsizesubsets$ feature subsets $\featuresubset_1, \ldots, \featuresubset_\tsizesubsets \subset \{1,\ldots,\tdim\}$ mentioned above with $|\featuresubset_i| = \tdimsubsets \leq \tdim$ for $i=1,\ldots,\tsizesubsets$ and $1 \leq \tdimsubsets \leq \tdim$, a number $1 \leq \numberfeaturesubsetstest \leq \tsizesubsets$ of feature subsets to be tested per iteration, and a constant~$\numberfeaturetested\in\{1,\ldots,\tdim\}$.\footnote{The parameters $\numberfeaturetested$ and $\numberfeaturesubsetstest$ denote the number of features/feature subsets tested per iteration of the top-down and bottom-up construction, respectively. %\red{For all our experiments, we set $\numberfeaturetested = \numberfeaturesubsetstest$.
}
%The algorithm outputs pairs of decision branches and $\tdim$-dimensional boxes with $\dtnumberbounded(\decisionbranchbox) = \tdimsubsets$ for $i=1,\ldots,\sizedecisionbranchset$.

After initializing the two sets $\trainsetnonrare = \{(\vec{x},y) \in \trainset | y=0\}$ and $\trainsetrare = \{(\vec{x},y) \in \trainset | y=1\}$, the algorithm incrementally removes positive instances from $\trainsetrare$. For each such positive instance $(\vec{x}', 1)$, a $\tdim$-dimensional box is found containing~$\vec{x'}$ and potentially more data points. This is achieved by considering a random subset 
$\featuresubsets=\{\featuresubset_{i_1},\ldots,\featuresubset_{i_\numberfeaturesubsetstest}\} \subseteq \{\featuresubset_1, \ldots, \featuresubset_\tsizesubsets\}$ of the given feature subsets $\featuresubset_1, \ldots, \featuresubset_\tsizesubsets$ and by selecting the box $\decisionbranchbox_{opt} \subset \RealsInfinity^\tdim$ that maximizes an information gain computed via the function \textsc{GreedyMaxGainBox} (see below).
Given the box~$\decisionbranchbox_{opt}$, the function \textsc{RemoveInstances} removes all instances from $\trainsetnonrare$ and $\trainsetrare$ whose data points are contained in $\decisionbranchbox_{opt}$~(see Algorithm~\ref{alg:remove_instances}). 
The instances removed from $\trainsetnonrare$ and $\trainsetrare$ are collected in the sets~$\trainsetnonrareremoved$ and~$\trainsetrareremoved$, respectively, and are subsequently used as input for the function $\textsc{TopDownConstruct}$, which simply builds a classical decision tree for this subset in a top-down manner, yielding the decision branch~$\decisionbranchmodel$ (which consists of a single node in case $\trainsetnonrareremoved \cup \trainsetrareremoved$ is pure). Both the box~$\decisionbranchbox_{opt}$ and the decision branch $\decisionbranchmodel$ are added to the set~$\decisionbranchset$, which forms the output of \textsc{DecisionBranches} after all iterations are done.
%\footnote{Note that the final boxes cover all rare (training) instances and correspond to the boxes associated with the rare instances for normal decision trees, see again Figure~\ref{fig:range_queries}.}

Each call of \textsc{DecisionBranches} returns a set of decision branches along with the associated boxes. 
In this context, $\decisionbranchbox$ is a $d$-dimensional box (with at most $D$ bounded dimensions) resulting from the bottom-up construction phase (Lines 4--14 in Algorithm \ref{alg:bottom_up_construct}), while $\decisionbranchmodel$ is a decision branch that corresponds to a small decision (sub-)tree (constructed in a top-down manner), which separates the subset $R_0 \cup R_1$ of the training points contained in $\decisionbranchbox$ (Line 15 of Algorithm \ref{alg:bottom_up_construct}).
Overall, the set of all decision branches corresponds to a standard decision tree built in a top-down manner for all data points. Note, however, that we do not build this complete decision tree, but only the decision branches. Some examples of decision branches are provided in Figure~\ref{fig:main_idea}. Each decision branch~$\decisionbranchmodel$ (lower part of a decision tree) in the figure corresponds to a feature space~$\featuresubset \in \featuresubsets$ and can be seen as a branch of an overall decision tree (outlined in gray; not built). The bounding box~$\decisionbranchbox$ associated with a given $\decisionbranchmodel$ is visualized as a dashed black rectangle. The leaves of the branches corresponding to the positive class are highlighted in yellow.

\subsubsection{Maximum Gain Boxes}
\begin{algorithm}[t]
\caption{\textsc{GreedyMaxGainBox($\greedytrainsubset$, $\vec{x}'$,  $\featuresubset$)}}
 \label{alg:greedy_max_gain_box}
\begin{algorithmic}[1]
\REQUIRE $\greedytrainsubset=\{(\vec{x}_1,y_1), \ldots, (\vec{x}_\greedytrainsubsetsize, y_\greedytrainsubsetsize)\} \subset \Reals^\tdim \times \{0,1\}$, $\vec{x}' \in \Reals^\tdim$, and $\featuresubset \subset \{1,\ldots,\tdim\}$ with $|\featuresubset| = \tdimsubsets$ for some $1 \leq \tdimsubsets \leq \tdim$.
\ENSURE Box $\decisionbranchbox \subset \RealsInfinity^\tdim$ with $\dtnumberbounded(\decisionbranchbox) \leq \tdimsubsets$ and gain $\greedygain\in\Reals$.
\STATE $\featuresubsetsequence = (i_1, \ldots, i_\tdimsubsets) \gets \textsc{RandomSequence}(\featuresubset)$ \label{alg:line:feature_sequence}
\STATE $ \decisionbranchbox\gets$ \textsc{InitialEmptyBox($\vec{x'}$, $\greedytrainsubset$, $\featuresubsetsequence$)}
\FOR{$j=1,\ldots,\tdimsubsets$}
%\STATE $P_L, P_R \gets$ \textsc{FilterPoints}($\vec{x}$, $\trainset$, $\decisionbranchbox$, $i_j$)
\STATE $\decisionbranchbox \gets \textsc{ExpandBox}(\vec{x'}, \greedytrainsubset, \decisionbranchbox, i_j)$ \label{line:expand_box}
\ENDFOR
\STATE $\greedygain \gets \textsc{Gain}(\greedytrainsubset, \decisionbranchbox)$
\STATE \textbf{return} $\decisionbranchbox, \greedygain$
\end{algorithmic}
\end{algorithm}
%\red{with $\dtnumberbounded(\decisionbranchbox_{opt}) \leq \tdimsubsets$}
%\red{induced by a $\featuresubset \in \featuresubsets$}
The function \textsc{GreedyMaxGainBox}, shown in Algorithm~\ref{alg:greedy_max_gain_box}, computes, for a given point $\vec{x'} \in \Reals^\tdim$, a subset $\greedytrainsubset \subseteq \trainset$ of the training instances, and a feature subset $\featuresubset \subset \{1,\ldots,\tdim\}$, a $\tdim$-dimensional box $\decisionbranchbox$ with $\dtnumberbounded(\decisionbranchbox) \leq \tdimsubsets$ along with an associated gain $\greedygain \in \Reals$. In a first step, a fixed yet random feature index sequence $\featuresubsetsequence$ is defined based on $\featuresubset$. In a second step, the function \textsc{InitialEmptyBox} computes an initial box~$\decisionbranchbox$ that only contains the point $\vec{x'}$ (and possibly duplicates of $\vec{x'}$) and heuristically covers as much empty space as possible %aims at maximizing the volume covered by the box 
in the dimensions specified by $\featuresubsetsequence$ (see Section~\ref{sec:implementation_details} for the implementation details).
%\footnote{\red{How do we handle duplicates, i.e., points being exactly the same as $\vec{x}$?}} 

\begin{figure}[t]
    \centering
    %\vspace{-0.3cm}
    \subfigure[Initial Box]{
        \includegraphics[width=0.22\columnwidth]{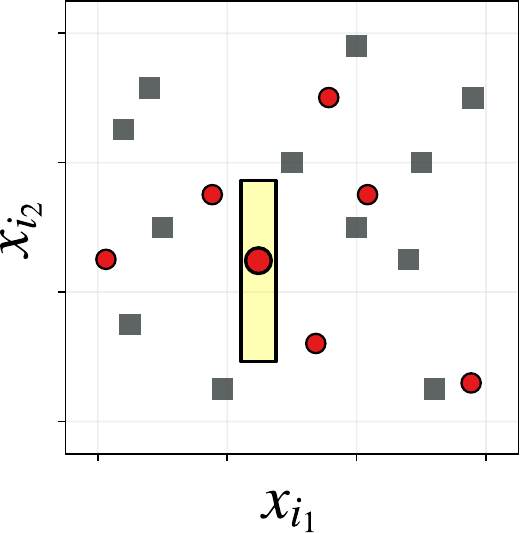}
        \label{fig:expand1}
    }
    \hfill
    \subfigure[$i_1$]{
        \includegraphics[width=0.22\columnwidth]{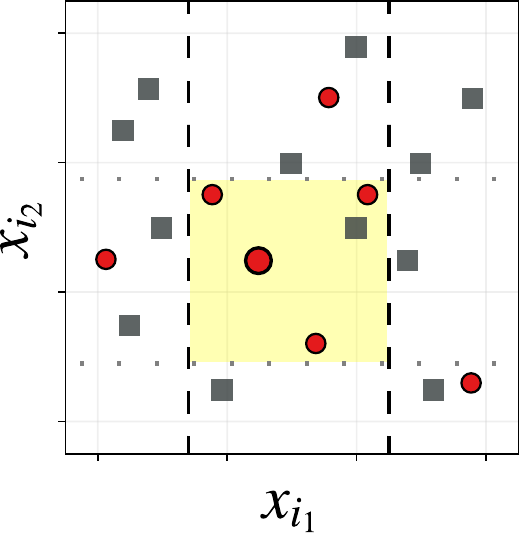}
        \label{fig:expand2}
    }
    \hfill
    \subfigure[$i_2$]{
        \includegraphics[width=0.22\columnwidth]{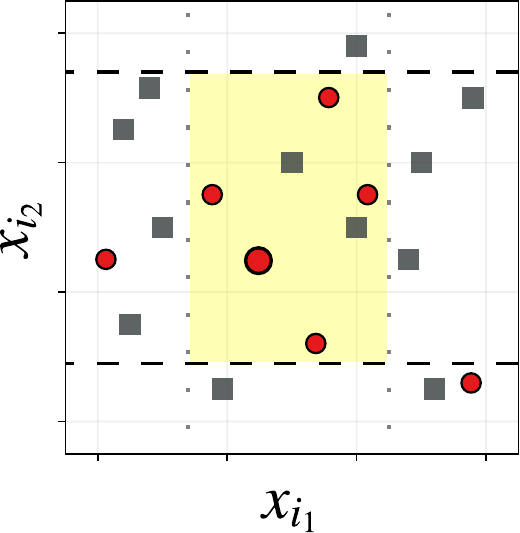}
        \label{fig:expand3}
    }
    \hfill
    \subfigure[Final Box]{
        \includegraphics[width=0.22\columnwidth]{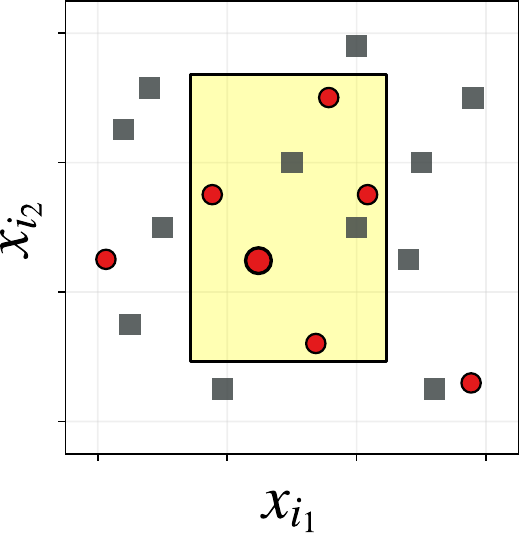}
        \label{fig:expand4}
    }
    %\vspace{-0.2cm}
    \caption{$\textsc{ExpandBox}$: The red points depict positive and the black points negative instances, respectively. The initially covered point corresponds to the positive instance $\vec{x}$.
    %The red points depict rare and the black points non-rare instances, respectively. The circled point corresponds to the rare instance $\vec{x}$.
    }
    \label{fig:expand}
\end{figure}

%\red{Marcos: Next part understandable?} 
Next, this box is expanded according to an information gain criterion defined below by iterating over the same feature index sequence~$\featuresubsetsequence$ and by applying the function \textsc{ExpandBox}, see Figure~\ref{fig:expand}.%The details are provided next.
% lincorporates the core idea of the box construction algorithm. In its essence, boxes are constructed for a given rare point $x$, a feature subset $S$ and training points $T$ that minimizes the gini impurity. The gini impurity, also known as gini index, is formally described in Equation \ref{eq:gini}. At first, the order of the features included in the considered feature subset $S$ are randomly shuffled. As the order on how the features are being processed plays an important role for resulting box, the order is randomized at every call. 
%is used to define the boundary in dimension $i_2$, \ie, the rare points $\vec{z} \in \greedytrainsubset - P$ with $z_{i_2}$ whose  .
%i.e. the points with equal values. In case there are no more points existing in one direction, the closest rare point is searched in that direction. By this heuristic, the chances are increased to find more rare instances during the expansion phase. Otherwise, there would be the risk that the initial boxes would be to small for finding any rare instance during the expansion. 
%The resulting box is characterized by only containing the center point $\vec{x}$ while being large enough for the subsequent expansion phase.
%\paragraph{Box Expansion}
%The initial box obtained via \textsc{InitialEmptyBox} is expanded in the next phase by iteratively expanding the dimensions in the same order as for $\textsc{InitialEmptyBox}$.
%
\begin{algorithm}[t]
 \caption{$\textsc{RemoveInstances}(T, \decisionbranchbox$)}
 \label{alg:remove_instances}
\begin{algorithmic}[1]
\REQUIRE Data set $T = \{(\vec{x}_1,y_1), \ldots, (\vec{x}_m, y_m)\} \subset \Reals^\tdim \times \{0,1\}$ and $\tdim$-dimensional box $\decisionbranchbox \subset \RealsInfinity^\tdim$ %$B \in \mathbb{R}^{2 \times \tdimsubsets}$
\ENSURE Data sets $\overline{T}\subset \Reals^\tdim \times \{0,1\}$ and $\overline{R}\subset \Reals^\tdim \times \{0,1\}$
\STATE $\overline{R} \gets \{(\vec{x},y) \in T \mid \vec{x} \in B\}$
\STATE $\overline{T} \gets \{(\vec{x},y) \in T \mid \vec{x} \notin B\}$
% \STATE $\overline{R} \gets \{\}; \overline{T} \gets \{\}$
% \FOR{$(\vec{x}, y) \in T$}
%     \IF{$\vec{x} \in B$}
%         \STATE $\overline{R} \gets \overline{R} \cup \{(\vec{x}, y)\}$
%     \ELSE
%         \STATE $\overline{T} \gets \overline{T} \cup \{(\vec{x}, y)\}$
%     \ENDIF
% \ENDFOR
\RETURN $\overline{T},\overline{R}$
\end{algorithmic}
\end{algorithm}
%}
More precisely, for each dimension~$i_j$ in Line~\ref{line:expand_box}, both the left and right boundary of the current box~$\decisionbranchbox$ are expanded w.r.t. dimension~$i_j$. This is done by only considering the points of $\overline{\greedytrainsubset} \subset \greedytrainsubset$ that are contained in a box $\overline{\decisionbranchbox}$, which is the same box as $\decisionbranchbox$ except that the left and right boundary in dimension $i_j$ are set $-\infty$ and $+\infty$, respectively. 
%The box~$\decisionbranchbox$ is then expanded in both directions. 
To expand the left boundary in dimension~$i_j$, the points $\vec{x} \in \overline{\greedytrainsubset}$ with $x_{i_j} < x'_{i_j}$ are sorted in increasing order w.r.t. their distance $|x_{i_j} - x'_{i_j}|$ to the point~$\vec{x'}$ in dimension~$i_j$. Starting with $I=\{\vec{x'}\}$, the sorted points are then processed incrementally and added to the set $I$. 
Let
$O=\greedytrainsubset \backslash I$ be the set of remaining ``outer'' points. To decide if the left boundary should be expanded, we compute the gain $G(\greedytrainsubset)$ via:
\begin{equation}
     G(\greedytrainsubset) = Q(\greedytrainsubset) - \frac{|I|}{|\greedytrainsubset|} Q(I) - \frac{|O|}{|\greedytrainsubset|}Q(O)
\end{equation}
As for decision trees, $Q$ is a classification impurity measure such as the Gini index. 
After all or a pre-defined number $\greedymaxpoints$ of points (\eg, $\greedymaxpoints=20$) have been processed, the new left boundary $l_{i_j}$ of $\decisionbranchbox$ in dimension $i_j$ is set to the value so that the gain $G(\greedytrainsubset)$ is maximized.
%by setting the left boundary so that the new box contains the corresponding point.
The right boundary $r_{i_j}$ of $\decisionbranchbox$ is computed similarly. 
The expansion is repeated for $j=1,\ldots,\tdimsubsets$.
%\footnote{The instances contained in such a box are removed via the function \textsc{RemoveInstances} in Algorithm~\ref{alg:bottom_up_construct}. To avoid overlapping boxes per model $\decisionbranchset$ computed by \textsc{BottomUpConstruct}, some of the boxes might have to be pruned in a post-processing step.}
In Figure~\ref{fig:expand}, the overall process is visualized for the case of $\tdimsubsets=2$.
%In the expansion phase, the dimensions of the box are also extended sequentially in the same order as in the initialization phase. An example is shown in Figure \ref{fig:expand}. At first, all points are filtered which reside inside the box area excluding the dimension of interest as illustrated in Figure \ref{fig:expand2}. Thereby, only the points remain with the potential of being covered by the box when expanding in this dimension. In the code, the filtering is conducted by the function \textsc{FilterPointsLeftRight()}. It returns the remaining points separated into the points smaller than the box $P_L$, i.e. left of the box, and the points greater than the box boundaries $P_R$, i.e. right from the box. In the next step, the boundaries of the box are expanded in \textsc{ExpandBox()}. For each direction, the remaining points are sorted from closest to furthest point to the respective boundary. Then, the border is expanded step-wise among the sorted points. At every step, the information gain is calculated and compared to the current best gain. In case the new boundary has a higher gain than the current best, the best gain is updated. 
% \begin{itemize}
%     \item max gain, because we want to have few (!) boxes covering the rare instances
% \end{itemize}

Overall, the goal of \textsc{ExpandBox} is to expand the initial box in such a way that information gain is maximized (similarly to top-down construction of decision trees). 
%Intuitively, we expect that a range search will be conducted for each of these boxes; hence, a minimal set of boxes that ``cover'' all the rare instances is desirable. 
At the end of Algorithm~\ref{alg:greedy_max_gain_box}, the function \textsc{Gain} is used to compute the final gain induced by the ``split'' of $\greedytrainsubset$ into $I=\greedytrainsubset \cap \decisionbranchbox$ and $O=\greedytrainsubset \backslash I$. Both the final box $\decisionbranchbox$ as well as the final gain $\greedygain$ are returned by \textsc{GreedyMaxGainBox}.
\subsubsection{Comparison with Decision Trees}
Our construction scheme is hypothesized to produce fragments of standard decision trees, as they share common characteristics such as orthogonal splits, splits respecting infinity, and training based on an impurity criterion. In principle, these models could be generated via a top-down approach, which suggests that they belong to the same model space and are expected to achieve comparable generalization (confirmed empirically in Section~\ref{sec:model_comparison}).
Yet, each box along with its associated decision branch only resort to a few features, which will enable fast spatial look-ups %of the corresponding instances 
in the database. Also note that, for each positive instance, a large set of $\tsizesubsets$ feature subsets is considered to identify an appropriate box containing mostly positive instances, which is motivated by the observation that, when constructing normal decision trees, care must (only) be taken when constructing the lower parts of the trees (\ie, essentially, arbitrary splits can be made in the top parts of trees, whereas discriminative feature splits are needed for the lower parts)~\cite{GiesekeI18}. 
%As we will show in our experimental evaluation, the decision branches exhibit a classification performance that matches the ones of standard decision trees and random forests.
\subsubsection{Decision Branch Ensembles}
\label{sec:ensembles}
The bottom-up construction scheme for the decision branches is driven by introducing randomness in various phases. %of the different algorithmic building blocks.
For instance, a random sequence~$\featuresubsetsequence$ is defined based on~$\featuresubset$ in $\textsc{GreedyMaxGainBox}$. While this use of randomness is not strictly needed for single calls to \textsc{DecisionBranches}, it is crucial when generating ensembles of decision branches, as is the case for standard tree ensembles such as random forests and extremely randomized trees.
Based on this observation, deriving 
%Extending decision branches to 
an ensemble consisting of decision branch models is then straightforward and can be done by simply combining the sets $\decisionbranchset_1,\ldots,\decisionbranchset_\ndbensemble$ obtained by $\ndbensemble$ calls of \textsc{DecisionBranches}, where only the instances of the database are returned that are classified as positive by the majority of the models.
% (\eg, selecting the rare instances $(\vec{x},y) \in \trainsetnonrare$ at random, randomizing order specified by a feature subset $\featuresubset$, \ldots). 
% \begin{itemize}
%     \item \red{ (since the construction is influenced by the order in which the features are processed)}
% \end{itemize}
% As for standard tree ensembles, such as random forests and extremely randomized trees, decision branches can be combined to obtain decision branch ensembles, which usually exhibit a better performance than the individual decision branch sets.
\subsection{Fast Query Processing}
%\subsection{Fast Rare-Object Query Processing}
\label{sec:fast_rare_object_query_processing}
\begin{figure}[t]
    \centering
    % \vspace{-0.5cm}
	%\includegraphics[width=1.0\columnwidth]{figures/figure1/Figure_1_part2.pdf}
 \def\svgwidth{\columnwidth}
\import{figures/figure1}{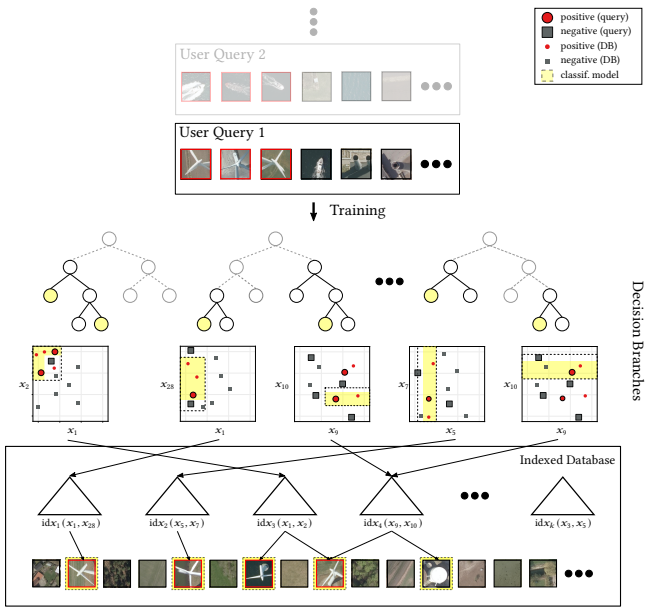}
%\vspace{-0.3cm}
	\caption{Answering user queries via decision branches and index structures pre-built for large databases.}
	\label{fig:main_idea}
 %\vspace{-0.2cm}
\end{figure}
The remaining task is efficient inferencing via range queries (Figure~\ref{fig:framework}).
Overall, we aim at processing many user queries with small latencies given a large database. In addition, we assume limited computational resources; in particular, our approach does not rely on the use of many servers. 
As mentioned above, we begin by pre-building a large number of index structures $idx_1, \ldots, idx_\tsizesubsets$ in the offline preprocessing phase.
%, one for each of the random feature subsets $\featuresubset_1,\ldots,\featuresubset_\tsizesubsets \subset \{1,\ldots,\tdim\}$ that are used by Algorithm~\ref{alg:bottom_up_construct}. 
Note again that these indexes have to be built only \emph{once} beforehand and are independent of a particular user query. 
The processing of a new user query, \ie, for a new data set $\trainset=\{(\vec{x}_1,y_1), \ldots, (\vec{x}_\tsize, y_\tsize)\} \subset \Reals^\tdim \times \{0,1\}$ consisting of a few positive and many negative instances, then proceeds as follows:
\begin{enumerate}[(1)]
    \item In the first step, a set of decision branches (corresponding to a decision tree) is built via the function \textsc{DecisionBranches}. 
%Thus, given a new user query, a set of decision branches is built based on the feature subsets $\featuresubset_1,\ldots,\featuresubset_\tsizesubsets$.
\item In the second step, we conduct, for each decision branch~$\decisionbranchmodel$ with associated box~$\decisionbranchbox$ and feature subset~$\featuresubset$, an orthogonal range query using the index structure built beforehand for~$\featuresubset$ to retrieve all instances of the database contained in $\decisionbranchbox$.
\item Finally, in the third step, these instances are classified into positive or negative by the corresponding decision branches, where only positive instances are returned to the user. %filtered by the corresponding decision branches to obtain the instances of the database being classified as positive \blue{(\ie, all instances that end up in the positive leaves of decision branches; see Figure \ref{fig:main_idea}).
\end{enumerate}
Note that a crucial ingredient for the efficiency of our query processing is that each resulting box $\decisionbranchbox \subset \Reals^\tdim$ associated with a decision branch is only (half-)bounded in a few dimensions, \ie, $\dtnumberbounded(\decisionbranchbox) \leq \tdimsubsets$, where $\tdimsubsets$ is a small constant (\eg, $\tdimsubsets=4$). This property stems from the fact that such boxes are constrained using only the features provided by a set $\featuresubset \subset \{1,\ldots,\tdim\}$ with $|\featuresubset| = \tdimsubsets$, see again Algorithm~\ref{alg:greedy_max_gain_box}.
Thus, the index that is available for that feature subset $\featuresubset$ can be used to efficiently conduct the corresponding range query. 
%The extension to decision branch ensembles $\decisionbranchset_1,\ldots,\decisionbranchset_\ndbensemble$ consisting of $\ndbensemble$ models is straightforward; here, only the instances of the database that are classified as positive by the majority of the models are returned.

The workflow is illustrated in Figure~\ref{fig:main_idea}. 
To sum up, for each user query given as a classification data set, a set of decision branches is constructed. Each decision branch corresponds to a box (black dashed rectangles), and the database instances being contained in one of the boxes can quickly be retrieved using the associated index. The resulting (small) answer sets are then filtered for positive instances via the corresponding decision branch for each box.

\subsection{Implementation Details}
\label{sec:implementation_details}
Here we provide some details related to \textsc{InitialEmptyBox} and \textsc{TopDownConstruct} used in Algorithm~\ref{alg:bottom_up_construct}, as well as to the memory layout and space consumption of the index structures. %in the offline preprocessing phase.
%The reader can skip these rather technical parts and to directly move on to Section~\ref{sec:experiments}.
%Next, we provide details related to the function \textsc{InitialEmptyBox} used in Algorithm~\ref{alg:greedy_max_gain_box}, to the hybrid memory scheme for \kdtrees, to different ways of implementing the function \textsc{TopDownConstruct} in Algorithm~\ref{alg:bottom_up_construct}, as well as to the extraction of features for the geospatial case study.
\subsubsection{Initial Box}
\begin{figure}[t]%[h]
    \centering
    %\vspace{-0.2cm}
    \subfigure[Start]{
        \includegraphics[width=0.22\columnwidth]{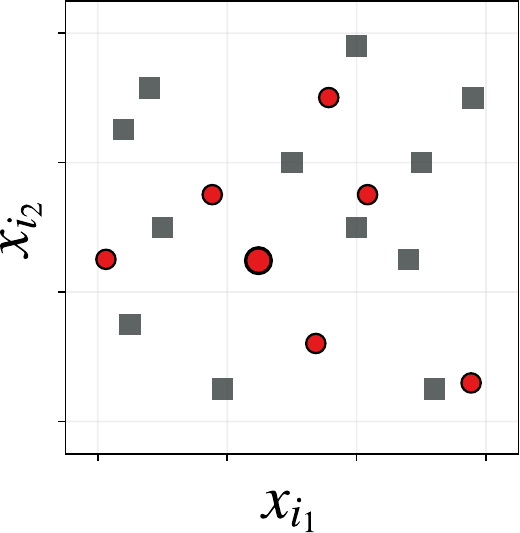}
        \label{fig:init1}
    }
    \hfill
     \subfigure[$i_1$]{
        \includegraphics[width=0.22\columnwidth]{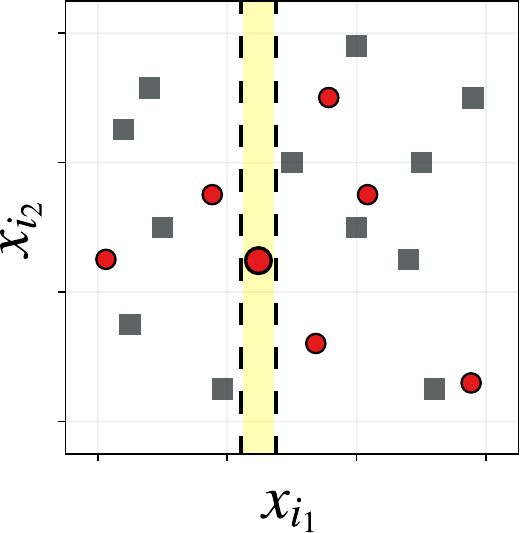}
        \label{fig:init3}
    }
    \hfill
    \subfigure[$i_2$]{
        \includegraphics[width=0.22\columnwidth]{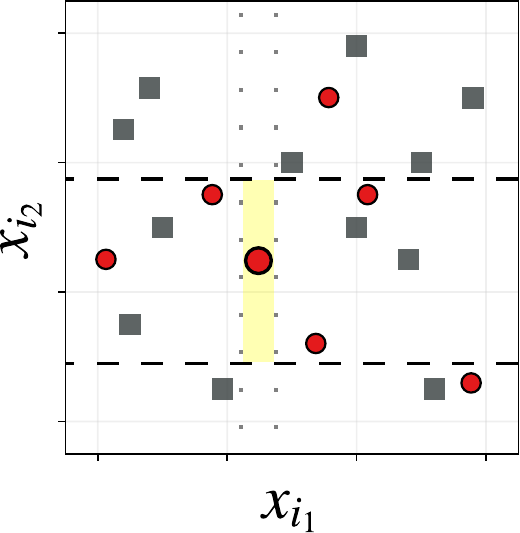}
        \label{fig:init2}
    }
    \hfill
    \subfigure[Box]{
        \includegraphics[width=0.22\columnwidth]{figures/box_construction/init/Init_4.pdf}
        \label{fig:init4}
    }
    %\vspace{-0.1cm}
    \caption{Illustration of \textsc{InitialEmptyBox}.
    %The outlined red point indicates the center point $\vec{x}$.
    }
    \label{fig:init}
%    \vspace{-0.2cm}
\end{figure}

%Afterwards, an initial box is created within the function \textsc{InitialEmptyBox}.
%\red{Marcos: this section ok?} 
For the sake of completeness, we provide details related to the function \textsc{InitialEmptyBox}. %; the reader is invited to skip these rather technical details.
For a given point $\vec{x'} \in \Reals^\tdim$, a subset $\greedytrainsubset$ of training instances, and a sequence $\featuresubsetsequence=(i_1, \ldots, i_\tdimsubsets)$, the procedure \textsc{InitialEmptyBox} constructs
%The procedure \textsc{InitialEmptyBox} constructs, for a given point $\vec{x'} \in \Reals^\tdim$, a subset $\greedytrainsubset$ of training instances, and a sequence $\featuresubsetsequence=(i_1, \ldots, i_\tdimsubsets)$,
%$\featuresubset \subset \{1,\ldots,\tdim\}$ with $|\featuresubset|=\tdimsubsets$ for some $1 \leq \tdimsubsets\leq \tdim$, 
a $\tdim$-dimensional box $\decisionbranchbox$
%with $\dtnumberbounded(\decisionbranchbox) \leq \tdimsubsets$ 
that only contains the query point $\vec{x'}$ and no other points. At the same time, this box also covers as much empty space as possible in the dimensions specified by $\featuresubsetsequence$. 
The box is computed by deriving bounds for the dimensions specified by $\featuresubsetsequence$; the left and right bounds for the other dimensions are set to $-\infty$ and $+\infty$, respectively. 
Identifying such a box is known to be computationally challenging, even for low-dimensional spaces.\footnote{%Strictly speaking, maximizing the volume is only well-defined for boxes that are bounded (\ie, not bound can be $-\infty$ or $\infty$). 
This problem is known as the maximal empty rectangle containing a query point, \ie, finding the largest axis-parallel rectangle $R$ over a set of $\tsize$ points such that $R$ contains only a given query point. In high-dimensional spaces, finding $R$ is non-trivial, with recent solutions in $\O(log^4\tsize )$ time complexity~\cite{kaplan2011finding}. %When no query point is given, recent research describes algorithms with 
%$\O(\tsize 2^{\O(\log^*\tsize)}\log \tsize)$ time for $\tdim = 2$ and $\tilde{O}(\tsize^{(5\tdim+2)/6})$ time for $\tdim \geq 4$~\cite{chan2021fast}.
We propose a simple yet effective heuristic that (a) efficiently yields good initial boxes fulfilling the desired properties and that (b) is also driven by some randomness such that different boxes are obtained across different runs. As for the other functions, introducing randomness is important for the construction of ensembles (see above). We illustrate the process of box initialization with an example in Figure \ref{fig:init}.}

\subsubsection{Decision Branch Models}
\label{sec:DecisionBranchModels}
%We present two approaches to implement the function \textsc{TopDownConstruct} in Algorithm~\ref{alg:bottom_up_construct}. 
In case $\trainsetnonrareremoved \cup \trainsetrareremoved$ is pure, \textsc{TopDownConstruct} in Algorithm~\ref{alg:bottom_up_construct} returns a single node. 
Given an impure set in $\decisionbranchbox_{opt}$ (see, \eg, Figure~\ref{fig:expand}d), the procedure \textsc{TopDownConstruct} builds a tree $\decisionbranchmodel$ for the instances given in $\trainsetnonrareremoved \cup \trainsetrareremoved$. In our experiments, we consider two versions to implement this function. 
The first one, T\textsubscript{a}, resorts to all $\tdim$ features. The second one, T\textsubscript{s}, only considers the $\tdimsubsets$ features specified by the feature subset $\featuresubset$ that corresponds to the box $\decisionbranchbox_{opt}$ (\ie, $\decisionbranchmodel \subset \decisionbranchbox_{opt}$ holds true). The benefit of the first implementation is that it generally yields slightly better models in terms of classification accuracy. However, it requires retrieving the full set of $\tdim$ features returned by the range query. %(in case the answer set of the search query is small, this is generally not a big issue).
The benefit of the second approach is that the~$\tdimsubsets$ features can directly be stored in the leaves of the index structures (using $\O(\testsize \tdimsubsets)$ additional space), which, in turn, allows for immediate retrieval of the features needed when processing the range queries. However, this variant generally yields slightly worse classification results than T\textsubscript{a}, see Section~\ref{sec:experiments}.

\subsubsection{Hybrid Memory K-D Trees}
\label{sec:hybrid_memory_kdtrees}
%A potential drawback of our approach is that a quite large number $\tsizesubsets$ of multidimensional index structures have to be built beforehand. While each such structure only needs $\O(\testsize)$ additional space for $\testsize$ data points, a large number $\tsizesubsets$ of index structures might have to be built. Hence, one might quickly run out of main memory of the given system. 
A large number $\tsizesubsets$ of index structures have to be pre-built for a large database. 
For \kdtrees, $\O(\testsize)$ additional space is needed, where $\testsize$ is the number of elements in the database. Hence, given a large $\tsizesubsets$, it might not be feasible to keep all these index structures in the main memory of a system.
In order to address this problem, we implement an efficient \kdtree-based index that supports multidimensional range queries and leverages disk storage. 
%Assuming the total amount of the data would exceed the memory capacity of the underlying machine, 
In particular, our \kdtree implementation only stores the tree structure up to the leaves in main memory. The leaves containing the instances with all their features are stored consecutively on disk. If a query is now forwarded to the index structure, only the leaves that intersect with the query rectangle need to be loaded from disk.
%\footnote{
A crucial hyperparameter for the success of our \kdtree index is the leaf size. The smaller the size of the leaves, the less data not contained in the query rectangle is falsely loaded. Since the bottleneck of this method will tend to be the data loading time from disk, it can be expected that a smaller leaf size leads to faster query times. On the other hand, the loading time is also influenced by the block size of the underlying (file) system: 
%\todot{or is it disk block size too?}. 
%In the most common file systems, data is loaded from disk in form of fixed size blocks. Therefore, 
leaves that are smaller than the actual block size cannot make proper use of device bandwidth. Additionally, the larger the leaf size, the smaller are the parts of the \kdtrees that have to be kept in main memory.
Ultimately, we meet a trade-off among all of these factors, making the optimal leaf size a critical hyperparameter that has to be determined before constructing the indexes. As we will show in our experiments, it is possible to keep the top parts of \kdtrees in main memory for a large number \tsizesubsets~of index structures.

\section{Experiments}
\label{sec:experiments}
% \begin{itemize}
%     \item \todot{Christian/Denis: We need to save space. We have to make the plots "smaller". For instance, we can group similar figures (e.g., 7 8, 9, 10) and use figure* for one big joint figure. The previous figures are subfigures. Then, for each figure, we only use 0.24 of the textwidth...if that is too small, we only combine three figures that way. Generally, I think one can group two figures next to each other also for the thin column figures. I also find the figures a bit too large at the moment...this would solve both problems. Figures 16 and 17 can also be made smaller I think. }
% \end{itemize}

\begin{table}[t]%[t]
    \caption{Summary of model notation.}
    \vspace{-5pt}
    \resizebox{1.0\columnwidth}{!}{

    \begin{tabular}{p{0.15\linewidth}p{0.85\linewidth}}
        \toprule
        Parameter     & Description\\
        \midrule
        \textit{B}    & Bottom-up (depth of $\decisionbranchmodel$ is 0) \\ 
        \textit{T\textsubscript{s}}  & Bottom-up + top-down (using $\tdimsubsets$ features)\\ 
        \textit{T\textsubscript{a}}  & Bottom-up + top-down (using all $\tdim$ features)\\ 
        \textit{<num>t}
        & \dbensemble: number of  estimators (default: \textit{25t})\\ 
        \textit{<num>} & \decisionbranch/\dbensemble: \tdimsubsets\@\xspace (default: 10)\\
        \textit{<num>} & \decisiontree/\randomforest: \# of features used (default: all)\\
        \bottomrule      
    \end{tabular}
    }
\label{tab:db_parameter}
\end{table}

We provide an extensive evaluation of our approach w.r.t. three aspects:
\begin{inparaenum}[(1)]
    \item efficiency of index-supported \tdecisionbranches and ensembles in terms of inference runtime against scan-based evaluation of decision trees and random forests as well as nearest neighbor search 
    %in a realistic geospatial search case study over a large collection of satellite imagery 
    (Section~\ref{sec:geospatial_search});
    \item sensitivity of \tdecisionbranches models to crucial hyperparameters and comparison to traditional tree-based algorithms in (imbalanced) binary classification tasks (Section~\ref{sec:modelbenchmark}); and 
    %\item sensitivity of \tdecisionbranches models to \blue{crucial} hyperparameters %such as~$\tdimsubsets$ and~$\tsizesubsets$ 
    %(Section~\ref{sec:sensitivity});  
    %\item empirical performance in comparison to traditional tree-based algorithms in (imbalanced) binary classification tasks (Section~\ref{sec:model_comparison}); and
    \item trade-off between storage consumption of the underlying index structures and performance (Section~\ref{sec:space_consumption}).
\end{inparaenum}
In particular, we consider a realistic case study 
%in the geospatial domain 
with more than a billion data points. The results show that our search-by-classification framework yields as accurate results as traditional search-by-classification schemes (that have to scan the entire database per user query), albeit at dramatically reduced query processing times. %, while returning the results as fast as traditional nearest neighbor search, see again Figure~\ref{fig:motivation}.

\subsection{Experimental Setup}
All experiments were conducted on an Ubuntu 18.04~server with 24~\textit{AMD EPYC 7402P} cores, 192~GB DDR4-RAM and 30~TB of NVMe storage. The construction schemes for \tdecisionbranches and for the hybrid memory \kdtree were implemented in Python~3.8, where Cython was used for computationally intensive parts.%\footnote{Our source code including our implementation for decision branches, the overall search framework, and scripts for all experiments is provided at \url{https://github.com/decisionbranches/sigmod_decisionbranches}}
% \begin{enumerate}
%     \item \textbf{Decision Tree (\decisiontree)}: decision tree classifier from the scikit-learn library~\cite{scikit-learn}.
%     \item \textbf{Random Forest (\randomforest)}: random forest classifier from the scikit-learn library~\cite{scikit-learn}.
%     \item \textbf{Extremely Randomized Trees (\extratrees)}: Extremly randomized trees classifier from the scikit-learn library~\cite{scikit-learn}. 
%     \item \textbf{Decision Branches (\decisionbranch)}: our core \tdecisionbranch approach as presented in Section~\ref{sec:approach}. %   box search algorithm in combination with the decision tree extension. Implemented in Numpy/Python.
%     \item \textbf{Decision Branches Ensemble (\dbensemble)}: an ensemble of \tdecisionbranch models as presented in Section~\ref{sec:approach}. %our box search algorithm as an ensemble of 25 models together with the decision tree extension. Implemented in Numpy/Python.
% \end{enumerate}
%\noindent \textbf{Our approach and variants}.
\subsubsection{Models}
We consider the Scikit-Learn~\cite{scikit-learn} implementations of decision trees~(\textbf{\decisiontree}), random forests~(\textbf{\randomforest}), and extremely randomized trees~(\textbf{\extratrees}) as baseline search-by-classification models (which retrieve the positive instances by scanning the entire database). These models are compared with our novel \tdecisionbranch approach~(\textbf{\decisionbranch}) as well as with a corresponding ensemble variant~(\textbf{\dbensemble}) introduced in Section~\ref{sec:ensembles}.
For both \decisionbranch and \dbensemble, three versions are considered. For the first version~(\textit{B}), \textsc{TopDownConstruct} returns a single node, \ie, all the decision branch models have depth zero. The other two versions, \textit{T\textsubscript{s}} and~\textit{T\textsubscript{a}}, are described in Section~\ref{sec:DecisionBranchModels}. For \textit{T\textsubscript{s}}, \textsc{TopDownConstruct} only uses the features $\featuresubset \in \featuresubsets$ that yielded the box $\decisionbranchbox_{opt}$ (see Algorithm~\ref{alg:bottom_up_construct}). In contrast, \textsc{TopDownConstruct} resorts all the $\tdim$ features for \textit{T\textsubscript{a}}.

An overview of the notation is provided in Table~\ref{tab:db_parameter}. 
In our evaluation, we aim to demonstrate that our decision tree and tree ensemble variants yield a competitive classification performance compared to the aforementioned baselines, without the need to scan the entire data catalog. Additionally, we include an efficient exact nearest neighbor search baseline (denoted by \textbf{\nnb}) for comparison, which uses our \kdtree implementation \cite{Bentley79} to accelerate the search. This baseline indicates the classification performance that users would experience when interacting with an NN-based search engine that only allows single-instance search queries, an approach commonly found in visual search engines \cite{KEISLER2019visualsearch}.
To establish a fair comparison, we treat the \nnb as a model, where all~$\knearestneighbors$ nearest neighbors produced given a user query are considered to be positive, while all other instances are considered to be negative.
\footnote{The value of $\knearestneighbors$ is set to the true number of corresponding positive instances per user query/training set. Note that such important information is actually not known in practice, which, hence, gives the \nnb an unfair advantage.}
We assess its classification performance as an average among all single-instance searches obtained from the positive instances in a training set, thus abstracting a notion of typical result quality.

%We already convey additional information to the search and which is not known to the compared classifiers.

%\noindent\textbf{Metrics.} 
\subsubsection{Metrics}
We assess the quality of the models using the $\fscore$-score, defined as the harmonic mean of precision~$\precision$ and recall~$\recall$, \ie, $\fscore= 2 \cdot \nicefrac{\precision \cdot \recall}{(\precision + \recall})$. 
The application-level impact of a high $\fscore$-score (\eg, \textasciitilde 0.8) is achieving a good trade-off between precision and recall, where the system misses only a few relevant objects during the search.
%Given an $\fscore$-score of 0.8 in image retrieval, it means that the system has achieved a balanced performance in terms of precision and recall. It suggests that \textasciitilde 80\% of the retrieved images are relevant (precision), while only \textasciitilde 20\% of relevant images are missed (recall).
In addition to measuring the total execution time~\totaltime for the approaches, we also measure the training time~\trainingtime and the query time~\querytime for the different models, where the former captures the time needed to build the classification models and the latter to retrieve instances from the database.

%evaluate the total execution time \totaltime of the index-supported \tdecisionbranch models per user query, we measure the training time~\trainingtime and the query time~\querytime for retrieving the data separately.

%For all considered models, we also measure the training time (i.e., time to construct the boxes) and the query time (i.e., time to retrieve the data)

%\noindent \textbf{Reproducibility}. 
%Additional pieces of information regarding the experimental setup are provided in the supplemental material.

%Both the bottom-up construction scheme for decision branches as well as the hybrid memory version of $k$-d trees (see below) are majorly implemented in Python~3.8, where Cython was used for the computationally intense parts. 
%Our source code including our implementation for decision branches and the overall search framework as well as the scripts for all experiments is provided via an anonymous GitHub repository. 
%\footnote{Available at \url{https://github.com/decisionbranches/sigmod_decisionbranches}}
%The repository includes detailed instructions on installation requirements and reproducibility of the experiments conducted over the publicly available data sets (see Section~\ref{sec:modelbenchmark}). 
%The geospatial data mentioned in Section~\ref{sec:geospatial_search} is not publicly available yet and, therefore, cannot be included in the code repository.

\subsection{Efficiency Study of Index-Aware Models}
\label{sec:geospatial_search}

\begingroup
\setlength{\intextsep}{8pt}%
\setlength{\columnsep}{5pt}%
\begin{wrapfigure}{r}{4.2cm}
    \centering
    %\vskip-0.1cm
    \subfigure[Chimney]{
        \mycolorbox{\includegraphics[width=0.1\columnwidth]{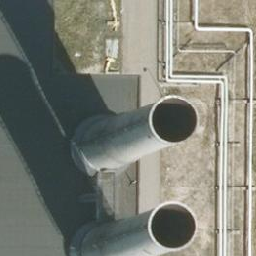}}
    }
    \subfigure[Plane]{
        \mycolorbox{\includegraphics[width=0.1\columnwidth]{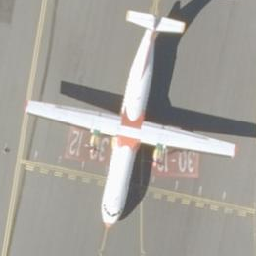}}
    }
    % \subfigure[Pool]{
    %     \includegraphics[width=0.1\textwidth]{figures/classes/pool.png}
    % }
    \subfigure[Ship]{
        \mycolorbox{\includegraphics[width=0.1\columnwidth]{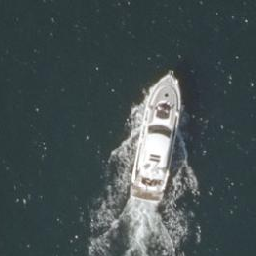}}
    }
    \\ %\vspace{-6pt}
    \subfigure[PV~Panel]{
        \mycolorbox{\includegraphics[width=0.1\columnwidth]{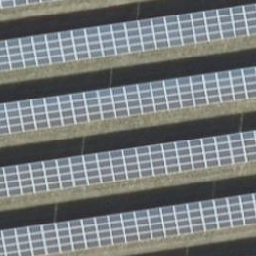}}
    }
    \subfigure[Tank]{
        \mycolorbox{\includegraphics[width=0.1\columnwidth]{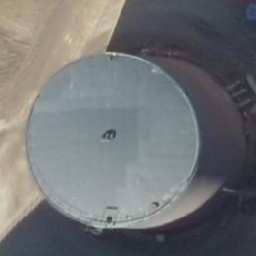}}
    }
    \subfigure[Turbine]{
        \mycolorbox{\includegraphics[width=0.1\columnwidth]{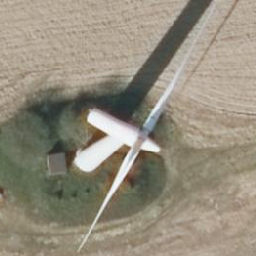}}
    }
    %\vspace{-0.5ex}
    \caption{Aerial Data Set}
   % \vspace{-1ex}
    \label{fig:classes}
\end{wrapfigure}
We begin by demonstrating the effectiveness of our approach through a geospatial case study, using 12.5 cm per pixel resolution aerial imagery of Denmark from 2018 (non-public data set). The images are decomposed into 256×256 pixel patches, resulting in a data set size of $\testsize=1,441,557,000$ patches.
% We first demonstrate the effectiveness of our approach for a realistic geospatial case study.
% We use aerial data with a resolution of $12.5$~cm per pixel and imagery of Denmark from 2018 (non-public data set), with the images being decomposed into patches of size $256\times256$ pixels.
% %with a stride of 12$\times$12 pixels. 
% %For evaluation, the data set is split into a training and a test area with a 75\% to 25\% ratio. 
% In total, the database contains $\testsize=1,441,557,000$ patches. 
We identified and manually collected labels for six classes of potentially interesting objects, namely chimneys, planes, ships, PV panels, storage tanks, and (wind) turbines, see Figure~\ref{fig:classes}.
Feature extraction was performed via a ResNet101 neural network~\cite{HeZRS15} pretrained on ImageNet~\cite{ILSVRC15}, where the feature embedding layer was adapted to yield $\tdim=50$ features. The model weights were fine-tuned by training the network for a few epochs using labeled patches over~7 classes (\ie, those depicted in Figure~\ref{fig:classes} and an additional class for ``other objects'') and the Adam optimizer~\cite{adam} with a learning rate of $0.0001$.\footnote{ResNet101 yielded satisfying results~\cite{HeZRS15}. Although several alternative methods could be applied for the purpose of feature extraction, we argue that the quality of features would affect all downstream models, including our competitors, random forests and decision trees. The extraction of ``optimal'' features depends on the data considered and is not the main goal of our study.} Note that the ResNet101 was used only as a pre-processing step, since its application would have required scanning the complete data set for every user query.

\endgroup
% Examples for each class are shown in Figure~\ref{fig:classes}. 
%For composing the individual user queries, we consider a data set consisting of labeled patches. 

To generate realistic queries, we considered, for each user query, a set of $30$ ``rare'' patches ($y=1$) belonging to the target class (\eg, PV panels). We also utilized $30,000$ ``non-rare'' patches ($y = 0$) for each query, which were selected uniformly at random from the catalog (containing $\testsize$ patches). This approach can be justified by (a) the assumption that non-rare patches are extremely more frequent than rare ones for the search tasks addressed (\eg, only about one out of one hundred thousand patches corresponds to a PV panel, which leads to only a few falsely labeled instances in our user query) and by (b) that a user query in practice might also contain a few falsely labeled instances (\eg, non-rare patches labeled as rare).
\footnote{
Note that machine learning models, including those considered in this work, can typically deal with a small amount of such ``label noise''. %Note that, in practice, large amounts of non-rare instances can generally also be easily collected in the context of such a geospatial search. For example, all non-rare patches being contained in a rectangular area can be labeled as ``non-rare''; such techniques are already implemented in our graphical search engine.
%Random forests/our decision branch ensembles are generally also robust enough to account for a few falsely labelled instances.
}

\subsubsection{Query Response Time and Classification Quality.}
%Building on top of the insights acquired from the previous experiments, we can now report the key findings of the case study related to both the efficiency and the effectiveness of our search-by-classification framework.
We compare the average total time~\totaltime of user queries induced by the six classes shown in Figure~\ref{fig:classes}.
%and evaluated the models on a large-scale setting with
%$\testsize =1,441,557,000$ instances. 
As detailed above, each query was composed of $30,030$ training instances. 
The quality in terms of $\fscore$-score was computed using a hold-out set consisting of $110,000$ labeled instances, whereas the runtime results were obtained using the entire database with $\testsize$ instances. We repeated the experiment five times with different seeds (accounting for randomness) and report averaged runtimes and scores.

%For the \tdecisionbranch models, we consider a cold setting for the \kdtrees to avoid benefiting from caching (a warm setting generally leads to even lower query times). 
%\blue{For the \tdecisionbranch models, we ensure no caching effect is present that could influence \kdtrees query times.}
The \kdtree indexes were built for $\tdimsubsets=3$ and different values for $\tsizesubsets \in \{50,150,300\}$ with a fixed leaf size of $5,632$. %For all models, we measure the training time \trainingtime, querying time \querytime as well as the test $\fscore$-score.
To implement a fast \nnb, we built another \kdtree using all the $\testsize$ instances in the database and three (random) features. The leaf size was set to 43 and the full tree was loaded into memory (including the leaves). We measured the query time for \nnb by searching for the $\knearestneighbors=1,000$ nearest neighbors of $100$ labeled interesting data instances and report the average query times. The $\fscore$-score was obtained via the hold-out set of $110,000$ labeled instances, where for each user query instance, $\knearestneighbors$ was set to the (true) number of positive instances in the entire database.

%To mimic a real use case, we pick a random rare instance from the $110,000$ labeled instances and search for $\knearestneighbors=1,000$ nearest neighbors in the query set ($30,030$ instances). 
%to assess the query time and report average results over 100 queries.
Table~\ref{tab:geospatial_results} summarizes the results obtained. For $\tsizesubsets=300$ (right column), the fastest \decisionbranch model (\decisionbranch\textsubscript{[B,3]}) achieved a 715$\times$ \emph{lower} \totaltime than the \decisiontree model, while exhibiting a comparable $\fscore$-score. In the ensemble case, ensembles with of~5 and~25 \tdecisionbranch models were considered, respectively. 
%\red{\dbensemble with all features is not listed in the results due to the large overhead originated from the additional loading and filtering phase, which can already be observed for a single \decisionbranch [B,T$_\text{a}$,3].} 
Here, the fastest \dbensemble model (\dbensemble\textsubscript{[T\textsubscript{s},5t,3]}) achieved a 195$\times$ \emph{lower} \totaltime than the \randomforest (25 trees) and \extratrees (25 trees) with a similar $\fscore$-score. 
The \nnb was substantially faster than the other approaches, yielded, however, a significantly worse $\fscore$-score.\footnote{Another \nnb based on $\tdim=50$ features also resulted in a similar $\fscore$-score, while exhibiting a significantly worse query time. The corresponding results were therefore disregarded. Note that we also excluded approximate nearest neighbor search algorithms since they would have generally led to even worse $\fscore$ scores, despite their potential to further reduce the query time.} 
%. As mentioned in Section~\ref{sec:approach}, the NN search strongly suffers from high dimensional data (\ie, the curse of dimensionality). Most significantly, 
%However, the $\fscore$-score of the NN model is significantly worse compared to the ones of the other models.
The table also provides insights w.r.t. space consumption and model quality. 
%However, such a remarkable speed-up comes at the expense of increased storage consumption. 
Specifically, 600~GB of disk storage were required to store the entire database, while the \kdtree indexes required 13~TB of additional storage for $\tsizesubsets=300$ (to store the associated leaves). 
These storage requirements could be reduced at a minor expense in $\fscore$-score by selecting a lower value of $\tsizesubsets$ (\eg, 50 or 150), see again Table~\ref{tab:geospatial_results}. 
%Other ways to further reduce the additional space needed are discussed in Section~\ref{sec:exps_discussion}.

\begin{table*}[t]
    \centering
    %\vspace{-0.2cm}
    \caption{Results on aerial image data set (size 0.6 TB). Time in seconds. Index size is reported next to each value of $k$.}
    %Storage consumption: $\tsizesubsets = 50 \rightarrow$ 2.17 TB, $\tsizesubsets = 150 \rightarrow$ 6.5 TB, $\tsizesubsets = 300 \rightarrow$ 13 TB.
    
    %\vspace{-0.1cm}
    \resizebox{1.0\textwidth}{!}{
    \begin{tabular}{lrrrrrrrrrrrr}
        \toprule
        \multirow{2}*{Model} & \multicolumn{4}{c}{$\tsizesubsets = 50 \rightarrow 2.17 $ TB} & \multicolumn{4}{c}{$\tsizesubsets = 150 \rightarrow 6.5$ TB} & \multicolumn{4}{c}{$\tsizesubsets = 300 \rightarrow 13$ TB}\\
        \cmidrule(r){2-5}\cmidrule(r){6-9}\cmidrule(r){10-13}
         &  \trainingtime &  \querytime &  \totaltime & $\fscore$-score & \trainingtime &  \querytime &  \totaltime & $\fscore$-score & \trainingtime &  \querytime &  \totaltime & $\fscore$-score \\
        \midrule

        \decisionbranch\textsubscript{[B,3]}                        &          0.307 &       1.556   &       1.863      &    0.800     &          0.398 &       1.047      &       1.445   &    0.833  &   0.583 & 0.971       &   1.554       &   0.850       \\
        \decisionbranch\textsubscript{[T\textsubscript{s},3]}     &          0.310 &       1.445   &       1.756      &    0.801     &          0.399 &       1.090      &       1.489   &    0.824  &   0.567 & 0.892       &   1.459       &   0.847       \\
        \decisionbranch\textsubscript{[T\textsubscript{a},3]}     &          0.335 &      16.618   &      16.953      &    0.818     &          0.420 &      14.685      &      15.105   &    0.833  &   0.672 & 13.844      &   14.516      &   0.854       \\
        \decisiontree                                               &          0.855 &     1,043.433 &     1,044.288    &    0.829     &          0.855 &   1,043.433      &     1,044.288 &    0.829  &   0.855 & 1,043.433   &   1,044.288   &   0.829       \\
        \nnb                                                          &          \textemdash     &     0.298    &     0.298       &    0.431        &        \textemdash     &     0.298       &     0.298    &    0.431 &   \textemdash     & 0.298      &   0.298      &   0.431       \\
        \midrule
        \dbensemble\textsubscript{[B,5t,3]}                         &          0.529 &       9.760   &      10.288      &    0.895     &          0.993 &       5.666      &       6.658   &    0.914  &   1.862 & 5.156       &   7.018       &   0.912       \\
        \dbensemble\textsubscript{[T\textsubscript{s},5t,3]}      &          0.508 &       8.418   &       8.926      &    0.884     &          1.013 &       5.455      &       6.468   &    0.904  &   1.886 & 4.895       &   6.780       &   0.904       \\
        \dbensemble\textsubscript{[B,25t,3]}                        &          0.891 &      28.607   &      29.497      &    0.915     &          1.543 &      22.639      &      24.182   &    0.925  &   2.729 & 19.716      &   22.445      &   0.930       \\
        \dbensemble\textsubscript{[T\textsubscript{s},25t,3]}     &          0.892 &      26.466   &      27.358      &    0.897     &          1.573 &      21.212      &      22.785   &    0.916  &   2.688 & 18.596      &   21.284      &   0.921       \\
        \randomforest                                               &          0.274 &     1,319.688 &     1,319.961    &    0.904     &          0.274 &     1,319.688    &     1,319.961 &    0.904  &   0.274 & 1,319.688   & 1,319.961     &   0.904       \\
        \extratrees                                                 &          0.122 &     1,332.026 &     1,332.148    &    0.950     &          0.122 &     1,332.026    &     1,332.148 &    0.950  &   0.122 & 1,332.026   & 1,332.148     &   0.950       \\
        \bottomrule
    \end{tabular}
    }
    \label{tab:geospatial_results}
%    \vspace{-1ex}
\end{table*}
\subsubsection{Optimal $k$-d Tree Leaf Size}
% \begin{itemize}
%     \item \todot{\url{https://dl.acm.org/doi/10.1145/502807.502809}}
%     \item Marcos: high-dimensional range queries, also fast?
% \end{itemize}
%\noindent\textbf{Index Structures.} 

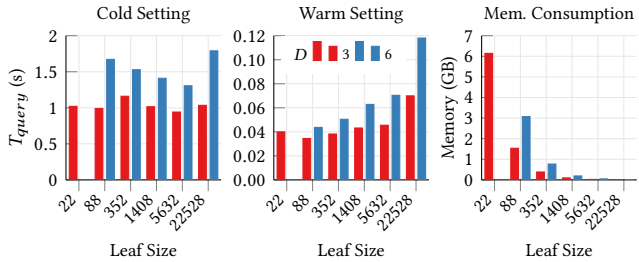
\begin{figure}
    \centering
     \pgfplotstableread[col sep=comma,]{data/kdtree_opt_leafsize_cold.csv}\leafsizecoldtable
\pgfplotstablegetcolsof{\leafsizecoldtable}
\pgfmathtruncatemacro{\coldNumColumns}{\pgfplotsretval-1}
\pgfplotstableread[col sep=comma,]{data/kdtree_opt_leafsize_warm.csv}\leafsizewarmtable
\pgfplotstablegetcolsof{\leafsizewarmtable}
\pgfmathtruncatemacro{\warmNumColumns}{\pgfplotsretval-1}
\pgfplotstableread[col sep=comma,]{data/kdtree_memory.csv}\kdtreememory
\pgfplotstablegetcolsof{\kdtreememory}
\pgfmathtruncatemacro{\memoryNumColumns}{\pgfplotsretval-1}
\begin{tikzpicture}
\begin{groupplot}[barPlotStyle,
        ybar = 0.05cm,
        width=0.43\linewidth, 
        xlabel = {Leaf Size}, 
        ylabel = {\querytime (s)},
        group style={%
            group size=3 by 1,
            horizontal sep=0.7cm,
            ylabels at=edge left,
        },%
        legend style={
            at = {(0.85, 1.0)}, 
            anchor = north east,
        },
    ]   
    \nextgroupplot[title={Cold Setting},
    	bar width = 0.6/\coldNumColumns,
    	ymin = 0,
    	ymax = 2,
    	ytick distance=0.5,
    	xtick=data,
    	xticklabels from table={\leafsizecoldtable}{leaf_size},
    ]
     % add the data rows (from: https://tex.stackexchange.com/a/364100)
    \foreach \i in {1,...,\coldNumColumns} {
        \addplot table [
            x expr=\coordindex,
            y index=\i,
            col sep=comma,
        ] {\leafsizecoldtable};
    }
    \nextgroupplot[title={Warm Setting},
    	bar width = 0.6/\warmNumColumns,
    	ymin = 0,
    	ymax = 0.12,
    	ytick distance=0.02,
    	xtick=data,
    	xticklabels from table={\leafsizewarmtable}{leaf_size},
    	y tick label style={
    	    /pgf/number format/fixed, 
    	    /pgf/number format/fixed zerofill, 
        	/pgf/number format/precision=2,
    	},
    ]
    \addlegendimage{empty legend}
    \addlegendentry{\textbf{$\tdimsubsets$}}
    
     % add the data rows (from: https://tex.stackexchange.com/a/364100)
    \foreach \i in {1,...,\warmNumColumns} {
        \addplot table [
            x expr=\coordindex,
            y index=\i,
            col sep=comma,
        ] {\leafsizewarmtable};
         % extract the column name and store it in `\colname'
        \pgfplotstablegetcolumnnamebyindex{\i}\of{\leafsizewarmtable}\to{\colname}
        % add column name to the legend
        \addlegendentryexpanded{\colname};
    }
    \nextgroupplot[title={Mem. Consumption},
    	bar width = 0.6/\memoryNumColumns,
    	ymin = 0,
    	ymax = 7,
    	ytick distance=1,
    	xtick=data,
    	xticklabels from table={\kdtreememory}{leaf_size},
    	xlabel = {Leaf Size}, 
    	ylabel = {Memory (GB)},
	]
     % add the data rows (from: https://tex.stackexchange.com/a/364100)
    \foreach \i in {1,...,\memoryNumColumns} {
        \addplot table [
            x expr=\coordindex,
            y index=\i,
            col sep=comma,
        ] {\kdtreememory};
    }
\end{groupplot}
\end{tikzpicture}
     %\vspace{-4ex}
    \caption{Impact of leaf size on query time and size of a single \kdtree. No results for $\tdimsubsets=6$ and leaf size 22 due to exceeding machine memory capacity.}
    \label{fig:kdtree_leafsize}
    \vspace{-1ex}
\end{figure}

Next, we investigate the influence of the leaf size of our \kdtrees by considering \kdtrees with varying leaf sizes.\footnote{The leaf size of a \kdtree impacts the runtime when executing range queries. Here, a smaller leaf size leads to a higher cost of traversing the tree, but a lower cost for scanning the points contained in the leaves intersecting with a query rectangle.} 
This experiment was conducted for two feature subset sizes, $\tdimsubsets=3$ and $\tdimsubsets=6$, and for each of the two cases, $\tsizesubsets=10$ random feature subsets were considered. For each of these feature subsets, \kdtrees with leaf sizes ranging from 22 to 22,582 were built.
%Similarly to the previous experiment, we utilized training sets ($30,030$ instances) based on our aerial data set to train decision branch models. 
For each of the six classes shown in Figure~\ref{fig:classes}, one user query was generated, each with 30 positive and 30,000 negative instances. Fitting a decision branch model for each such query led to a set of range queries (\ie, the $B_{opt}$ boxes), which were processed by a corresponding \kdtree index. To assess the impact of the leaf sizes on the query time, we report the average query time needed to process such user queries for each case and for each leaf size.
%needed to process the range queries. \todot{how many user queries? how can range queries?}}
% We consider \kdtrees built for
% feature subsets of size $\tdimsubsets=3$ and $\tdimsubsets=6$, respectively, and with varying leaf sizes from 22 to 22, 528 to determine the optimal leaf size for the \kdtrees. For several user queries based on our six classes, decision branch models are trained using 30 positive samples and 30,000 negative samples from our aerial data set. We extract range predicates from the trained decision branch models (\ie, boxes $B_opt$) and execute the queries using our \kdtree index structures. We report the average query time of the range queries to measure the effect of varying leaf sizes on the query time.}
% We consider \kdtrees built for feature subsets of size $\tdimsubsets=3$ and $\tdimsubsets=6$, respectively,
% %, to measure the impact on the query time~\querytime. We pre-build 300 \kdtree indexes for $\tdimsubsets=3$ and 150 for $\tdimsubsets=6$.
% %, due to the increased expressive (classification) power of 6-dimensional boxes. 
% and vary the leaf size of the \kdtrees from $22$ to $22,528$. 
% To measure the impact on the query time (\ie, the time needed to conduct the range queries), we consider a series of range queries induced by trained decision branch models for the data described above and report the average query time.
We repeated the experiments for two settings, which we call ``cold'' and ``warm''. For the cold setting, underlying caches (\eg disk, OS) were emptied, whereas for the warm setting the loaded objects of a query were already cached, which led to a reduced loading time. 
We report results for both settings in Figure~\ref{fig:kdtree_leafsize}.
The minimal query time for the cold setting was achieved with a leaf size of $5,632$. Note that \querytime was slightly larger for $\tdimsubsets=6$, which is in line with the runtime bounds and performance issues in high-dimensional spaces of index structures summarized in Section~\ref{sec:background}, \ie, longer query times for growing~$\tdimsubsets$. For the warm setting, the query times were significantly reduced. This behavior suggests that, if redundant boxes occur during the search, 
%(especially for ensemble models), 
smaller leaf sizes may be considered to make better use of caching opportunities.
Figure~\ref{fig:kdtree_leafsize} also shows the memory footprint of a single \kdtree index structure for different leaf sizes and~$\tdimsubsets$. As expected, memory requirements are inversely proportional to leaf size. Hence, when halving the leaf size, memory requirements double due to the larger tree that needs to be loaded into memory (the leaves are stored on disk). 
For the runtime evaluation provided below, we therefore selected the optimal leaf size of $5,632$, since it not only achieved the fastest query time but also exhibited a small memory footprint. The latter allows for loading \emph{hundreds of index structures in memory at once}, even on commodity hardware. More specifically, one index requires only 24~MB for $\tdimsubsets = 3$ and 48~MB for $\tdimsubsets = 6$.

\subsubsection{Scaling Behavior of $k$-d Tree.} 
\begin{figure}
    \centering
    \begin{tikzpicture}
    \begin{groupplot}[
        linePlotStyleSet1,
        width=0.54\linewidth, 
        group style={%
            group size=2 by 1,
            horizontal sep=4em,
            ylabels at=edge left,
        },
        legend image post style={scale=0.57},
    ]   
    \nextgroupplot[
        title={\kdtree ($\tdimsubsets = 3$)},
        xmin=0, xmax=1150000000,
        xtick distance=250000000,
        ymin=0, ymax=0.008,
        ytick distance = 0.002,
        yticklabel style={
            /pgf/number format/fixed,
            /pgf/number format/precision=5
        },
        scaled y ticks=false,
        ylabel = {\querytime (s)},
        xtick scale label code/.code={
            \pgfmathparse{int(#1)}Number of Points  $\testsize$ ($\cdot 10^{\pgfmathresult}$)
        },
        every x tick scale label/.style={
            at={(0.5,-0.15)}, 
            anchor = north
        },
        legend style={
            legend columns = -1,
            at = {(0.54, 1.5)}, 
            anchor ={north},
        },
        mark repeat=1,
        every axis plot/.append style={ultra thick},
    ]
    \addplot%[teal,every mark/.append style={solid,fill=teal},mark=*]
    table {%
    1024000 0.000209518215420628
    2048000 0.000368218223909611
    4095000 0.000945751147796046
    8190000 0.00152348758712629
    16380000 0.00210122402645653
    32760000 0.00267896046578677
    65520000 0.00325669690511701
    131040000 0.00383443334444725
    262079000 0.00441216660345436
    524160000 0.00498990622310773
    1048320000 0.00556764266243797
    };
    \addplot%[purple, densely dashed, every mark/.append style={solid, fill=purple},mark=square*]
    table {%
    1024000 0.00128708489769159
    2048000 0.00132143806942156
    4095000 0.00137594634358625
    8190000 0.00146249682211416
    16380000 0.00159988714277772
    32760000 0.00181798068232929
    65520000 0.00216418259644093
    131040000 0.00271374387909515
    262079000 0.00358611204035912
    524160000 0.00497092569374801
    1048320000 0.0071691708243649
    };
    \addplot coordinates {
        (0, 0)
        (1000000000, 0.008)
    };
    \addplot[black, mark=*, mark size=2, only marks]
    table {%
    1024000 0.00046941373083323
    2048000 0.000695082876417332
    4095000 0.000915844440460164
    8190000 0.00124985747867156
    16380000 0.0017584383487701
    32760000 0.00215005848142831
    65520000 0.00276748564508221
    131040000 0.00356882903310983
    262079000 0.00425617840554974
    524160000 0.00518068273862198
    1048320000 0.00645709779527447
    };
    \legend{
        $\O(log\testsize)$,
        $\O(\testsize^{1-1/3})$,
        $\O(\testsize)$,
        Runtime,
    }
    \nextgroupplot[
        title={tree-based models},
        ymode=log,
        xmode=log,
        xlabel={Number of Points  $\testsize$},
        xmin=900000, xmax=1600000000,
        xtick distance=10,
        minor xtick=1600000000,
        xminorgrids=true,
        ylabel={\totaltime (s)},
        ymax=10000,
        ytick distance=10,
        legend style={
            legend columns = -1,
            at = {(0.58, 1.49)}, 
            anchor = north,
        },
        every axis plot/.append style={very thick},
    ]
    \addplot+[mark repeat=2]
    table {%
    0 0
    %%%%%%%%%%%
    693056 0.490254375
    1386112  0.98050875
    2772225 1.9610175
    5544450 3.922035
    11088900 7.84407
    %%%%%%%%%%
    22177800 15.6881407499313
    44355600 32.2738643487295
    66533400 46.9968742132187
    88711200 63.0633991956711
    110889000 79.3237023353577
    133066800 95.5303068558375
    155244600 111.676837007205
    177422400 127.819137295087
    199600200 144.150699496269
    221778000 160.396802544594
    243955800 176.573944330215
    266133600 192.788883248965
    288311400 208.883568882942
    310489200 224.950405836105
    332667000 241.055558840434
    354844800 257.117547472318
    377022600 273.324816505114
    399200400 289.622140645981
    421378200 305.804950753848
    443556000 321.565957943598
    465733800 337.667472521464
    487911600 353.543159127235
    510089400 369.610521197319
    532267200 385.447725931803
    554445000 401.63829155763
    576622800 417.380051016808
    598800600 433.44786632061
    620978400 449.546942631404
    643156200 466.062178810438
    665334000 481.976032535235
    687511800 498.098501443863
    709689600 514.039359092712
    731867400 530.131708184878
    754045200 545.965303063393
    776223000 561.967464447021
    798400800 577.793252786001
    820578600 593.757898012797
    842756400 609.694495201111
    864934200 625.624557614326
    887112000 641.717275341352
    909289800 657.979188323021
    931467600 674.453241229057
    953645400 690.653652747472
    975823200 706.867778420448
    998001000 723.073185682297
    1020178800 739.33832859993
    1042356600 755.422105113665
    1064534400 771.58371078968
    1086712200 787.619551936785
    1108890000 803.606212496758
    1131067800 819.722052534421
    1153245600 835.751526435216
    1175423400 851.638791124026
    1197601200 867.6153977712
    1219779000 883.558284084002
    1241956800 899.480789581935
    1264134600 915.396790464719
    1286312400 931.353985269864
    1308490200 947.303832411766
    1330668000 963.176265279452
    1352845800 979.143195271492
    1375023600 995.162915468216
    1397201400 1011.25219031175
    1419379200 1027.35018189748
    1441557000 1043.43280104796
    };
    \addplot+[mark repeat=1]
    table {%
    1024000 0.586589238696628
    2048000 0.593392777760823
    4095000 0.602441557566325
    8190000 0.618843616697523
    16380000 0.643171875
    32760000 0.673959051556057
    65520000 0.729809530893962
    131040000 0.823964815987481
    262079000 0.913977684232924
    524160000 1.07906337081061
    1048320000 1.44392835595873
    1441557000 1.459
    };
    \legend {
        \decisiontree,
        % \extratrees,
        % \randomforest,
        \decisionbranch\textsubscript{[T\textsubscript{s},3]},
    }
\end{groupplot}
\end{tikzpicture}
    
    %\vspace{-1ex}
    \caption{Scaling behavior of our \kdtree search for $\tdimsubsets = 3$~(left) and  traditional tree-based methods.} %on growing data showing linear scaling behavior~(right, axes are shown in logarithmic scale).}
    \label{fig:scaling}
\end{figure}
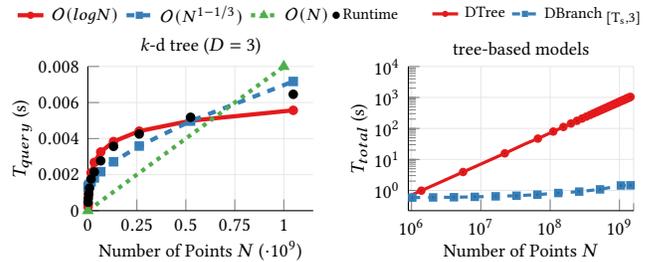

Next, we analyze if the theoretical time complexity $\O(\tdim \testsize^{1-\nicefrac{1}{\tdim}} + \indexStructuresQueryPoints)$ of range queries over a \kdtree can be 
%the \kdtree range search can be 
achieved in practice with our implementation. To do so, we measured the query time~\querytime of the \kdtree for increasing sizes of $\testsize$. We stopped the search after 10~leaves had been visited to keep $\indexStructuresQueryPoints$ (number of points returned) at a constant level. Thereby, the focus was on the tree traversal time instead of the time needed to return the contained data instances.
%load the data points stored in the leaves. 
%Without such a stopping criterion, a growing $\testsize$ would lead to linearly increasing $\indexStructuresQueryPoints$, which would eventually dominate the query time. 
In our analysis, we built \kdtree indexes for different subset sizes $\testsize$ and $\tdimsubsets=3$, scaled the number of instances by selecting instances at random from the database, and report the query times averaged over five runs. To measure those query times, we generated 300 random boxes ensuring no overlap, so as to minimize the chance of caching when leaves were loaded multiple times. Figure~\ref{fig:scaling}~(left) shows that the scaling behavior of our \kdtree implementation falls in-between the asymptotic curves for $\O(log \testsize)$ and $\O(\testsize^{1-1/3})$, which, again, is in line with the runtime bounds provided in Section~\ref{sec:background}.

%We build five \kdtree index structures for different subset sizes $\testsize \in$ \texttt{\{1024000, 2048000, 4095000, 8190000, 16380000, 32760000, 65520000, 131040000, 262079000, 524160000, 1048320000\}}. We also generate 300 random boxes in $\Reals^\tdimsubsets$ to measure query time, and ensure no overlap to minimize the chance of caching when leaves are loaded multiple times. Figure \ref{fig:qtime_dbranch} shows that our \kdtree implementation stays between the asymptotic curves for $\O(log(\tsize))$ and $\O(\tsize^{1-1/\tdimsubsets})$.

% \begin{figure}[t]
%     \centering
%     % \includegraphics[width=0.4\textwidth]{figures/experiments/section 2/scaling_behaviour_dim3.pdf}
%     \input{figures/results/scaling_behaviour_dim3}
%     \vspace{-0.5ex}
%     \caption{Scaling behavior of our \kdtree search ($\tdimsubsets = 3$).}
%      \label{fig:qtime_dbranch}
%      \vspace{-1ex}
% \end{figure}

% \begin{figure}[t]
%     \centering
%     % \includegraphics[width=0.4\textwidth]{figures/experiments/section 2/trees_scaling.pdf}
%     \input{figures/results/execution_time}
%     \vspace{-0.5ex}
%     \caption{Execution time of traditional tree-based methods on growing data showing linear scaling behavior. Axes are shown in logarithmic scale.}
%     \label{fig:qtime_full_scaling}
%     \vspace{-1ex}
% \end{figure}

\subsubsection{Scaling Behavior of Decision Branches.}
Finally, we assessed the scaling behavior of the two search-by-classification models with the lowest \totaltime in Table~\ref{tab:geospatial_results} for $\tsizesubsets=300$. Figure~\ref{fig:scaling} (right) shows the behavior of the models as we increased the number of points $\testsize$ in the data set. Note that, similarly to the \kdtree experiments above, data instances were selected at random for each value of $\testsize$. Each measurement for \decisionbranch\textsubscript{[T\textsubscript{s},3]} was averaged over three repetitions with different seeds to account for randomness in the method, while a single repetition was used for the \decisiontree. \decisionbranch\textsubscript{[T\textsubscript{s},3]} exhibits very slow growth in \totaltime (\eg, almost no increase for the last three measurements). 
%, but an upward trend as $\testsize$ exceeds $10^8$. %The latter is due to the increase in query answer size. 
By contrast, \totaltime for \decisiontree grows as expected, linearly with $\testsize$. When $\testsize > 10^9$, \decisionbranch\textsubscript{[T\textsubscript{s},3]} outperforms \decisiontree by almost three orders of magnitude, yielding the query results in a few seconds only.  

To conclude, our search framework enables quick processing of such user queries (\eg, search for planes in satellite imagery) and can report the (complete) answer set per query in a few seconds only, compared to minutes needed by scanning-based approaches. It is worth stressing that the runtime benefits will become more and more prominent and important in case the database would contain even more instances (\eg, hundreds of billions of instances).

\subsection{Sensitivity Analysis and Benchmark}\label{sec:modelbenchmark}

\begingroup
\setlength{\intextsep}{8pt}%
\setlength{\columnsep}{5pt}%
\begin{wraptable}{r}{4.1cm}
%\vspace{-2ex}% \vskip-1cm
\centering
\caption{Data Sets}
%\vspace{-1ex}
%\resizebox{\linewidth}{!}{
%\vspace{-1ex}
    \begin{tabular}{lrrr}
    \toprule
    Data set       & $\bar{\tsize}$    & $\tdim$   & $\nclasses$  \\ 
    \midrule
    iris       & 150     & 4   & 3  \\
    satimage   & 6,430   & 36  & 6  \\
    letter    & 20,000  & 16  & 26 \\
    mnist      & 60,000  & 784 & 10 \\
    senseit    & 98,528  & 100 & 3  \\
    %Artificial* & 100,000 & 20   & 2  \\
    covtype    & 581,012 & 54  & 7 \\
    %\midrule
    %ImageNet    & 1,281,167     & 128   & 1,000  \\
    %Aerial Dataset   & 1,441,557,000   & 50  & 8  \\
    \bottomrule
    %\multicolumn{4}{l}{\tiny{* generated via  \textit{sklearn make\_classification} function.}}\\
    \end{tabular}
%}
\label{tab:datasets}
%\vspace{-4ex}
\end{wraptable}
Next, we consider six more data sets listed in Table~\ref{tab:datasets} with $\tdim$ features, $\nclasses$ classes, and $\bar{\tsize}$ instances. Most of them are widely used for comparing binary classification models. 
However, our setup differs from known benchmarks in order to mimic a use case similar to a search engine for imbalanced data. To resemble such a case, we converted the original classification tasks to binary ``one-vs-all'' problems. That is, for each data set, we selected a class $c_p$ to represent the positive class, while the remaining ones were regarded as negative. We fixed the number of positive objects in the training set to~$30$ and picked $\left (\nicefrac{30}{\bar{\tsize}_{c_p}}\right ) \cdot \bar{\tsize}_{c_n}$ instances from each other class $c_n$, where $\bar{\tsize}_{c_p}$ and $\bar{\tsize}_{c_n}$ were the number of instances belonging to the positive class $c_p$ and the other class $c_n$, respectively.
% of the other classes as non-rare elements
%elements from each of the other classes $c$  while the number of non-rare training objects is selected %in proportion to the fraction of the % corresponds to the same extent as the 
%number of rare training objects in relation to the total number of rare objects. 
The remaining instances were split up to form (almost) equal-sized validation and test sets. The hyperparameters of each model were optimized via grid search on the validation set and the final model qualities were assessed using the test set.
%Training and test data are standardized before evaluation. 
%\noindent\textbf{Runs}. 
Each run was repeated three times with different (random) seeds. 

\endgroup
%For this purpose, we split the test data into an equal-sized validation and test sets. 
%More details regarding the experimental setup are provided in the supplemental material as well as in our source code. %\footnote{An anonymous GitHub repository with the corresponding code be found at \url{https://github.com/decisionbranches/sigkdd_decisionbranches}.}

\subsubsection{Influence of Parameters}
\label{sec:sensitivity}

The sensitivity of the different \tdecisionbranch variants w.r.t. the feature subset size~$\tdimsubsets$ and the number of feature subsets $\numberfeaturesubsetstest$ tested is shown in Figure~\ref{fig:nfeat_fscore_time}. The results for $\fscore$-scores are averaged across all one-vs-all classification tasks and data sets, while the training time results are shown for covtype only (since this data set is dominant in terms of number of data instances, and, hence, runtime). We observe that all models benefit from larger $\tdimsubsets$, while $\fscore$-scores increase only slightly beyond $\tdimsubsets=6$. The $T_a$ model variants (\ie, including both bottom-up and top-down constructions using all available features) are less dependent on large $\tdimsubsets$ while exhibiting comparable training time to $T_s$ variants (\ie, including subsets of $\tdimsubsets$ features only). This effect indicates that $T_a$ variants can more effectively compensate for suboptimal boxes learned during the bottom-up construction phase by a subsequent fine-tuning (top-down) phase on all available features.
For the number of feature subsets $\numberfeaturesubsetstest$ tested per box, we considered different fractions of the total number of feature subsets $\tsizesubsets$ ranging from a single subset ($\numberfeaturesubsetstest=1$) to all ($\numberfeaturesubsetstest=\tsizesubsets$) available subsets. We can observe that a small amount of subsets (\ie, $\numberfeaturesubsetstest=\sqrt{\tsizesubsets}$) is already sufficient to achieve a good $\fscore$-score with minimal training times.
%, also see the discussion in Section~\ref{sec:exps_discussion}. 
%However, it can be seen that the models with a $T_a$ construction phase are not so dependent on large $\tdimsubsets$ as they can still compensate sub-optimal boxes coming from the bottom-up phase by fine-tuning on all available features. \todot{add something regarding the increasing training times}
To evaluate the influence of the total number of feature subsets~$\tsizesubsets$, we introduce a factor $\multiplicationfactor$ s.t.~$\tsizesubsets = \tdim \cdot \multiplicationfactor$.
%, where $\tdim$ is the dimensionality of each data set. 
This factor $\multiplicationfactor$ acts as a proxy, allowing us to adjust~$\tsizesubsets$ to heterogeneity in $\tdim$. Figure~\ref{fig:nind} shows that the overall impact of~$\tsizesubsets$ on model performance is moderate, independently of $\tdimsubsets$. As it can be seen, the training time grows linearly w.r.t.~$\tsizesubsets$.%, especially for the ensemble models with fine-tuning.
%. For each data set with the dimensionality $\tdim$, we determine the number of feature subsets with $\tsizesubsets = \tdim \cdot \multiplicationfactor$. 
%Due to the different dimensionality among the data sets, we evaluate different values of $\multiplicationfactor$ as a proxy for measuring the influence of $\tsizesubsets$ on the performance of the \tdecisionbranch models. 
%Figure~\ref{fig:nind} shows the results. In contrast to $\tdimsubsets$, the overall impact of $\tsizesubsets$ on model performance is moderate independently of $\tdimsubsets$. 
%
%
%Let us consider all four variants of \tdecisionbranch models to analyze their individual sensitivity to important parameters. For this, we compare the test $\fscore$-score averaged among all six datasets. We start with the size of the feature subsets $\tdimsubsets$. The results are shown in Figure~\ref{fig:nfeat_fscore_time}. With respect to the shown results, it can be stated the optimal size of feature subsets ranges between 8-15 depending on the chosen model. Comparing the individual model graphs, it is striking that the models that are only grown bottom-up require a minimum of features per subset to work properly. While the full-grown models also work on smaller sized feature subsets since the potentially underfitting hyperrectangles coming from the bottom-up construction are refined during the top-down construction. 
%
\begin{table*}[t]
    \centering
    \caption{Mean test $\fscore$-scores for single models on each data set, averaged among all classes.% and repeated three times with different random seeds.
    }
    \label{tab:res_small_single}
\resizebox{\textwidth}{!}{%
 \begin{tabular}{lcccccccccc}
\toprule
% & \multicolumn{9}{c}{Single Models}\\\cmidrule{2-10}
Data set &  DBranch\textsubscript{[B,4]} &  DBranch\textsubscript{[T\textsubscript{s},4]} &  DBranch\textsubscript{[T\textsubscript{a},4]} &  DBranch\textsubscript{[B,10]} &  DBranch\textsubscript{[T\textsubscript{s},10]} &  DBranch\textsubscript{[T\textsubscript{a},10]} &  DTree &  DTree\textsubscript{[4]} &  DTree\textsubscript{[10]}  &
\nnb \\
\midrule
covtype  &         $0.252 \pm .100$ &             $0.321 \pm .027$ &             $0.363 \pm .029$ &          $0.339 \pm .033$ &              $0.339 \pm .040$ &              $0.355 \pm .030$ &  $0.420 \pm .022$ &         $0.059 \pm .078$ &          $0.217 \pm .120$ & $0.198 \pm .054$  \\
iris     &         $0.948 \pm .049$ &             $0.948 \pm .049$ &             $0.953 \pm .041$ &          $0.967 \pm .044$ &              $0.967 \pm .044$ &              $0.972 \pm .036$&  $0.947 \pm .052$ &         $0.947 \pm .052$ &          $0.947 \pm .052$ & $0.789 \pm .146$\\
letter   &         $0.441 \pm .058$ &             $0.472 \pm .060$ &             $0.518 \pm .055$ &          $0.525 \pm .053$ &              $0.578 \pm .053$ &              $0.585 \pm .049$ &  $0.580 \pm .051$ &         $0.244 \pm .107$ &          $0.462 \pm .062$ & $0.217 \pm .081$ \\
mnist    &         $0.298 \pm .115$ &             $0.350 \pm .084$ &             $0.463 \pm .049$ &          $0.447 \pm .063$ &              $0.477 \pm .051$ &              $0.515 \pm .047$ &  $0.525 \pm .057$ &         $0.067 \pm .066$ &          $0.134 \pm .091$ & $0.226 \pm .121$ \\
satimage &         $0.647 \pm .043$ &             $0.659 \pm .031$ &             $0.667 \pm .035$ &          $0.659 \pm .040$ &              $0.669 \pm .046$ &              $0.675 \pm .057$ &  $0.653 \pm .053$ &         $0.632 \pm .051$ &          $0.660 \pm .032$ & $0.608 \pm .069$ \\
senseit  &         $0.465 \pm .037$ &             $0.480 \pm .031$ &             $0.499 \pm .035$ &          $0.488 \pm .056$ &              $0.485 \pm .041$ &              $0.481 \pm .034$ &  $0.512 \pm .040$ &         $0.441 \pm .050$ &          $0.439 \pm .047$ & $0.230 \pm .100$ \\ \hline
\textbf{Total}    &         $0.508 \pm .067$ &             $0.538 \pm .048$ &             $0.577 \pm .041$ &    $0.571 \pm .048$ &              $0.586 \pm .046$ &              $0.597 \pm .042$ &  $0.606 \pm .046$ &         $0.398 \pm .067$ &          $0.477 \pm .067$ & $0.378 \pm .095$\\
\bottomrule
\end{tabular}
}
\end{table*}
\begin{table*}[t]
    \centering
    \caption{Mean test $\fscore$-scores for ensemble models on each data set averaged among all classes.
    %and repeated three times with different random seeds.
    }
    \label{tab:res_small_ensemble}
    \resizebox{\textwidth}{!}{%
    \begin{tabular}{lcccccccccc}
\toprule
Data set &  DBEns\textsubscript{[B,4]} &  DBEns\textsubscript{[T\textsubscript{s},4]} &  DBEns\textsubscript{[T\textsubscript{a},4]} &  DBEns\textsubscript{[B,10]} &  DBEns\textsubscript{[T\textsubscript{s},10]} & DBEns\textsubscript{[T\textsubscript{a},10]} &  RForest &  RForest\textsubscript{[4]} &  RForest\textsubscript{[10]} &  ExTrees \\
\midrule
covtype  &       $0.371 \pm .047$ &           $0.367 \pm .027$ &           $0.428 \pm .035$ &        $0.425 \pm .029$ &            $0.402 \pm .034$ &            $0.437 \pm .035$ &    $0.431 \pm .027$ &           $0.067 \pm .007$ &           $0.157 \pm .025$ &    $0.449 \pm .043$ \\
iris     &       $0.979 \pm .027$ &           $0.972 \pm .038$ &           $0.972 \pm .038$ &        $0.979 \pm .027$ &            $0.972 \pm .038$ &            $0.972 \pm .038$ &    $0.967 \pm .027$ &           $0.972 \pm .038$ &            $0.972 \pm .038$ &    $0.977 \pm .030$ \\
letter   &       $0.678 \pm .041$  &           $0.696 \pm .033$ &           $0.730 \pm .031$ &        $0.721 \pm .036$ &            $0.753 \pm .030$ &            $0.751 \pm .031$ &    $0.677 \pm .054$ &           $0.380 \pm .067$ &            $0.701 \pm .031$ &    $0.735 \pm .028$ \\
mnist    &       $0.583 \pm .061$ &           $0.390 \pm .134$ &           $0.661 \pm .034$ &        $0.721 \pm .036$ &            $0.634 \pm .045$ &            $0.705 \pm .026$ &    $0.612 \pm .068$ &           $0.000 \pm .000$ &            $0.068 \pm .037$ &    $0.669 \pm .032$ \\
satimage &       $0.740 \pm .029$ &           $0.763 \pm .024$ &           $0.767 \pm .024$ &        $0.742 \pm .028$ &            $0.767 \pm .021$ &            $0.775 \pm .020$ &    $0.755 \pm .031$ &           $0.764 \pm .031$ &            $0.778 \pm .021$ &    $0.781 \pm .022$ \\
senseit  &       $0.497 \pm .022$ &           $0.565 \pm .039$ &           $0.556 \pm .049$ &        $0.509 \pm .021$ &            $0.581 \pm .041$ &            $0.573 \pm .036$ &    $0.586 \pm .043$ &           $0.529 \pm .058$ &            $0.570 \pm .050$ &    $0.614 \pm .033$ \\ \hline
\textbf{Total}    &       $0.641 \pm .038$ &           $0.626 \pm .049$ &           $0.686 \pm .035$ &        $0.683 \pm .030$ &            $0.685 \pm .035$ &            $0.702 \pm .031$ &    $0.671 \pm .042$ &           $0.452 \pm .033$ &            $0.541 \pm .034$ &    $0.704 \pm .031$ \\
\bottomrule
\end{tabular}
%\vspace{-1ex}
}
\end{table*}
\begin{figure*}[ht!]
    \centering
    \pgfplotstableread[col sep=comma,]{data/test_f1_nfeat_ind10.csv}\testfscorenfeat
\pgfplotstableread[col sep=comma,]{data/training_time_nfeat_ind10.csv}\trainingtimenfeat
%
% \pgfplotstableread[col sep=comma,]{data/test_precision_nfeat_ind10.csv}\testprecision
% %
% \pgfplotstableread[col sep=comma,]{data/test_recall_nfeat_ind10.csv}\testrecall
% %

\begin{tikzpicture}
 \begin{groupplot}[linePlotStyle,
        width=0.27\linewidth,
        xmax=20,
        xtick=data,
        xlabel = {$\tdimsubsets$},
        group style={%
            group name={precrecall},
            group size=4 by 1,
            horizontal sep=1.2cm,
        },%
        legend style={
            at = {(0.5, 1.4)}, 
            anchor = north west,
        },
    ]   
    \nextgroupplot[
    	ymin=0.4,
    	ymax = 0.71,
    	ytick distance = 0.1,
    	ylabel = {Test $\fscore$-Score},
	]
    \addplot table [
        x=n_feat,
        y=DB BU,
        col sep=comma,
    ] {\testfscorenfeat};
    \addplot table [
        x=n_feat,
        y=DB BU+TD,
        col sep=comma,
    ] {\testfscorenfeat};
    \addplot table [
        x=n_feat,
        y=DB Full,
        col sep=comma,
    ] {\testfscorenfeat};
    \addplot table [
        x=n_feat,
        y=ES BU,
        col sep=comma,
    ] {\testfscorenfeat};
    \addplot table [
        x=n_feat,
        y=ES BU+TD,
        col sep=comma,
    ] {\testfscorenfeat};
    \addplot table [
        x=n_feat,
        y=ES Full,
        col sep=comma,
    ] {\testfscorenfeat};
    \legend {
        \decisionbranch\textsubscript{[B]},
        \dbbts,
        \dbbta,
        \dbensemble\textsubscript{[B]},
        \esbts,
        \esbta,
    }
    \nextgroupplot[title={covtype},
        ymin = 0,
    	ymax = 3.1,
    	ytick distance = 1,
    	ylabel = {Training Time (s)},
	]
\addplot
table {%
2 0.160097939627511
4 0.189844472067697
6 0.282994258971441
8 0.190405652636573
10 0.408117430550711
15 0.659701415470668
20 0.916604155585879
};
\addplot
table {%
2 0.243411575044904
4 0.372064556394305
6 0.502202181589036
8 0.564703952698481
10 0.750140008472261
15 0.690331413632348
20 0.936006943384806
};
\addplot
table {%
2 0.197390783400763
4 0.321304173696609
6 0.457714705240159
8 0.471803483508882
10 0.712021941230411
15 0.703313759395054
20 1.03535938262939
};
\addplot
table {%
2 0.423833256676084
4 0.532193422317505
6 0.675423372359503
8 0.67603638058617
10 0.767269804364159
15 1.29990986415318
20 2.03157666751317
};
\addplot
table {%
2 0.729347263063703
4 0.98441363516308
6 1.2028758979979
8 1.26383371580215
10 1.46248001144046
15 1.97639018013364
20 2.45221451350621
};
\addplot
table {%
2 0.716077373141334
4 0.971253429140363
6 1.21506605829511
8 1.17859012739999
10 1.38942340442113
15 1.99027046703157
20 2.71689595494952
};

     \nextgroupplot[
        ymin=0.4, 
        ymax=0.71, 
        ytick distance=0.1,
        xlabel = {$\numberfeaturesubsetstest$},
         ylabel = {Test $\fscore$-Score},
        xtick={-0.5,0,1,2,3,4},
    xticklabels={,$1$,$\sqrt{\tsizesubsets}$,$0.5\tsizesubsets$,$0.75\tsizesubsets$,$\tsizesubsets$},
    xmax = 4,
    ]
\addplot 
table {%
0 0.458312191872699
1 0.528741377056496
2 0.545273369967509
3 0.552163641602464
4 0.549122323064396
};
%\addlegendentry{DB BU}
\addplot
table {%
0 0.461555579095393
1 0.523370198785521
2 0.546149920565832
3 0.549371672442299
4 0.550812542841496
};
%\addlegendentry{DB BU+TD}
\addplot 
table {%
0 0.487388198612323
1 0.553223607725712
2 0.563432017981823
3 0.57024641327287
4 0.566939164992984
};
%\addlegendentry{DB Full}
\addplot 
table {%
0 0.616916137017805
1 0.656160590423243
2 0.668632847486244
3 0.667606054876059
4 0.669871953978519
};
%\addlegendentry{ES BU}
\addplot 
table {%
0 0.642641481262458
1 0.670186855162639
2 0.67653952272733
3 0.673578658738874
4 0.673901108223664
};
%\addlegendentry{ES BU+TD}
\addplot 
table {%
0 0.621708920479298
1 0.678151154221623
2 0.69295502533886
3 0.690204097577433
4 0.689560348242806
};

     \nextgroupplot[
     title={covtype},
        ymin=0, 
        ymax=1.55, 
        ytick distance=0.5,
         ylabel = {Training Time (s)},
        xlabel = {        $\numberfeaturesubsetstest$},
         xtick={-0.5,0,1,2,3,4},
         xmax=4,
    xticklabels={,$1$,$\sqrt{\tsizesubsets}$,$0.5\tsizesubsets$,$0.75\tsizesubsets$,$\tsizesubsets$},
    ]

\addplot
table {%
0 0.00936818122863765
1 0.0320557299114408
2 0.164766005107335
3 0.26372310093471
4 0.300807998293922
};
%\addlegendentry{DB BU}
\addplot
table {%
0 0.0127016135624476
1 0.0352211679731096
2 0.176092806316557
3 0.265906072798229
4 0.29468822479248
};
%\addlegendentry{DB BU+TD}
\addplot 
table {%
0 0.0132715588524228
1 0.0353373118809291
2 0.168453511737642
3 0.259484245663597
4 0.288921254021781
};
%\addlegendentry{DB Full}
\addplot 
table {%
0 0.243678388214228
1 0.248425624003777
2 0.355216185004429
3 0.427786580081097
4 0.482248967718583
};
%\addlegendentry{ES BU}
\addplot 
table {%
0 0.295628419167927
1 0.355057911558466
2 0.805491412670864
3 0.991775533844409
4 1.20170392271557
};
%\addlegendentry{ES BU+TD}
\addplot 
table {%
0 0.39402486275598
1 0.427336954861867
2 0.886598190622792
3 1.07945914770172
4 1.30377073635354
};
\end{groupplot}
\end{tikzpicture}
    %\vspace{-1ex}
    \caption{Influence of feature subset size $\tdimsubsets$ and feature subsets tested $\numberfeaturesubsetstest$ on test $\fscore$-score and training time.}
    \label{fig:nfeat_fscore_time}
    %\vspace{-1ex}
\end{figure*}
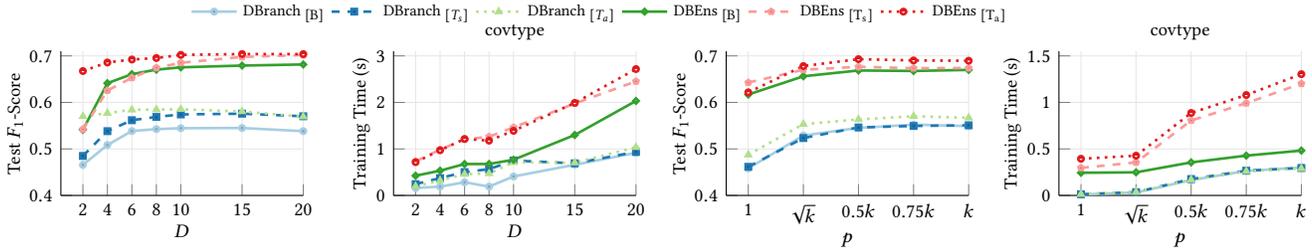
\begin{figure*}[ht!]
    %\vspace{-1.5ex}
    \centering
    \begin{tikzpicture}
 \begin{groupplot}[linePlotStyle,
        width=0.27\linewidth, 
        xmin = 0,
        xmax=10,
        xtick distance=2,
        xlabel = {$\multiplicationfactor$},
        group style={%
            group name={precrecall},
            group size=4 by 1,
            horizontal sep=1.2cm,
            ylabels at=edge left,
        },%
        legend style={
            at = {(0.5, 1.7)}, 
            anchor = north west,
            legend columns = 3,
        }
    ] 
    \nextgroupplot[title={$\tdimsubsets=3$},
    	ymin=0.4,
    	ymax = 0.71,
    	ytick distance = 0.1,
    	ylabel = {Test $\fscore$-Score},
	]
    \addplot
    table {%
    0.25 0.435798487520122
    0.5 0.453305315198938
    1 0.468758573937779
    2 0.472818668753206
    4 0.474420947680044
    6 0.488284094125823
    8 0.486042605992312
    10 0.491936069293395
    };
    \addplot
    table {%
    0.25 0.464251678878555
    0.5 0.483421835049378
    1 0.492989840236959
    2 0.499484204694599
    4 0.5144511989544
    6 0.520751730183108
    8 0.5246098567595
    10 0.520166606261922
    };
    \addplot
    table {%
    0.25 0.538580139174496
    0.5 0.549363684183972
    1 0.556306220527038
    2 0.562847071846547
    4 0.55939442535009
    6 0.565149572205756
    8 0.568680355234829
    10 0.56526496962049
    };
    \addplot
    table {%
    0.25 0.5033738295304
    0.5 0.538790610453553
    1 0.552923631623875
    2 0.572973579131943
    4 0.581279100836344
    6 0.584382008079991
    8 0.584004493224135
    10 0.589771771623285
    };
    \addplot
    table {%
    0.25 0.511540062085264
    0.5 0.538401166414285
    1 0.551065241528058
    2 0.572713213738175
    4 0.583894669262796
    6 0.587436678956472
    8 0.58837035693622
    10 0.592422635538599
    };
    \addplot
    table {%
    0.25 0.626728184312404
    0.5 0.651213344752756
    1 0.657778145954136
    2 0.669586117066812
    4 0.674114583018034
    6 0.674779235940487
    8 0.679131196642056
    10 0.677972822140294
    };
    \nextgroupplot[title={$\tdimsubsets=3$, covtype},
        	ymin=0.0,
        	ymax = 1.25,
        	ytick distance = 0.4,
        	ylabel = {Training Time (s)},
        ]
    \addplot
    table {%
    0.25 0.0247261637733096
    0.5 0.0258442674364362
    1 0.0200833933694022
    2 0.0723260811397007
    4 0.0600181647709437
    6 0.092915137608846
    8 0.234749521527971
    10 0.183271964391073
    };
    \addplot
    table {%
    0.25 0.0318285964784168
    0.5 0.0299612567538306
    1 0.0334475153968447
    2 0.0831273964473179
    4 0.144039630889893
    6 0.170815445127941
    8 0.30242083186195
    10 0.325461762292044
    };
    \addplot
    table {%
    0.25 0.0244679337456112
    0.5 0.0283347538539341
    1 0.025489228112357
    2 0.0632375762576148
    4 0.147161642710368
    6 0.146325701758975
    8 0.18732678322565
    10 0.255398137228829
    };
    \addplot
    table {%
    0.25 0.360218207041422
    0.5 0.364282880510603
    1 0.383978105726696
    2 0.430203494571504
    4 0.444757257189069
    6 0.547864879880633
    8 0.50868642897833
    10 0.506524324417114
    };
    \addplot
    table {%
    0.25 0.428991635640462
    0.5 0.420201312927973
    1 0.464657363437471
    2 0.507105804625012
    4 0.642734141576858
    6 0.751912196477254
    8 0.815691788991292
    10 0.939672288440523
    };
    \addplot
    table {%
    0.25 0.3841598374503
    0.5 0.450931753431048
    1 0.492515450432187
    2 0.543794518425351
    4 0.660975478944324
    6 0.778046414965675
    8 0.825447729655674
    10 0.943560622987293
    };

   \nextgroupplot[title={$\tdimsubsets=10$},
    	ymin=0.4,
    	ymax = 0.71,
    	ytick distance = 0.1,
    	ylabel = {Test $\fscore$-Score},
	]
    \addplot
    table {%
    0.25 0.519989811681099
    0.5 0.524718723039187
    1 0.536588442940986
    2 0.533629091933109
    4 0.551393074688825
    6 0.545124787445528
    8 0.548296728495671
    10 0.544515037478979
    };
    \addplot
    table {%
    0.25 0.537546555853481
    0.5 0.546131200590557
    1 0.561050642450492
    2 0.558113822119908
    4 0.569915199464317
    6 0.569951661501158
    8 0.565675898258568
    10 0.574155968677949
    };
    \addplot
    table {%
    0.25 0.562112403321369
    0.5 0.572065837169325
    1 0.574071644799511
    2 0.574176248762615
    4 0.587586948502986
    6 0.5819624058738
    8 0.582223945028223
    10 0.585282316573496
    };
    \addplot
    table {%
    0.25 0.652695023747096
    0.5 0.655750881104199
    1 0.668071571383415
    2 0.674953973918687
    4 0.67170513446999
    6 0.677294679478413
    8 0.677110787925273
    10 0.674207709328241
    };
    \addplot
    table {%
    0.25 0.649822350181788
    0.5 0.659729412589519
    1 0.669268837597231
    2 0.672902873418446
    4 0.680178505769695
    6 0.685049470642259
    8 0.685255648791168
    10 0.685198683250591
    };
    \addplot
    table {%
    0.25 0.686988808275277
    0.5 0.690934616457221
    1 0.698467185821729
    2 0.697079000607587
    4 0.69945461995348
    6 0.699714820426214
    8 0.700542605583313
    10 0.701219663918012
    };
    \nextgroupplot[title={$\tdimsubsets=10$, covtype},
    	ymin=0.0,
    	ymax = 1.85,
    	ytick distance = 0.6,
    	ylabel = {Training Time (s)},
    ]
    \addplot
    table {%
    0.25 0.0207129773639497
    0.5 0.0356924533843994
    1 0.07504669825236
    2 0.114923216047741
    4 0.25348687171936
    6 0.329805669330415
    8 0.437901099522909
    10 0.428215628578549
    };
    \addplot
    table {%
    0.25 0.0344236578260149
    0.5 0.0519121487935384
    1 0.0755714007786342
    2 0.150869278680711
    4 0.271455356052944
    6 0.427207345054263
    8 0.541587193806966
    10 0.836793411345709
    };
    \addplot
    table {%
    0.25 0.0275480293092273
    0.5 0.0516485827309744
    1 0.077699343363444
    2 0.142219815935407
    4 0.296445869264149
    6 0.345427217937651
    8 0.507924136661348
    10 0.761826640083676
    };
    \addplot
    table {%
    0.25 0.401037817909604
    0.5 0.433830828893752
    1 0.506276107969738
    2 0.550406149455479
    4 0.598411321640015
    6 0.788011278424944
    8 0.831114871161325
    10 0.879608108883812
    };
    \addplot
    table {%
    0.25 0.404411883581252
    0.5 0.450322605314709
    1 0.559664011001587
    2 0.717510325568063
    4 1.02800788198199
    6 1.22400825364249
    8 1.39953769956316
    10 1.6081421261742
    };
    \addplot
    table {%
    0.25 0.390168008350191
    0.5 0.450425874619257
    1 0.553237267902919
    2 0.656478291466123
    4 1.06814651262192
    6 1.24405070713588
    8 1.29719622929891
    10 1.41441743714469
    };
\end{groupplot}
\end{tikzpicture}
    %\vspace{-1ex}
    \caption{Influence of number of feature subsets $\tsizesubsets$ on both test $\fscore$-score and training time via proxy parameter $\multiplicationfactor$. 
    %Results are shown for different values of $\tdimsubsets$.
    }
    \label{fig:nind}
    %\vspace{-1ex}
\end{figure*}
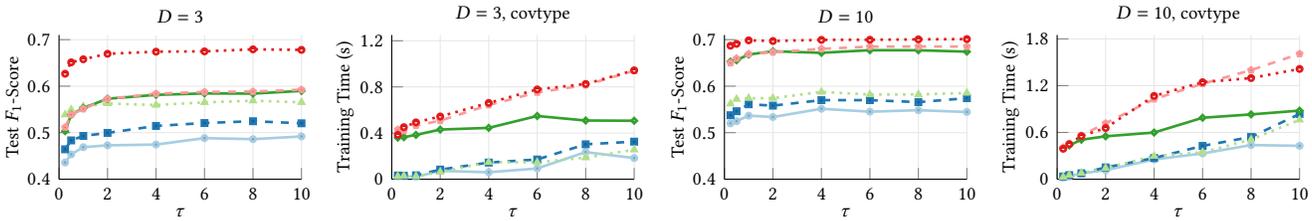

\subsubsection{Comparison with Baselines.}
\label{sec:model_comparison}
We compare \tdecisionbranch models ($\tdimsubsets = 4$ and $\tdimsubsets = 10$) with the other tree-based models and the \nnb.\footnote{For this experiment, all $\tdim$ features were used for the \nnb.}
Here, we also evaluate versions of the \decisiontree and \randomforest that resort to a limited number of features per tree, see again Section~\ref{sec:rare_leaves_range_queries}. Thereby, we can contrast the performance of our methods with naively constraining \decisiontree or \randomforest to the same number of features. All ensembles are based on 25 individual trees/decision branches. Tables~\ref{tab:res_small_single} and~\ref{tab:res_small_ensemble} summarize the results, which confirm that our alternative tree construction framework leads to models that exhibit a similar classification performance as those obtained via standard top-down construction schemes, while requiring only a fraction of the time to answer incoming user queries.
%We show that our alternative tree construction framework yields fragments of traditional decision trees that, in practice, exhibit a similar classification performance than those obtained via standard top-down construction schemes. 
%
For $\tdimsubsets=10$, the \decisionbranch variants achieved $\fscore$-scores similar to \decisiontree. With the same $\tdimsubsets$, our \dbensemble models slightly outperformed the \randomforest model, while achieving similarly results in comparison to \extratrees. Note that with a smaller feature subset size of $\tdimsubsets=4$, the classification performance of all \decisionbranch models slightly decreased. Yet, \decisionbranch and \dbensemble performed similarly to their classical counterparts and outperformed the constrained versions of \decisiontree and \randomforest. Our \tdecisionbranch models outperformed the \nnb under the considered assumptions and experimental settings.  %Our search-by-classification framework yields results in 0.3 seconds for one billion objects, which is suitable for the application scenarios we target in this work.
% For a small feature subset size $\tdimsubsets=2$, the initial boxes yield suboptimal (\eg, too large) coverings due to the minimal feature space they are constructed in. The more features are considered, the more precise the boxes become when no post-processing is applied. This can be seen in Figure~\ref{fig:nfeat_fscore_time}, where the recall for these models decreases and the precision increases. 
% In contrast to regular decision trees, which are usually grown until pure leaves are found, the \decisionbranch{\textsubscript{[B,3]}} model learns boxes that are not refined until complete purity. That is, its alternative tree construction \emph{does not} perform an additional post-processing step to fine-tune the initial boxes found, as the other approaches proposed in our work do. Hence, if a suboptimal box is constructed in the first step, it cannot be refined during the post-processing step. % unlike in the other approaches. 
% However, when combining the alternative tree construction scheme with a subsequent top-down refinement as in the fully-grown variant (\eg, \decisionbranch\textsubscript{[T\textsubscript{a},3]}), the classification results are significantly improved. Furthermore, combining fully-grown variants in an ensemble (\eg, \dbensemble\textsubscript{[T\textsubscript{s},25t,3]}) turns these individual weak but diverse classifiers into a single strong model that yields similar results as the ones produces by random forests.

\subsection{Storage and Performance Trade-Off}
\label{sec:space_consumption}
%Our experimental results reveal an intrinsic trade-off within our search-by-classification framework. 
Selecting suitable parameters for our framework entails finding a balance between storage capacity and performance, including response time and result accuracy. Figures \ref{fig:nfeat_fscore_time} and \ref{fig:nind} reveal that the more we increase the complexity of a decision branch model concerning crucial parameters $D$, $k$, and $p$, the more accurate the search results become in terms of $\fscore$-score. However, this comes at the cost of either (a) additional storage (for $\tsizesubsets$;
see Table 2), (b) longer query times (for $\tdimsubsets$; see Figure 8 for the difference between $\tdimsubsets=3$ and $\tdimsubsets=6$), or (c) extended training times (for $\tsizesubsets$, $\tdimsubsets$ and $\numberfeaturesubsetstest$; see again Figures \ref{fig:nfeat_fscore_time} and \ref{fig:nind}). 
Generally, the optimal parameter settings depend on the particular data catalog. 
%We advise comparing the outcomes for various parameter configurations, as demonstrated in our experiments, before deploying in a production setting. 
In our case study, we prove that even with a small storage overhead (\eg, only 3-4 times more storage for $\tsizesubsets=50$, $\tdimsubsets=3$ and $p=\sqrt{\tsizesubsets}$), we already achieve substantial speedup gains of approximately 590 times faster compared to scan-based approaches, while maintaining nearly the same $\fscore$-score. 

\section{Related Work}\label{sec:related-work}

%\red{Marcos: please check related work, ok?}
To our knowledge, this is the first work transforming the inference phase of decision trees and tree ensembles into a set of range queries in low-dimensional spaces. 
%Our approach is related to efficient decision tree algorithms, hyper-box-based machine learning, and nearest neighbor search.%, as detailed next.
% Training accurate classifiers over extremely imbalanced data has been the goal of a rich research corpus, focusing primarily on data-enhancing methods or model-enhancing methods. Data-enhancing methods alter the training data distribution via sampling strategies~\cite{marchant2021imbalance} or data augmentation techniques~\cite{bellinger2015synthetic}. By contrast, model-enhancing methods include meta-learning~\cite{wang2019metarare} and cost-sensitive learning techniques~\cite{khan2018costsensitive}.
% %, and tailored modeling of probability priors~\cite{bailerjones2008rare}. 
% Our approach falls into the second category, but differs from prior work by leveraging the synergies of decision-tree-based models with multidimensional index structures to accelerate query processing, without altering the training data. % nor making use of meta or cost information.
% % \todot{general ML problem, search for rare objects, unbalanced classes $\rightarrow$ speed up k-d tree traversal}
% %\subsubsection{Fast Inference.} 
Previous work has focused on the design of efficient decision-tree-based models, with early approaches such as SPRINT~\cite{shafer1996sprint} or BOAT~\cite{gehrke1999boat} aiming to speed up decision tree construction and inference over disk-resident data.
Substantial previous work has also been dedicated to deriving efficient approaches to construct and apply random forests~\cite{Louppe2014}, including distributed frameworks~\cite{PandaHBB2009,delRioLBH2014}.
%SPRINT~\cite{shafer1996sprint} leverages a sorted attribute list and a class list to efficiently determine optimal node splits. BOAT~\cite{gehrke1999boat} employs bootstrapping to efficiently find optimal node splitting criteria under a small training data sample.
Recently, fast decision tree construction has been studied in online settings, such as streaming data analysis~\cite{ke2017lightgbm,manapragada2018exfastdt}. Yet, all these approaches require \emph{scanning the entire database} during inference.
%These approaches focus on the traditional top-down decision tree induction scheme and assume % under the assumption 
%that the class distribution is similar among the different classes. However, traditional decision tree models tend to underperform under the imbalanced setting~\cite{Cieslak2008imbalanced}, which is our focus. Additionally, and in contrast to existing online approaches, our work proposes an alternative tree construction scheme specifically tailored for extremely imbalanced datasets. 
PRIM~\cite{Friedman1999} is an alternative top-down technique for finding boxes that cover regions of the feature space where the response average (\eg, weighted relative accuracy~\cite{Arzamasov2021reds}) is high. 
%Unlike our approach, boxes found by PRIM are not described by a binary tree, which hinders model interpretation~\cite{HastieTF2009}. 
In contrast to PRIM, our bottom-up construction approach mimics the construction of classical decision trees, with important algorithmic ingredients such as the use of randomness at various stages as well as boxes maximizing the (information) gain based on the Gini index. %Nevertheless, our index-aware framework could also be adapted to accelerate the retrieval of data covered by boxes constructed via PRIM. % in massive data catalogs.
Numerous fast (approximate) nearest neighbor search methods have been developed for handling single-object queries~\cite{johnson2019billion,Wang2021VLDB,Ram2013whichtree}.
%, which can be broadly classified as exact or approximate nearest neighbor search methods. 
Despite their efficiency, these methods are inferior in F1-score compared to search-by-classification approaches. 

\section{Conclusion}
\label{sec:conclusions}
%\begin{sloppypar}
We propose a novel approach to efficiently support search-by-classification tasks in large-scale databases. Our framework leverages a co-design of multidimensional indexes and decision trees and random forests that can rapidly process user queries formulated as binary classification data sets. 
%Our index-aware model construction scheme called \tdecisionbranches focused on covering positive data instances via boxes in lower dimensional spaces. 
The decision branches and associated boxes introduced allow for transforming the inference phase into a set of range queries, which can be efficiently supported by pre-built multidimensional indexes. 
Our experiments show that our framework achieves a similar classification performance in comparison to traditional decision trees and random forests, while drastically reducing the inference time.
%on both tabular and image data. 
We believe that this work will pave the way for novel index-aware machine learning models that will, in turn, lead to conceptually novel search engines in remote sensing, astrophysics, and many other data-intensive domains. 
In the future, we will explore techniques to reduce the storage overhead and optimize for incremental data updates, including LSM-trees~\cite{lsmtrees}, compressed~\cite{Arroyuelo2022compridx} and learned indexes~\cite{ding2020alex}.
%In particular, we believe that our work will inspire further research w.r.t. accelerating the inference phase of machine learning models using index-aware construction schemes.
\begin{acks}
    This research is supported by the Independent Research Fund Denmark (grant number 9131-00110B) and by Nvidia (hardware donations).
%     This research is supported by 
% %FG would like to acknowledge support from 
% the Independent Research Fund Denmark (grant number 9131-00110B) and Nvidia (for hardware aid). % \emph{Monitoring Changes in Big Satellite Data via Massively-Parallel Artificial Intelligence} 
\end{acks}

\bibliographystyle{ACM-Reference-Format}
\balance
\bibliography{literature}

%%% -*-BibTeX-*-
%%% Do NOT edit. File created by BibTeX with style
%%% ACM-Reference-Format-Journals [18-Jan-2012].

\begin{thebibliography}{44}

%%% ====================================================================
%%% NOTE TO THE USER: you can override these defaults by providing
%%% customized versions of any of these macros before the \bibliography
%%% command.  Each of them MUST provide its own final punctuation,
%%% except for \shownote{}, \showDOI{}, and \showURL{}.  The latter two
%%% do not use final punctuation, in order to avoid confusing it with
%%% the Web address.
%%%
%%% To suppress output of a particular field, define its macro to expand
%%% to an empty string, or better, \unskip, like this:
%%%
%%% \newcommand{\showDOI}[1]{\unskip}   % LaTeX syntax
%%%
%%% \def \showDOI #1{\unskip}           % plain TeX syntax
%%%
%%% ====================================================================

\ifx \showCODEN    \undefined \def \showCODEN     #1{\unskip}     \fi
\ifx \showDOI      \undefined \def \showDOI       #1{#1}\fi
\ifx \showISBNx    \undefined \def \showISBNx     #1{\unskip}     \fi
\ifx \showISBNxiii \undefined \def \showISBNxiii  #1{\unskip}     \fi
\ifx \showISSN     \undefined \def \showISSN      #1{\unskip}     \fi
\ifx \showLCCN     \undefined \def \showLCCN      #1{\unskip}     \fi
\ifx \shownote     \undefined \def \shownote      #1{#1}          \fi
\ifx \showarticletitle \undefined \def \showarticletitle #1{#1}   \fi
\ifx \showURL      \undefined \def \showURL       {\relax}        \fi
% The following commands are used for tagged output and should be
% invisible to TeX
\providecommand\bibfield[2]{#2}
\providecommand\bibinfo[2]{#2}
\providecommand\natexlab[1]{#1}
\providecommand\showeprint[2][]{arXiv:#2}

\bibitem[\protect\citeauthoryear{Arroyuelo, Navarro, Reutter, and
  Rojas-Ledesma}{Arroyuelo et~al\mbox{.}}{2022}]%
        {Arroyuelo2022compridx}
\bibfield{author}{\bibinfo{person}{D. Arroyuelo}, \bibinfo{person}{G. Navarro},
  \bibinfo{person}{J.~L. Reutter}, {and} \bibinfo{person}{J. Rojas-Ledesma}.}
  \bibinfo{year}{2022}\natexlab{}.
\newblock \showarticletitle{Optimal Joins Using Compressed Quadtrees}.
\newblock \bibinfo{journal}{\emph{ACM Transactions on Database Systems}}
  \bibinfo{volume}{47}, \bibinfo{number}{2} (\bibinfo{year}{2022}).
\newblock
\showISSN{0362-5915}


\bibitem[\protect\citeauthoryear{Arzamasov and B\"{o}hm}{Arzamasov and
  B\"{o}hm}{2021}]%
        {Arzamasov2021reds}
\bibfield{author}{\bibinfo{person}{V. Arzamasov} {and} \bibinfo{person}{K.
  B\"{o}hm}.} \bibinfo{year}{2021}\natexlab{}.
\newblock \showarticletitle{REDS: Rule Extraction for Discovering Scenarios}.
  In \bibinfo{booktitle}{\emph{Proceedings of the 2021 ACM SIGMOD International
  Conference on Management of Data}}. \bibinfo{pages}{115–128}.
\newblock
\showISBNx{9781450383431}


\bibitem[\protect\citeauthoryear{Bailer-Jones, Smith, Tiede, Sordo, and
  Vallenari}{Bailer-Jones et~al\mbox{.}}{2008}]%
        {bailerjones2008rare}
\bibfield{author}{\bibinfo{person}{C.~A.~L. Bailer-Jones},
  \bibinfo{person}{K.~W. Smith}, \bibinfo{person}{C. Tiede},
  \bibinfo{person}{R. Sordo}, {and} \bibinfo{person}{A. Vallenari}.}
  \bibinfo{year}{2008}\natexlab{}.
\newblock \showarticletitle{{Finding rare objects and building pure samples:
  probabilistic quasar classification from low-resolution Gaia spectra}}.
\newblock \bibinfo{journal}{\emph{Monthly Notices of the Royal Astronomical
  Society}} \bibinfo{volume}{391}, \bibinfo{number}{4} (\bibinfo{year}{2008}),
  \bibinfo{pages}{1838--1853}.
\newblock


\bibitem[\protect\citeauthoryear{Bentley}{Bentley}{1975}]%
        {Bentley1975}
\bibfield{author}{\bibinfo{person}{J.~L. Bentley}.}
  \bibinfo{year}{1975}\natexlab{}.
\newblock \showarticletitle{Multidimensional Binary Search Trees Used For
  Associative Searching}.
\newblock \bibinfo{journal}{\emph{Commun. ACM}} \bibinfo{volume}{18},
  \bibinfo{number}{9} (\bibinfo{year}{1975}), \bibinfo{pages}{509--517}.
\newblock


\bibitem[\protect\citeauthoryear{Bentley}{Bentley}{1979}]%
        {Bentley79}
\bibfield{author}{\bibinfo{person}{J.~L. Bentley}.}
  \bibinfo{year}{1979}\natexlab{}.
\newblock \showarticletitle{Multidimensional Binary Search Trees in Database
  Applications}.
\newblock \bibinfo{journal}{\emph{{IEEE} Transactions on Software Engineering}}
  \bibinfo{volume}{5}, \bibinfo{number}{4} (\bibinfo{year}{1979}),
  \bibinfo{pages}{333--340}.
\newblock


\bibitem[\protect\citeauthoryear{Beyer, Goldstein, Ramakrishnan, and
  Shaft}{Beyer et~al\mbox{.}}{1999}]%
        {BeyerGRS99}
\bibfield{author}{\bibinfo{person}{K.~S. Beyer}, \bibinfo{person}{J.
  Goldstein}, \bibinfo{person}{R. Ramakrishnan}, {and} \bibinfo{person}{U.
  Shaft}.} \bibinfo{year}{1999}\natexlab{}.
\newblock \showarticletitle{When Is ''Nearest Neighbor'' Meaningful?}. In
  \bibinfo{booktitle}{\emph{Proceedings of the 7th International Conference on
  Database Theory, {ICDT}}}. \bibinfo{pages}{217--235}.
\newblock


\bibitem[\protect\citeauthoryear{Branco, Torgo, and Ribeiro}{Branco
  et~al\mbox{.}}{2016}]%
        {branco2016surveyimbalanced}
\bibfield{author}{\bibinfo{person}{P. Branco}, \bibinfo{person}{L. Torgo},
  {and} \bibinfo{person}{R.~P. Ribeiro}.} \bibinfo{year}{2016}\natexlab{}.
\newblock \showarticletitle{A Survey of Predictive Modeling on Imbalanced
  Domains}.
\newblock \bibinfo{journal}{\emph{Comput. Surveys}} \bibinfo{volume}{49},
  \bibinfo{number}{2} (\bibinfo{year}{2016}).
\newblock


\bibitem[\protect\citeauthoryear{Breiman}{Breiman}{2001}]%
        {Breiman2001}
\bibfield{author}{\bibinfo{person}{L. Breiman}.}
  \bibinfo{year}{2001}\natexlab{}.
\newblock \showarticletitle{Random Forests}.
\newblock \bibinfo{journal}{\emph{Machine Learning}} \bibinfo{volume}{45},
  \bibinfo{number}{1} (\bibinfo{year}{2001}), \bibinfo{pages}{5--32}.
\newblock


\bibitem[\protect\citeauthoryear{Cheng, Xie, Han, Guo, and Xia}{Cheng
  et~al\mbox{.}}{2020}]%
        {cheng2020remotesensing}
\bibfield{author}{\bibinfo{person}{G. Cheng}, \bibinfo{person}{X. Xie},
  \bibinfo{person}{J. Han}, \bibinfo{person}{L. Guo}, {and} \bibinfo{person}{G.
  Xia}.} \bibinfo{year}{2020}\natexlab{}.
\newblock \showarticletitle{Remote Sensing Image Scene Classification Meets
  Deep Learning: Challenges, Methods, Benchmarks, and Opportunities}.
\newblock \bibinfo{journal}{\emph{IEEE Journal of Selected Topics in Applied
  Earth Observations and Remote Sensing}}  \bibinfo{volume}{13}
  (\bibinfo{year}{2020}), \bibinfo{pages}{3735--3756}.
\newblock


\bibitem[\protect\citeauthoryear{d.~Berg, Cheong, v.~Kreveld, and
  Overmars}{d.~Berg et~al\mbox{.}}{2008}]%
        {BergCKO08}
\bibfield{author}{\bibinfo{person}{M. d. Berg}, \bibinfo{person}{O. Cheong},
  \bibinfo{person}{M.~J. v. Kreveld}, {and} \bibinfo{person}{M.~H. Overmars}.}
  \bibinfo{year}{2008}\natexlab{}.
\newblock \bibinfo{booktitle}{\emph{Computational geometry: algorithms and
  applications, 3rd Edition}}.
\newblock \bibinfo{publisher}{Springer}.
\newblock


\bibitem[\protect\citeauthoryear{del R{\'i}o, L{\'o}pez, Ben{\'i}tez, and
  Herrera}{del R{\'i}o et~al\mbox{.}}{2014}]%
        {delRioLBH2014}
\bibfield{author}{\bibinfo{person}{S. del R{\'i}o}, \bibinfo{person}{V.
  L{\'o}pez}, \bibinfo{person}{J.~M. Ben{\'i}tez}, {and} \bibinfo{person}{F.
  Herrera}.} \bibinfo{year}{2014}\natexlab{}.
\newblock \showarticletitle{On the use of MapReduce for imbalanced big data
  using Random Forest}.
\newblock \bibinfo{journal}{\emph{Information Sciences}}  \bibinfo{volume}{285}
  (\bibinfo{year}{2014}), \bibinfo{pages}{112--137}.
\newblock


\bibitem[\protect\citeauthoryear{Ding, Minhas, Yu, Wang, Do, Li, Zhang,
  Chandramouli, Gehrke, Kossmann, et~al\mbox{.}}{Ding et~al\mbox{.}}{2020}]%
        {ding2020alex}
\bibfield{author}{\bibinfo{person}{Jialin Ding}, \bibinfo{person}{Umar~Farooq
  Minhas}, \bibinfo{person}{Jia Yu}, \bibinfo{person}{Chi Wang},
  \bibinfo{person}{Jaeyoung Do}, \bibinfo{person}{Yinan Li},
  \bibinfo{person}{Hantian Zhang}, \bibinfo{person}{Badrish Chandramouli},
  \bibinfo{person}{Johannes Gehrke}, \bibinfo{person}{Donald Kossmann},
  {et~al\mbox{.}}} \bibinfo{year}{2020}\natexlab{}.
\newblock \showarticletitle{ALEX: an updatable adaptive learned index}. In
  \bibinfo{booktitle}{\emph{Proceedings of the 2020 ACM SIGMOD International
  Conference on Management of Data}}. \bibinfo{pages}{969--984}.
\newblock


\bibitem[\protect\citeauthoryear{Drusch, {Del Bello}, Carlier, Colin,
  Fernandez, Gascon, Hoersch, Isola, Laberinti, Martimort, Meygret, Spoto, Sy,
  Marchese, and Bargellini}{Drusch et~al\mbox{.}}{2012}]%
        {DRUSCH201225}
\bibfield{author}{\bibinfo{person}{M. Drusch}, \bibinfo{person}{U. {Del
  Bello}}, \bibinfo{person}{S. Carlier}, \bibinfo{person}{O. Colin},
  \bibinfo{person}{V. Fernandez}, \bibinfo{person}{F. Gascon},
  \bibinfo{person}{B. Hoersch}, \bibinfo{person}{C. Isola}, \bibinfo{person}{P.
  Laberinti}, \bibinfo{person}{P. Martimort}, \bibinfo{person}{A. Meygret},
  \bibinfo{person}{F. Spoto}, \bibinfo{person}{O. Sy}, \bibinfo{person}{F.
  Marchese}, {and} \bibinfo{person}{P. Bargellini}.}
  \bibinfo{year}{2012}\natexlab{}.
\newblock \showarticletitle{Sentinel-2: ESA's Optical High-Resolution Mission
  for GMES Operational Services}.
\newblock \bibinfo{journal}{\emph{Remote Sensing of Environment}}
  \bibinfo{volume}{120} (\bibinfo{year}{2012}), \bibinfo{pages}{25--36}.
\newblock
\showISSN{0034-4257}


\bibitem[\protect\citeauthoryear{Fern\'{a}ndez-Delgado, Cernadas, Barro, and
  Amorim}{Fern\'{a}ndez-Delgado et~al\mbox{.}}{2014}]%
        {delgado:14}
\bibfield{author}{\bibinfo{person}{M. Fern\'{a}ndez-Delgado},
  \bibinfo{person}{E. Cernadas}, \bibinfo{person}{S. Barro}, {and}
  \bibinfo{person}{D. Amorim}.} \bibinfo{year}{2014}\natexlab{}.
\newblock \showarticletitle{Do we Need Hundreds of Classifiers to Solve Real
  World Classification Problems?}
\newblock \bibinfo{journal}{\emph{Journal of Machine Learning Research}}
  \bibinfo{volume}{15} (\bibinfo{year}{2014}), \bibinfo{pages}{3133--3181}.
\newblock


\bibitem[\protect\citeauthoryear{Friedman and Fisher}{Friedman and
  Fisher}{1999}]%
        {Friedman1999}
\bibfield{author}{\bibinfo{person}{J.~H. Friedman} {and} \bibinfo{person}{N.~I.
  Fisher}.} \bibinfo{year}{1999}\natexlab{}.
\newblock \showarticletitle{Bump hunting in high-dimensional data}.
\newblock \bibinfo{journal}{\emph{Statistics and Computing}}
  \bibinfo{volume}{9} (\bibinfo{year}{1999}), \bibinfo{pages}{123--143}.
\newblock


\bibitem[\protect\citeauthoryear{Gaede and G{\"{u}}nther}{Gaede and
  G{\"{u}}nther}{1998}]%
        {GaedeG98}
\bibfield{author}{\bibinfo{person}{V. Gaede} {and} \bibinfo{person}{O.
  G{\"{u}}nther}.} \bibinfo{year}{1998}\natexlab{}.
\newblock \showarticletitle{Multidimensional Access Methods}.
\newblock \bibinfo{journal}{\emph{Comput. Surveys}} \bibinfo{volume}{30},
  \bibinfo{number}{2} (\bibinfo{year}{1998}), \bibinfo{pages}{170--231}.
\newblock


\bibitem[\protect\citeauthoryear{Gehrke, Ganti, Ramakrishnan, and Loh}{Gehrke
  et~al\mbox{.}}{1999}]%
        {gehrke1999boat}
\bibfield{author}{\bibinfo{person}{J. Gehrke}, \bibinfo{person}{V. Ganti},
  \bibinfo{person}{R. Ramakrishnan}, {and} \bibinfo{person}{W.-Y. Loh}.}
  \bibinfo{year}{1999}\natexlab{}.
\newblock \showarticletitle{BOAT--Optimistic Decision Tree Construction}. In
  \bibinfo{booktitle}{\emph{Proceedings of the 1999 ACM SIGMOD International
  Conference on Management of Data}}. \bibinfo{pages}{169--180}.
\newblock


\bibitem[\protect\citeauthoryear{Geurts, Ernst, and Wehenkel}{Geurts
  et~al\mbox{.}}{2006}]%
        {GeurtsEW2006}
\bibfield{author}{\bibinfo{person}{P. Geurts}, \bibinfo{person}{D. Ernst},
  {and} \bibinfo{person}{L. Wehenkel}.} \bibinfo{year}{2006}\natexlab{}.
\newblock \showarticletitle{Extremely randomized trees}.
\newblock \bibinfo{journal}{\emph{Machine Learning}} \bibinfo{volume}{63},
  \bibinfo{number}{1} (\bibinfo{year}{2006}), \bibinfo{pages}{3--42}.
\newblock


\bibitem[\protect\citeauthoryear{Gieseke and Igel}{Gieseke and Igel}{2018}]%
        {GiesekeI18}
\bibfield{author}{\bibinfo{person}{F. Gieseke} {and} \bibinfo{person}{C.
  Igel}.} \bibinfo{year}{2018}\natexlab{}.
\newblock \showarticletitle{Training Big Random Forests with Little Resources}.
  In \bibinfo{booktitle}{\emph{Proceedings of the 24th ACM SIGKDD International
  Conference on Knowledge Discovery and Data Mining, {KDD}}}.
  \bibinfo{pages}{1445--1454}.
\newblock


\bibitem[\protect\citeauthoryear{Hastie, Tibshirani, and Friedman}{Hastie
  et~al\mbox{.}}{2009}]%
        {HastieTF2009}
\bibfield{author}{\bibinfo{person}{T. Hastie}, \bibinfo{person}{R. Tibshirani},
  {and} \bibinfo{person}{J.~H. Friedman}.} \bibinfo{year}{2009}\natexlab{}.
\newblock \bibinfo{booktitle}{\emph{The Elements of Statistical Learning}}.
\newblock \bibinfo{publisher}{Springer}.
\newblock


\bibitem[\protect\citeauthoryear{He, Zhang, Ren, and Sun}{He
  et~al\mbox{.}}{2016}]%
        {HeZRS15}
\bibfield{author}{\bibinfo{person}{K. He}, \bibinfo{person}{X. Zhang},
  \bibinfo{person}{S. Ren}, {and} \bibinfo{person}{J. Sun}.}
  \bibinfo{year}{2016}\natexlab{}.
\newblock \showarticletitle{Deep Residual Learning for Image Recognition}. In
  \bibinfo{booktitle}{\emph{2016 IEEE Conference on Computer Vision and Pattern
  Recognition, {CVPR}}}. \bibinfo{pages}{770--778}.
\newblock


\bibitem[\protect\citeauthoryear{Ivezi{\'c}, Kahn, Tyson, Abel, Acosta,
  Allsman, Alonso, AlSayyad, Anderson, Andrew, et~al\mbox{.}}{Ivezi{\'c}
  et~al\mbox{.}}{2019}]%
        {ivezic2019lsst}
\bibfield{author}{\bibinfo{person}{{\v{Z}}. Ivezi{\'c}}, \bibinfo{person}{S.~M.
  Kahn}, \bibinfo{person}{J.~A. Tyson}, \bibinfo{person}{B. Abel},
  \bibinfo{person}{E. Acosta}, \bibinfo{person}{R. Allsman},
  \bibinfo{person}{D. Alonso}, \bibinfo{person}{Y. AlSayyad},
  \bibinfo{person}{S.~F. Anderson}, \bibinfo{person}{J. Andrew},
  {et~al\mbox{.}}} \bibinfo{year}{2019}\natexlab{}.
\newblock \showarticletitle{LSST: From Science Drivers to Reference Design and
  Anticipated Data Products}.
\newblock \bibinfo{journal}{\emph{The Astrophysical Journal}}
  \bibinfo{volume}{873}, \bibinfo{number}{2} (\bibinfo{year}{2019}),
  \bibinfo{pages}{111}.
\newblock


\bibitem[\protect\citeauthoryear{Johnson, Douze, and J{\'e}gou}{Johnson
  et~al\mbox{.}}{2019}]%
        {johnson2019billion}
\bibfield{author}{\bibinfo{person}{J. Johnson}, \bibinfo{person}{M. Douze},
  {and} \bibinfo{person}{H. J{\'e}gou}.} \bibinfo{year}{2019}\natexlab{}.
\newblock \showarticletitle{Billion-scale similarity search with {GPUs}}.
\newblock \bibinfo{journal}{\emph{IEEE Transactions on Big Data}}
  \bibinfo{volume}{7}, \bibinfo{number}{3} (\bibinfo{year}{2019}),
  \bibinfo{pages}{535--547}.
\newblock


\bibitem[\protect\citeauthoryear{Kaplan and Sharir}{Kaplan and Sharir}{2011}]%
        {kaplan2011finding}
\bibfield{author}{\bibinfo{person}{H. Kaplan} {and} \bibinfo{person}{M.
  Sharir}.} \bibinfo{year}{2011}\natexlab{}.
\newblock \bibinfo{title}{Finding the Maximal Empty Rectangle Containing a
  Query Point}.
\newblock
\newblock
\showeprint[arxiv]{1106.3628}


\bibitem[\protect\citeauthoryear{Ke, Meng, Finley, Wang, Chen, Ma, Ye, and
  Liu}{Ke et~al\mbox{.}}{2017}]%
        {ke2017lightgbm}
\bibfield{author}{\bibinfo{person}{G. Ke}, \bibinfo{person}{Q. Meng},
  \bibinfo{person}{T. Finley}, \bibinfo{person}{T. Wang}, \bibinfo{person}{W.
  Chen}, \bibinfo{person}{W. Ma}, \bibinfo{person}{Q. Ye}, {and}
  \bibinfo{person}{T.-Y. Liu}.} \bibinfo{year}{2017}\natexlab{}.
\newblock \showarticletitle{LightGBM: A Highly Efficient Gradient Boosting
  Decision Tree}. In \bibinfo{booktitle}{\emph{Advances in Neural Information
  Processing Systems, {NeurIPS}}}, Vol.~\bibinfo{volume}{30}.
  \bibinfo{pages}{3149–3157}.
\newblock


\bibitem[\protect\citeauthoryear{Keisler, Skillman, Gonnabathula, Poehnelt,
  Rudelis, and Warren}{Keisler et~al\mbox{.}}{2019}]%
        {KEISLER2019visualsearch}
\bibfield{author}{\bibinfo{person}{R. Keisler}, \bibinfo{person}{S.~W.
  Skillman}, \bibinfo{person}{S. Gonnabathula}, \bibinfo{person}{J. Poehnelt},
  \bibinfo{person}{X. Rudelis}, {and} \bibinfo{person}{M.~S. Warren}.}
  \bibinfo{year}{2019}\natexlab{}.
\newblock \showarticletitle{Visual search over billions of aerial and satellite
  images}.
\newblock \bibinfo{journal}{\emph{Computer Vision and Image Understanding}}
  \bibinfo{volume}{187} (\bibinfo{year}{2019}), \bibinfo{pages}{102790}.
\newblock
\showISSN{1077-3142}


\bibitem[\protect\citeauthoryear{Keivani and Sinha}{Keivani and Sinha}{2018}]%
        {pmlr-v80-keivani18a}
\bibfield{author}{\bibinfo{person}{O. Keivani} {and} \bibinfo{person}{K.
  Sinha}.} \bibinfo{year}{2018}\natexlab{}.
\newblock \showarticletitle{Improved nearest neighbor search using auxiliary
  information and priority functions}. In \bibinfo{booktitle}{\emph{Proceedings
  of the 35th International Conference on Machine Learning, {ICML}}},
  Vol.~\bibinfo{volume}{80}. \bibinfo{pages}{2573--2581}.
\newblock


\bibitem[\protect\citeauthoryear{Kingma and Ba}{Kingma and Ba}{2015}]%
        {adam}
\bibfield{author}{\bibinfo{person}{D.~P. Kingma} {and} \bibinfo{person}{J.
  Ba}.} \bibinfo{year}{2015}\natexlab{}.
\newblock \showarticletitle{Adam: {A} Method for Stochastic Optimization}. In
  \bibinfo{booktitle}{\emph{3rd International Conference on Learning
  Representations, {ICLR}}}.
\newblock


\bibitem[\protect\citeauthoryear{Lee and Wong}{Lee and Wong}{1977}]%
        {Lee77worstcase}
\bibfield{author}{\bibinfo{person}{D.~T. Lee} {and} \bibinfo{person}{C.~K.
  Wong}.} \bibinfo{year}{1977}\natexlab{}.
\newblock \showarticletitle{Worst-Case Analysis for Region and Partial Region
  Searches in Multidimensional Binary Search Trees and Balanced Quad Trees}.
\newblock \bibinfo{journal}{\emph{Acta Informatica}} \bibinfo{volume}{9},
  \bibinfo{number}{1} (\bibinfo{year}{1977}), \bibinfo{pages}{23–29}.
\newblock


\bibitem[\protect\citeauthoryear{Li, Dragicevic, Castro, Sester, Winter,
  Coltekin, Pettit, Jiang, Haworth, Stein, and Cheng}{Li et~al\mbox{.}}{2016}]%
        {songnian2016biggeodata}
\bibfield{author}{\bibinfo{person}{S. Li}, \bibinfo{person}{S. Dragicevic},
  \bibinfo{person}{F.~A. Castro}, \bibinfo{person}{M. Sester},
  \bibinfo{person}{S. Winter}, \bibinfo{person}{A. Coltekin},
  \bibinfo{person}{C. Pettit}, \bibinfo{person}{B. Jiang}, \bibinfo{person}{J.
  Haworth}, \bibinfo{person}{A. Stein}, {and} \bibinfo{person}{T. Cheng}.}
  \bibinfo{year}{2016}\natexlab{}.
\newblock \showarticletitle{Geospatial big data handling theory and methods: A
  review and research challenges}.
\newblock \bibinfo{journal}{\emph{ISPRS Journal of Photogrammetry and Remote
  Sensing}}  \bibinfo{volume}{115} (\bibinfo{year}{2016}),
  \bibinfo{pages}{119--133}.
\newblock


\bibitem[\protect\citeauthoryear{Louppe}{Louppe}{2014}]%
        {Louppe2014}
\bibfield{author}{\bibinfo{person}{G. Louppe}.}
  \bibinfo{year}{2014}\natexlab{}.
\newblock \emph{\bibinfo{title}{Understanding Random Forests}}.
\newblock \bibinfo{thesistype}{Ph.D. Dissertation}. \bibinfo{school}{University
  of Li\`ege, Faculty of App. Sciences, Dep. of Electrical Engineering \& Comp.
  Science}.
\newblock


\bibitem[\protect\citeauthoryear{Manapragada, Webb, and Salehi}{Manapragada
  et~al\mbox{.}}{2018}]%
        {manapragada2018exfastdt}
\bibfield{author}{\bibinfo{person}{C. Manapragada}, \bibinfo{person}{G.~I.
  Webb}, {and} \bibinfo{person}{M. Salehi}.} \bibinfo{year}{2018}\natexlab{}.
\newblock \showarticletitle{Extremely Fast Decision Tree}. In
  \bibinfo{booktitle}{\emph{Proceedings of the 24th ACM SIGKDD International
  Conference on Knowledge Discovery and Data Mining, {KDD}}}.
  \bibinfo{pages}{1953–1962}.
\newblock


\bibitem[\protect\citeauthoryear{Metcalf, Meneghetti, Avestruz, Bellagamba,
  Bom, Bertin, Cabanac, Courbin, Davies, Decenci{\`e}re, et~al\mbox{.}}{Metcalf
  et~al\mbox{.}}{2019}]%
        {metcalf2019strong}
\bibfield{author}{\bibinfo{person}{R.~B. Metcalf}, \bibinfo{person}{M.
  Meneghetti}, \bibinfo{person}{C. Avestruz}, \bibinfo{person}{F. Bellagamba},
  \bibinfo{person}{C.~R. Bom}, \bibinfo{person}{E. Bertin}, \bibinfo{person}{R.
  Cabanac}, \bibinfo{person}{F. Courbin}, \bibinfo{person}{A. Davies},
  \bibinfo{person}{E. Decenci{\`e}re}, {et~al\mbox{.}}}
  \bibinfo{year}{2019}\natexlab{}.
\newblock \showarticletitle{The strong gravitational lens finding challenge}.
\newblock \bibinfo{journal}{\emph{Astronomy \& Astrophysics}}
  \bibinfo{volume}{625} (\bibinfo{year}{2019}), \bibinfo{pages}{A119}.
\newblock


\bibitem[\protect\citeauthoryear{O’Neil, Cheng, Gawlick, and
  O’Neil}{O’Neil et~al\mbox{.}}{1996}]%
        {lsmtrees}
\bibfield{author}{\bibinfo{person}{Patrick O’Neil}, \bibinfo{person}{Edward
  Cheng}, \bibinfo{person}{Dieter Gawlick}, {and} \bibinfo{person}{Elizabeth
  O’Neil}.} \bibinfo{year}{1996}\natexlab{}.
\newblock \showarticletitle{The Log-Structured Merge-Tree (LSM-Tree)}.
\newblock \bibinfo{journal}{\emph{Acta Inf.}} \bibinfo{volume}{33},
  \bibinfo{number}{4} (\bibinfo{date}{jun} \bibinfo{year}{1996}),
  \bibinfo{pages}{351–385}.
\newblock
\showISSN{0001-5903}
\urldef\tempurl%
\url{https://doi.org/10.1007/s002360050048}
\showDOI{\tempurl}


\bibitem[\protect\citeauthoryear{Panda, Herbach, Basu, and Bayardo}{Panda
  et~al\mbox{.}}{2009}]%
        {PandaHBB2009}
\bibfield{author}{\bibinfo{person}{B. Panda}, \bibinfo{person}{J.~S. Herbach},
  \bibinfo{person}{S. Basu}, {and} \bibinfo{person}{R.~J. Bayardo}.}
  \bibinfo{year}{2009}\natexlab{}.
\newblock \showarticletitle{PLANET: Massively Parallel Learning of Tree
  Ensembles with MapReduce}.
\newblock \bibinfo{journal}{\emph{Proceedings of the VLDB Endowment}}
  \bibinfo{volume}{2}, \bibinfo{number}{2} (\bibinfo{year}{2009}),
  \bibinfo{pages}{1426–1437}.
\newblock
\showISSN{2150-8097}


\bibitem[\protect\citeauthoryear{Pedregosa, Varoquaux, Gramfort, Michel,
  Thirion, Grisel, Blondel, Prettenhofer, Weiss, Dubourg, Vanderplas, Passos,
  Cournapeau, Brucher, Perrot, and Duchesnay}{Pedregosa et~al\mbox{.}}{2011}]%
        {scikit-learn}
\bibfield{author}{\bibinfo{person}{F. Pedregosa}, \bibinfo{person}{G.
  Varoquaux}, \bibinfo{person}{A. Gramfort}, \bibinfo{person}{V. Michel},
  \bibinfo{person}{B. Thirion}, \bibinfo{person}{O. Grisel},
  \bibinfo{person}{M. Blondel}, \bibinfo{person}{P. Prettenhofer},
  \bibinfo{person}{R. Weiss}, \bibinfo{person}{V. Dubourg}, \bibinfo{person}{J.
  Vanderplas}, \bibinfo{person}{A. Passos}, \bibinfo{person}{D. Cournapeau},
  \bibinfo{person}{M. Brucher}, \bibinfo{person}{M. Perrot}, {and}
  \bibinfo{person}{{\'E}.~D. Duchesnay}.} \bibinfo{year}{2011}\natexlab{}.
\newblock \showarticletitle{Scikit-learn: Machine Learning in {P}ython}.
\newblock \bibinfo{journal}{\emph{Journal of Machine Learning Research}}
  \bibinfo{volume}{12} (\bibinfo{year}{2011}), \bibinfo{pages}{2825--2830}.
\newblock


\bibitem[\protect\citeauthoryear{Ram and Gray}{Ram and Gray}{2013}]%
        {Ram2013whichtree}
\bibfield{author}{\bibinfo{person}{P. Ram} {and} \bibinfo{person}{A. Gray}.}
  \bibinfo{year}{2013}\natexlab{}.
\newblock \showarticletitle{Which Space Partitioning Tree to Use for Search?}.
  In \bibinfo{booktitle}{\emph{Advances in Neural Information Processing
  Systems, {NeurIPS}}}, Vol.~\bibinfo{volume}{26}. \bibinfo{pages}{656–664}.
\newblock


\bibitem[\protect\citeauthoryear{Russakovsky, Deng, Su, Krause, Satheesh, Ma,
  Huang, Karpathy, Khosla, Bernstein, Berg, and Fei-Fei}{Russakovsky
  et~al\mbox{.}}{2015}]%
        {ILSVRC15}
\bibfield{author}{\bibinfo{person}{O. Russakovsky}, \bibinfo{person}{J. Deng},
  \bibinfo{person}{H. Su}, \bibinfo{person}{J. Krause}, \bibinfo{person}{S.
  Satheesh}, \bibinfo{person}{S. Ma}, \bibinfo{person}{Z. Huang},
  \bibinfo{person}{A. Karpathy}, \bibinfo{person}{A. Khosla},
  \bibinfo{person}{M. Bernstein}, \bibinfo{person}{A.~C. Berg}, {and}
  \bibinfo{person}{L. Fei-Fei}.} \bibinfo{year}{2015}\natexlab{}.
\newblock \showarticletitle{{ImageNet Large Scale Visual Recognition
  Challenge}}.
\newblock \bibinfo{journal}{\emph{International Journal of Computer Vision}}
  \bibinfo{volume}{115}, \bibinfo{number}{3} (\bibinfo{year}{2015}),
  \bibinfo{pages}{211--252}.
\newblock


\bibitem[\protect\citeauthoryear{Shafer, Agrawal, and Mehta}{Shafer
  et~al\mbox{.}}{1996}]%
        {shafer1996sprint}
\bibfield{author}{\bibinfo{person}{J.~C. Shafer}, \bibinfo{person}{R. Agrawal},
  {and} \bibinfo{person}{M. Mehta}.} \bibinfo{year}{1996}\natexlab{}.
\newblock \showarticletitle{SPRINT: A Scalable Parallel Classifier for Data
  Mining}. In \bibinfo{booktitle}{\emph{Proceedings of the 22th International
  Conference on Very Large Data Bases, {VLDB}}}. \bibinfo{pages}{544–555}.
\newblock
\showISBNx{1558603824}


\bibitem[\protect\citeauthoryear{Wan, Wang, Hoi, Wu, Zhu, Zhang, and Li}{Wan
  et~al\mbox{.}}{2014}]%
        {wan2014cbir}
\bibfield{author}{\bibinfo{person}{J. Wan}, \bibinfo{person}{D. Wang},
  \bibinfo{person}{S.~C.~H. Hoi}, \bibinfo{person}{P. Wu}, \bibinfo{person}{J.
  Zhu}, \bibinfo{person}{Y. Zhang}, {and} \bibinfo{person}{J. Li}.}
  \bibinfo{year}{2014}\natexlab{}.
\newblock \showarticletitle{Deep Learning for Content-Based Image Retrieval: A
  Comprehensive Study}. In \bibinfo{booktitle}{\emph{Proceedings of the 22nd
  ACM International Conference on Multimedia, {ACMMM}}}.
  \bibinfo{pages}{157–166}.
\newblock


\bibitem[\protect\citeauthoryear{Wang, Xu, Yue, and Wang}{Wang
  et~al\mbox{.}}{2021}]%
        {Wang2021VLDB}
\bibfield{author}{\bibinfo{person}{M. Wang}, \bibinfo{person}{X. Xu},
  \bibinfo{person}{Q. Yue}, {and} \bibinfo{person}{Y. Wang}.}
  \bibinfo{year}{2021}\natexlab{}.
\newblock \showarticletitle{A Comprehensive Survey and Experimental Comparison
  of Graph-Based Approximate Nearest Neighbor Search}.
\newblock \bibinfo{journal}{\emph{Proceedings of the VLDB Endowment}}
  \bibinfo{volume}{14}, \bibinfo{number}{11} (\bibinfo{year}{2021}),
  \bibinfo{pages}{1964–1978}.
\newblock


\bibitem[\protect\citeauthoryear{Wang, Ramanan, and Hebert}{Wang
  et~al\mbox{.}}{2019}]%
        {wang2019metarare}
\bibfield{author}{\bibinfo{person}{Y.-X. Wang}, \bibinfo{person}{D. Ramanan},
  {and} \bibinfo{person}{M. Hebert}.} \bibinfo{year}{2019}\natexlab{}.
\newblock \showarticletitle{Meta-Learning to Detect Rare Objects}. In
  \bibinfo{booktitle}{\emph{2019 IEEE/CVF International Conference on Computer
  Vision, {ICCV}}}. \bibinfo{pages}{9924--9933}.
\newblock


\bibitem[\protect\citeauthoryear{Weiss}{Weiss}{2004}]%
        {weiss2004miningrare}
\bibfield{author}{\bibinfo{person}{G.~M. Weiss}.}
  \bibinfo{year}{2004}\natexlab{}.
\newblock \showarticletitle{Mining with Rarity: A Unifying Framework}.
\newblock \bibinfo{journal}{\emph{SIGKDD Explorations Newsletter}}
  \bibinfo{volume}{6}, \bibinfo{number}{1} (\bibinfo{year}{2004}),
  \bibinfo{pages}{7–19}.
\newblock


\bibitem[\protect\citeauthoryear{Zhang and Zhao}{Zhang and Zhao}{2015}]%
        {zhang2015astronomy}
\bibfield{author}{\bibinfo{person}{Y. Zhang} {and} \bibinfo{person}{Y. Zhao}.}
  \bibinfo{year}{2015}\natexlab{}.
\newblock \showarticletitle{Astronomy in the big data era}.
\newblock \bibinfo{journal}{\emph{Data Science Journal}}  \bibinfo{volume}{14}
  (\bibinfo{year}{2015}), \bibinfo{pages}{11}.
\newblock


\end{thebibliography}

\end{document}